\documentclass[a4paper,11pt]{article}
\usepackage{jheppub}
\usepackage{lineno}
\usepackage{xcolor}
\usepackage{cancel}
\usepackage[normalem]{ulem}
\usepackage{subcaption}
\usepackage{bm}
\usepackage{wrapfig,lipsum,booktabs}

\newcommand{\fig}[1]{{Fig.~\ref{#1}}}

\newcommand{\LL}{{\textrm{L}}}
\newcommand{\HH}{{\textrm{H}}}

\newcommand{\etH}{e_{*}^{\HH}}
\newcommand{\etL}{e_{*}^{\LL}}
\newcommand{\stH}{s_{*}^{\HH}}
\newcommand{\stL}{s_{*}^{\LL}}
\newcommand{\geff}{N}

\title{\boldmath  Non-conformal obstructions to bubble expansion}

\author[a,b,c]{David Mateos,}
\author[d]{Mikel Sanchez-Garitaonandia,}
\author[a,b]{Pedro Taranc\'on-\'Alvarez.}

\affiliation[a]{Departament de F\'\i sica Qu\'antica i Astrof\'\i sica,  Universitat de Barcelona, Mart\'\i\  i Franqu\`es 1, 
\mbox{ES-08028}, Barcelona, Spain.}
\affiliation[b]{Institut de Ci\`encies del Cosmos (ICC),  Universitat de Barcelona, Mart\'\i\  i Franqu\`es 1, 
\mbox{ES-08028}, Barcelona, Spain.}
\affiliation[c]{Instituci\'o Catalana de Recerca i Estudis Avan\c cats (ICREA), Passeig Llu\'\i s Companys 23, \\
\mbox{ES-08010}, Barcelona, Spain.}
\affiliation[d]{School of Mathematics and Statistics, University College Dublin, Dublin, Ireland}

\emailAdd{dmateos@fqa.ub.edu; 
mikel.sanchezgaritaonandia@ucd.ie;
pedro.tarancon@fqa.ub.edu}

\abstract{
We investigate the hydrodynamics of expanding bubbles in first-order phase transitions with non-conformal thermodynamics. We analyze a broad class of equations of state interpolating between bag-model descriptions, commonly used for electroweak transitions, and QCD-like theories. As a concrete benchmark, we determine the bubble solutions for pure SU(3) Yang-Mills theory. We uncover a new set of hydrodynamic obstructions to bubble expansion. These obstructions arise both at the bubble wall and along the fluid flow, and can partially or completely eliminate otherwise allowed solutions. We also identify a new class of solutions, which we dub ``shocked detonations'', consisting of ordinary detonations with an additional shock inserted in the rarefaction wave. As a consequence of these obstructions, the space of admissible bubble wall velocities is significantly constrained, with gaps appearing between different expansion regimes. For example, for QCD-like theories, all detonation solutions, including shocked detonations, are excluded. We show that these effects can strongly impact the kinetic energy budget of the fluid and, therefore, the resulting gravitational-wave signal, potentially suppressing the most efficient configurations. Our results highlight the importance of non-conformal dynamics for accurately modeling phase transitions and the resulting gravitational-wave spectrum. The code used to construct the bubble solutions is publicly available. For completeness, we also show how the interpolation between the bag-model and QCD-like limits can be realized holographically by varying the backreaction of matter fields on the geometry.
}

\begin{document}
\maketitle
\flushbottom
\section{Introduction}
First-order phase transitions (FOPT) provide a unique window into probing physics beyond the Standard Model (SM), since the latter  does not exhibit any phase transitions (PTs) relevant to the early universe \cite{Aoki:2006we,Kajantie:1996mn, Laine:1998vn, Rummukainen:1998as}. In an astrophysical context, neutron star mergers \cite{Casalderrey-Solana:2022rrn} and core-collapse supernovae \cite{Bleau:2026ala} provide promising environments in which nuclear matter may undergo a phase transition. Such transitions typically proceed through the nucleation, expansion, and collision of bubbles, converting released energy into fluid motion that can generate gravitational radiation potentially detectable by future experiments \cite{Witten:1984rs, Kosowsky:1991ua, Kosowsky:1992rz, Kamionkowski:1993fg}. Reviews on these processes can be found in \cite{Caprini:2019egz, Weir:2017wfa, Hindmarsh:2020hop, Athron:2023xlk}.

Extracting meaningful information from a gravitational-wave (GW) signal requires a detailed understanding of its spectrum. Extensive work, combining analytical methods and hydrodynamic simulations, has established a well-developed framework to estimate the signal amplitude. The key parameters are the nucleation temperature, transition rate, bubble wall velocity, speed of sound, and kinetic energy fraction \cite{Caprini:2019egz, Hindmarsh:2016lnk, Hindmarsh:2019phv, Jinno:2020eqg, Jinno:2022mie, RoperPol:2023dzg}. 

Most studies of gravitational radiation from FOPT assume a constant speed of sound. This is  a good approximation for electroweak-like theories, where only a small fraction of the degrees of freedom participate in the transition. In this case, the remaining degrees of freedom effectively act as a thermal bath at nearly constant temperature, leading to an approximately constant speed of sound. Moreover, since most of these degrees of freedom are effectively massless at the transition temperature, the speed of sound is approximately
$c_s^2 \simeq 1/3$ \cite{Giese:2020znk, Tenkanen:2022tly}. We will refer to this type of scenario as characterised by a  ``bag-model-type'' equation of state (EoS). 

In contrast, when a large fraction of the degrees of freedom participates in the transition, the speed of sound can develop a strong dependence on temperature and other thermodynamic variables. A prototypical example arises in Yang--Mills (YM) type theories, for which the region of the phase diagram that contains the phase transition, including metastable and unstable phases, has recently been mapped using the density-of-states method \cite{Springer:2021liy,Mason:2022trc,Mason:2022aka,Springer:2022qos,Springer:2023wok,Lucini:2023irm,Mason:2023ixv,Springer:2023hcc,Bennett:2024bhy,Mason:2024dve,Bennett:2025whm,Zierler:2026hyt}. Similar behavior is expected to arise in Quantum Chromodynamics (QCD) at high baryon density \cite{Stephanov:2004wx,Alford:2007xm,Fukushima:2010bq,Guenther:2022wcr}, as well as in potential dark sectors \cite{Halverson:2020xpg,Reichert:2022naa,Morgante:2022zvc,Zu:2023olm} and more general scenarios beyond the Standard Model \cite{Caprini:2015zlo,Caprini:2019egz,Athron:2023xlk}. In what follows, we refer to the equations of state in all these cases as ``QCD-like'' EoS. A proper understanding of the consequences of a non-constant speed of sound is therefore essential for probing such transitions. In fact, the analysis of \cite{Springer:2021liy,Mason:2022trc,Mason:2022aka,Springer:2022qos,Springer:2023wok,Lucini:2023irm,Mason:2023ixv,Springer:2023hcc,Bennett:2024bhy,Mason:2024dve,Bennett:2025whm,Zierler:2026hyt} reveals two more features of QCD-like transitions that will  play a key role in the dynamics of bubbles nucleated in these transitions. One is a fairly restricted range of allowed supercooling or superheating. The other is the fact that thermodynamic observables are single-valued as a function of the energy density. 

The hydrodynamics of bubble expansion in the presence of these three features  remains largely unexplored. Preliminary studies include analyses based on template models with constant $c_s$  \cite{Giese:2020rtr, Giese:2020znk, Leitao:2014pda}, while small variations in the speed of sound were considered in \cite{Tenkanen:2022tly}. Related features have also been explored within specific holographic models \cite{ Gursoy:2008za,Gursoy:2009jd, Janik:2017ykj, Bea2018,Bea:2021zol, Bea:2022mfb, Janik:2022wsx, Sanchez-Garitaonandia:2023zqz}.

In this work, we perform a systematic study of the novel qualitative features that arise during the expansion of individual bubbles for a broad class of EoS interpolating between the bag-model and QCD-like limits---see Fig.~\ref{fig: g_eff_figure}. Tracking how bubble solutions evolve as the EoS is varied provides new insight into the role played by its different thermodynamic properties. As a concrete benchmark, we determine the bubble solutions for pure $SU(3)$ Yang--Mills (YM) theory, using the EoS of \cite{Lucini:2023irm}. We focus primarily on supercooled bubbles, except in Sec.~\ref{qcdeos}, where some of the results also apply to superheated bubbles \cite{Bea:2024bxu,Barni:2024lkj}.

Our main result is the identification of new hydrodynamic obstructions to bubble expansion that are absent in theories with a weak temperature dependence of the speed of sound, a large available range of supercooling or superheating, and thermodynamic observables that are multi-valued functions of the energy density. As we will show, these obstructions fall into two categories. The first are local obstructions, which originate from the matching conditions at the bubble wall; we will refer to these as ``wall obstructions''. The second are global obstructions, which arise from the impossibility of constructing a continuous fluid profile connecting the wall to the required asymptotic states; we will refer to these as ``flow obstructions''.

The physical role of flow obstructions depends on the type of solution. They play no role for deflagrations. For ordinary detonations, a flow obstruction marks the point beyond which the continuous rarefaction wave cannot be extended; however, the solution can always be continued by inserting a shock into the rarefaction wave, giving rise to a new class of solutions that we dub ``shocked detonations''. For hybrids, by contrast, the flow obstruction genuinely prevents the construction of solutions beyond it, although in practice the second law is more restrictive: once positive entropy production is imposed, the corresponding solutions are excluded before the flow obstruction is reached. 

Depending on the EoS, the combination of wall and flow obstructions can partially or completely eliminate entire classes of solutions. Consequently, they can significantly restrict the range of allowed bubble wall velocities and strongly affect the resulting gravitational-wave (GW) signal, since hybrid solutions, which typically maximize the kinetic energy fraction at fixed transition strength, are among the most severely impacted.

The paper is organized as follows. In Sec.~\ref{sec: section 2} we introduce the family of non-conformal equations of state considered throughout, interpolating between QCD- and bag-model-like behaviour. In Sec.~\ref{sec: section 3} we review the construction of single-bubble, self-similar flows. In Sec.~\ref{qcdeos} we analyze the bubble solutions for QCD-like equations of state, using a holographic model and pure $SU(3)$ Yang--Mills theory as concrete benchmarks, and show that detonations are excluded. In Sec.~\ref{sec: section 4} we identify the new hydrodynamic obstructions to bubble expansion, which arise both at the wall and along the flow. We then show that a flow obstruction of this kind gives rise to a new class of solutions, which we dub shocked detonations. In Sec.~\ref{space_of_sols} we study the impact of these obstructions on the space of allowed bubble solutions across the family of EoS, and in Sec.~\ref{sec: section 6} we analyze the resulting kinetic energy fraction. We conclude in Sec.~\ref{sec: section 7} with a discussion and an outlook.

Although holography plays no role in our analysis, we explain for completeness in Appendix~\ref{app: backreaction} how, in a holographic model, varying the backreaction of the matter sector on the five-dimensional metric interpolates between a QCD-like EoS and a bag model.

The numerical results presented in this work were obtained using \href{https://github.com/pedrota2000/SNOBEX}{\texttt{SNOBEX}} (\textbf{S}elf-similar \textbf{N}on-conformal \textbf{O}bstructions to \textbf{B}ubble \textbf{EX}pansion), a Python code developed alongside this paper and publicly available at the following \href{https://github.com/pedrota2000/SNOBEX}{GitHub repository}. The code solves for self-similar, single-bubble flows across a broad class of equations of state, computing deflagration, detonation, hybrid and shocked detonation solutions together with the hydrodynamic obstructions that constrain them. A detailed description of its structure and numerical implementation is provided in Appendix~\ref{app: snobex}.

\section{Non-conformal equations of state}
\label{sec: section 2}
The class of EoS that we use to draw our conclusions interpolates between QCD-like and bag-model-like EoS. We therefore begin by describing the former.

\subsection{QCD-like equations of state}
The holographic model of \cite{Bea2018}, with parameters $\phi_Q = 10$ and $\phi_M = 1.0$, provides an example of a QCD-like EoS \cite{Springer:2021liy,Mason:2022trc,Mason:2022aka,Springer:2022qos,Springer:2023wok,Lucini:2023irm,Mason:2023ixv,Springer:2023hcc,Bennett:2024bhy,Mason:2024dve,Bennett:2025whm,Zierler:2026hyt}, up to possible differences at low temperatures discussed below. We emphasize, however, that this model merely serves as a convenient example exhibiting the relevant features that we now describe, and that holography plays no role in our results. 

\begin{figure}[tp]
    \centering
    \includegraphics[width = 1.0\linewidth]{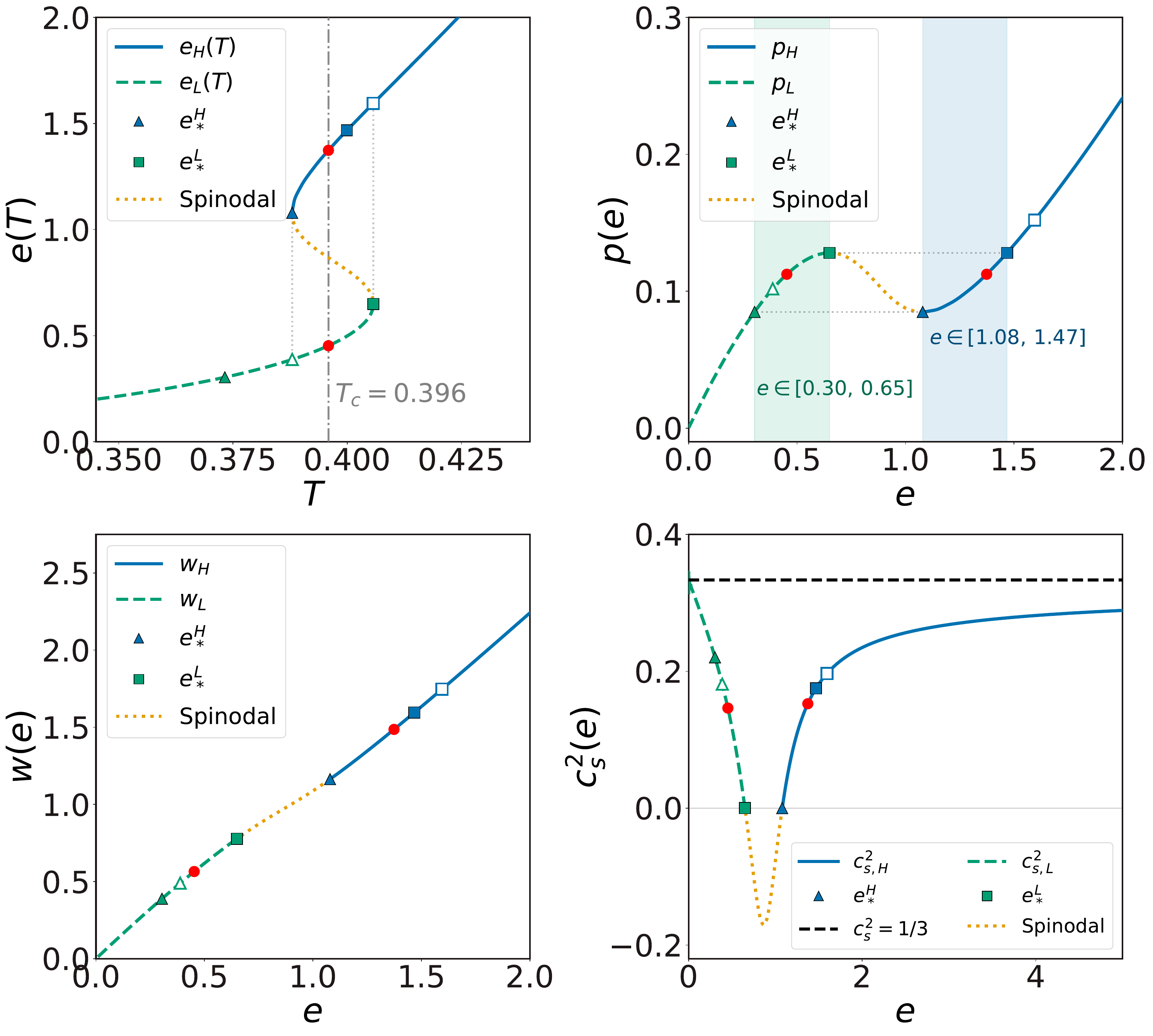}
    \caption{QCD-like EoS extracted from the model of \cite{Bea2018}, with parameters $\phi_Q = 10$ and $\phi_M = 1.0$. Top-left: energy density as a
    function of temperature. Top-right: pressure as a function of energy density. Bottom-left: enthalpy as a function of energy density. Bottom-right: speed of sound squared as a function of energy density; the black, horizontal dashed line marks the conformal value $c_s^2 = 1/3$. States shown in solid blue and dashed green are locally stable, whereas those in dotted orange are locally unstable ($c_s^2 < 0$). The critical temperature is $T_c = 0.396 \Lambda$.}
    \label{fig: EoS-holography}
\end{figure}

\begin{figure}[tp]
    \centering
    \includegraphics[width=0.95\linewidth]{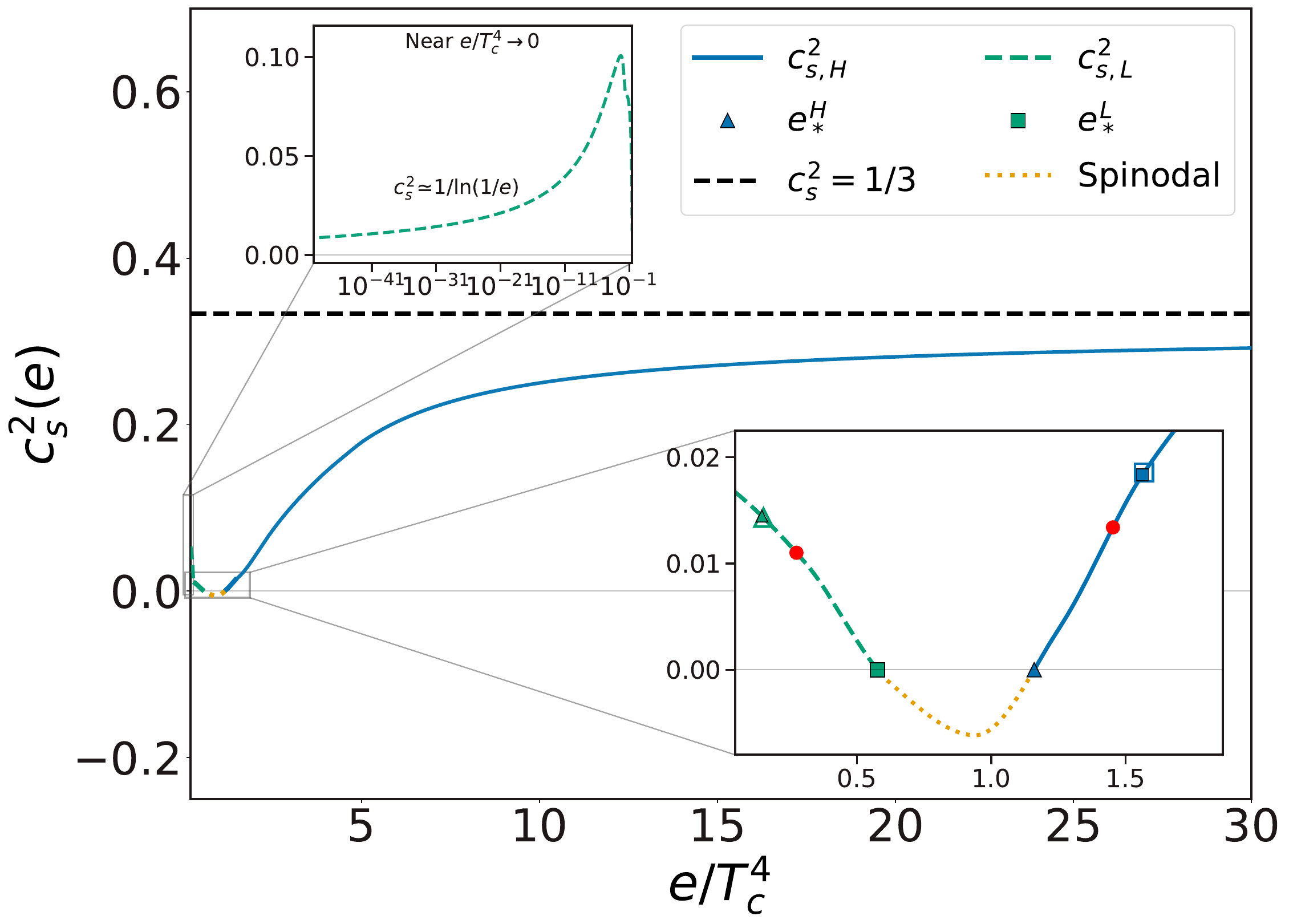}
    \caption{Speed of sound squared $c_s^2$ as a function of energy density normalized to the critical temperature in $SU(3)$ pure YM theory. States shown in solid blue and dashed green are locally stable, whereas those in dotted orange are locally unstable ($c_s^2 < 0$). Insets: small-energy behaviour $c_s^2\approx 1/\ln(1/e)$ (top-left) and a zoom of the transition (bottom-right).}
    \label{fig: cs2_holography}
\end{figure}

The energy density $e$, the pressure $p$,  the enthalpy $w$, and the speed of sound squared of the model $c_s^2$, are shown in \fig{fig: EoS-holography}. The EoS consists of two locally stable branches, shown as a solid blue curve and a dashed green curve, connected by an intermediate locally unstable branch, shown as a dotted orange curve, known as the spinodal branch. The critical temperature is $T_c=0.396 \Lambda$, where $\Lambda$ is an intrinsic scale of the theory. From now on we set
\begin{equation}
\label{lamda}
\Lambda = 1 \,,
\end{equation}
which means that all dimensionful quantities are measured in units of this scale, except in
\fig{fig: cs2_holography}, where we normalize the energy density to the critical temperature. 

The existence of multiple states at a given temperature over a finite temperature range is a universal feature of first-order phase transitions. By contrast, the behaviour as a function of the energy density distinguishes a QCD-like EoS from a bag model. In the QCD-like case, there is a single equilibrium state for each value of the energy density. As a consequence, all thermodynamic quantities are single-valued functions of the energy density, as illustrated in \fig{fig: EoS-holography}. The pressure as a function of the energy density, shown in the top-right panel, is often referred to as the ``reduced EoS'', since it contains less information than the full EoS---see e.g.~the discussion in \cite{Bea:2024bxu}. We will return to this point below.

The speed of sound squared is defined as 
\begin{equation}
c_s^2 = \frac{dp}{de} = \frac{s}{c_v} \,,
\end{equation}
where $s$ is the entropy density and $c_v$ the specific heat. As expected, $c_s^2$ becomes negative along the intermediate unstable branch, where $c_v<0$, and asymptotically approaches the conformal value $c_s^2=1/3$ at both high and low energies. In the holographic EoS, these limiting values arise because the model interpolates between ultraviolet (UV) and infrared (IR) fixed points.

It is instructive to compare this behaviour with that expected in a QCD-like theory. To this end, in \fig{fig: cs2_holography} we also show the speed of sound in pure $SU(3)$ YM theory \cite{Castorina:2009de,Vovchenko:2014pka,Khaidukov:2018lor,Giusti:2025fxu,Mason:2022trc,Mason:2022aka,Lucini:2023irm}. The high-energy part of the curve is taken from \cite{Giusti:2025fxu}, while the region around the transition, shown in detail in the inset on the right-hand side, is taken from \cite{Lucini:2023irm,private}.\footnote{These results still suffer from some lattice effects \cite{Lucini:2023irm}.} We fix the undetermined additive constant in the entropy density of \cite{Lucini:2023irm} by requiring that the speed of sound at the critical temperature on the high-energy branch match that of \cite{Giusti:2025fxu}.
Although the qualitative forms are similar, both the extent of the transition region and the magnitude of the dip in YM are much smaller than in the model of \fig{fig: EoS-holography} \cite{Springer:2021liy,Mason:2022trc,Mason:2022aka,Springer:2022qos,Springer:2023wok,Lucini:2023irm,Mason:2023ixv,Springer:2023hcc,Bennett:2024bhy,Mason:2024dve,Bennett:2025whm,Zierler:2026hyt}---we will return to this point below. The low-energy part of the YM curve corresponds to a physical glueball resonance gas \cite{Agasian:2017tag}, including four resonances with masses taken from \cite{Athenodorou:2020ani}.

The physics behind the speed of sound in pure $SU(3)$ YM is as follows. At high energies, asymptotic freedom implies that the theory approaches a Gaussian conformal fixed point, and hence $c_s^2\to 1/3$. As the energy density decreases, the speed of sound decreases and becomes negative along the spinodal branch, as shown in the inset on the right-hand side of \fig{fig: cs2_holography}. As the energy density decreases further and the system enters the confining phase, the speed of sound rises again, reaches a maximum, and eventually vanishes as the energy density approaches zero, as illustrated in the upper inset. This behaviour reflects the presence of a mass gap, $m_{\rm gap}$: at sufficiently low energies, the thermodynamics is dominated by a dilute gas of massive, non-relativistic excitations, for which the pressure is parametrically suppressed relative to the energy density. In pure YM theory, these excitations are glueballs, and the mass gap is the mass of the lightest glueball. As the temperature approaches zero, the speed of sound vanishes linearly with $T$ \cite{Castorina:2009de,Khaidukov:2018lor}. Since \mbox{$e\sim m_{\rm gap}^2 T^2 \exp(-m_{\rm gap}/T)$}, this implies that $c_s^2$ vanishes  logarithmically slowly as a function of the energy density, as illustrated by the upper inset of \fig{fig: cs2_holography}.

Crucially for our analysis, in the hydrodynamic limit the possible bubble solutions depend only on the properties of the EoS over a limited range of energy densities around the phase transition. First, the locally unstable branch plays no role: because of the spinodal instability, no steady-state bubble flow described by ideal hydrodynamics can probe such states (see e.g.~\cite{Bea:2021zol}). Although the microscopic wall profile may pass through this region, the self-similar hydrodynamic description treats the wall as a discontinuity. The spinodal branch is therefore irrelevant for our analysis. Accordingly, in some of the plots below we omit this branch, in particular the portions of the speed-of-sound curves for which $c_s^2<0$.

Second, only the properties of the EoS in the shaded regions of \fig{fig: EoS-holography} top-right can affect the possible bubble solutions. To see this, let us first define the special points shown in \fig{fig: EoS-holography}. The red dots mark the phase transition: a pair of points, one on the high-energy branch and one on the low-energy branch, with the same critical temperature and pressure, as illustrated in the two top panels of the figure. The filled blue triangle and filled green square indicate the turning points of the EoS in the $e-T$ plane, as shown in the top-left panel, or equivalently the maximum and minimum of the pressure in the $p-e$ plane, as shown in the top-right panel. These two points can be projected onto the opposite branches either at constant temperature or at constant pressure. The constant-temperature projections are the hollow green triangle and the hollow blue square in the top-left panel, while the constant-pressure projections are the filled green triangle and the filled blue square in the top-right panel.

With these definitions in hand, we can now understand which part of the EoS can affect the possible bubble solutions in the hydrodynamic description. This follows from two observations. The first is that, in order to construct mathematical solutions, only the reduced EoS $p(e)$ is needed. (Verifying that these solutions obey the second law requires additional information, as discussed below.) However, if only $p(e)$  in the locally stable states is specified, then  the critical temperature and corresponding pressure, namely the location of the red dots, are not fixed. The reduced EoS only constrains these points to lie between the filled green triangle and green square, and between the filled blue triangle and blue square, respectively, since only pairs of points in these regions can have the same pressure. These are precisely the points for which a bubble with zero wall velocity is possible, corresponding to a phase-coexistence state. The range of energies covered by these points is indicated by the shaded green and blue bands in the top-right panel of \fig{fig: EoS-holography}.

For QCD-like EoS, the second observation is that the largest range is explored by bubbles with arbitrarily small wall velocity. In other words, the relevant region of the EoS shrinks as the wall velocity increases. This is not obvious a priori, but it will emerge from our numerical analysis. It therefore follows that, in this case, the maximum range of the EoS relevant for bubble solutions is the same as the range that determines phase-coexistence solutions. Strictly speaking, there will be some deflagration and hybrid flows that will explore energies larger than the one given by the blue square. However, in the high energy region our parameterization of the EoS is qualitatively similar to pure YM $SU(3)$. There will be no supercooled bubble flow exploring energies smaller than the one given by the green triangle. The bag-model-like case will be discussed in Sec.~\ref{space_of_sols}.

These ranges for the EoS of pure $SU(3)$ YM \cite{Lucini:2023irm} are indicated by the blue squares and green triangles in the inset on the right-hand side of \fig{fig: cs2_holography}. Note that the hollow and filled markers lie almost on top of one another. Their proximity to the points where the speed of sound becomes negative implies that the form of the EoS elsewhere plays no role in determining the possible bubble solutions for this theory. Thus, the fact that in our models the speed of sound rises again to $1/3$ at low energies, whereas in pure $SU(3)$ YM it reaches a maximum and then decreases again, is irrelevant for understanding the possible bubble solutions in the latter theory. This difference could play a role, however, in more general QCD-like theories. Although our analysis could be straightforwardly adapted to such cases, we leave this extension to future work.

For completeness, we recall that real-world QCD at zero chemical potential does not exhibit a true phase transition, but rather a crossover. In this case, the speed of sound has the same qualitative form as in pure YM, with two important differences: first, it develops a dip but never becomes negative; second, its maximum in the hadronic phase is somewhat higher than in YM, due to the presence of light pions.

\subsection{Bag-model-like equations of state}
The phase transition presented above differs qualitatively from an electroweak-like phase transition. In the latter, there is  a large number of degrees of freedom  that are effectively massless around, and do not participate in, the phase transition. They thus act as a thermal bath whose contribution to the pressure dominates and drives the speed of sound toward $c_s^2 \simeq 1/3$ \cite{Tenkanen:2022tly}. In other words, the pressure can be split as
\begin{equation}
    p = \mathcal{P}(T) +\frac{1}{3} N T^4,
    \label{eq: Radiation + potential EoS}
\end{equation}
where $\mathcal{P}$ is the contribution to the pressure of the sector participating in the phase transition and $N$ is (proportional to) the number of relativistic degrees of freedom weakly coupled to the previous ones, not participating in the transition. Qualitatively, $\mathcal{P}$  takes the  form described in the previous section and can be understood as the on-shell effective potential for the order parameter which, in an electroweak-like theory, can be identified with the Higgs expectation value.
When $N\gg 1$, as in electroweak-like theories, the combination \eqref{eq: Radiation + potential EoS} approaches the EoS of a bag model. The latter is defined by the fact that the pressures in each branch take the form
\begin{equation}
    p_{\LL}(T) = \frac{1}{3}a_{\LL} \, T^4,\quad p_{\HH}(T) = \frac{1}{3}a_{\HH}  \, T^4 -\epsilon \,,
    \label{eq: Bag-Model-pressure}
\end{equation}
with $\epsilon$ the bag constant. 
This leads to the reduced EoS
\begin{equation}
\label{redu}
    p_{\LL}(e) = \frac{1}{3}e,\quad p_{\HH}(e) = \frac{1}{3}e -\frac{4\epsilon}{3}\,,
\end{equation}
which implies a constant speed of sound in both branches given by $c_s^2=1/3$. Note that the true EoS \eqref{eq: Bag-Model-pressure} contains information about the number of degrees of freedom in each branch, which is proportional to $a_{\LL,\HH}$, whereas this information is absent in the reduced EoS \eqref{redu}. 

\begin{figure}[tp]
    \centering
    \includegraphics[width=1.0\linewidth]{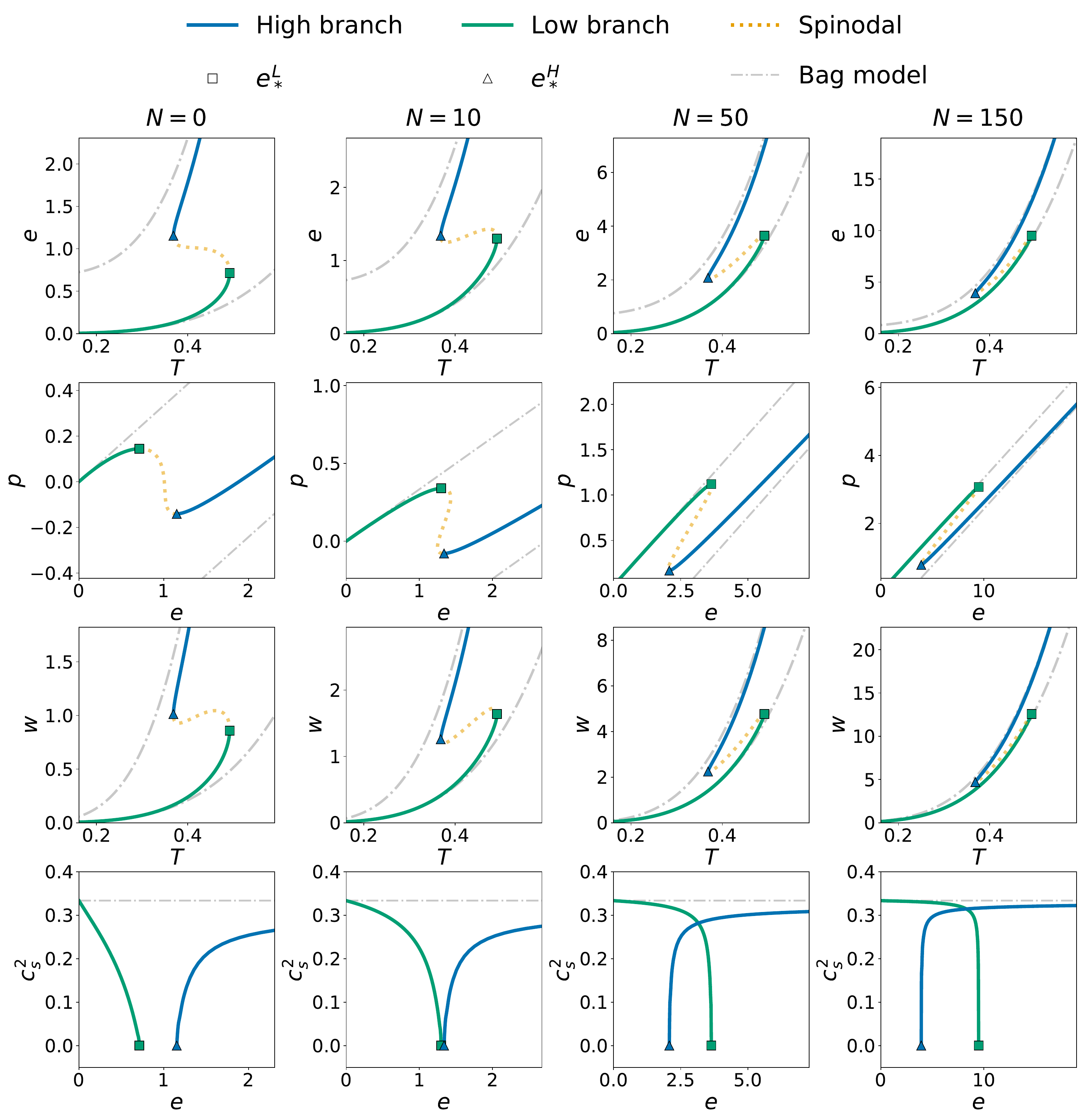}
    \caption{Thermodynamic relations for the family of EoS defined in \eqref{eq: Radiation + potential EoS}, taking $\mathcal{P}(T)$ to be  QCD-like. The top row shows the equation of state, $e(T)$, while the second row displays the reduced equation of state, $p(e)$. The third row presents the enthalpy density as a function of temperature, $w(T)=e(T)+p(T)$, and the bottom row shows the speed of sound squared, $c_s^2 = \partial p/\partial e$, as a function of the energy density, excluding the part of the curves where the speed of sound becomes negative. Results are shown for different values of the number of degrees of freedom that do not participate in the phase transition, $N$. As $N$ increases, the system smoothly interpolates between a strongly non-conformal regime and behaviour characteristic of a bag-model equation of state. This is illustrated by the dot-dashed curves, which are obtained by fitting the constants $a_\LL, a_\HH$ and $\epsilon$ so that the high- and low-energy branches of the bag model agree with the high- and low-temperature limits of the EoS \eqref{eq: Radiation + potential EoS}.}
    \label{fig: g_eff_figure}
\end{figure}

In \fig{fig: g_eff_figure} we show the EoS \eqref{eq: Radiation + potential EoS} for different values of $N$, taking $\mathcal{P}(T)$ to correspond to the pressure of a QCD-like theory. To improve the visualization in the  plots, we extract $\mathcal{P}(T)$ from the same holographic model of \cite{Bea2018}, but using the parameters $\phi_Q = 10$ and $\phi_M = 0.85$, which differ from those employed in \fig{fig: EoS-holography}.

As anticipated, for $N\gg 1$ the high- and low-energy branches of the full EoS approach those of a bag model. We also observe that, in contrast to the QCD-like case, in this limit there are multiple equilibrium states for certain values of the energy density. We will come back to this feature below, as it plays a crucial role in determining the space of allowed bubble solutions. 

We close this section with a comment on the connection to holography. The same splitting as in Eq.~\eqref{eq: Radiation + potential EoS} can be implemented on the gravitational side by varying the backreaction of the matter sector on the five-dimensional metric. Since this connection is not central to the present paper, we defer a detailed discussion to Appendix~\ref{app: backreaction}.

\subsection{Equation of state ansatz}
As we are interested in exploring the new features that arise in bubble expansion as a consequence of non-conformality, we wish to scan over models that interpolate between strongly non-conformal behaviour and bag-model-like EoS.
To this end, we will work with an analytic ansatz for the EoS, rather than deriving it from first principles, while incorporating a number of general features inspired by the example in \fig{fig: EoS-holography}. Specifically, we require a finite amount of supercooling and superheating, a vanishing speed of sound at the points of maximal supercooling and superheating, and an asymptotically conformal equation of state at both high and low temperatures. The latter property is also present in the model of \fig{fig: EoS-holography}. Here it is primarily adopted to constrain the general ansatz for the EoS, but it is not essential and plays no role in the results that follow. Finally, we will disregard the locally unstable branch and instead work with two stable branches: a high-energy (\(\HH\)) branch and a low-energy (\(\LL\)) branch.\footnote{It is common to refer to the high-energy phase as the symmetric phase and to the low-energy phase as the broken phase. Since our discussion does not rely on the presence of a symmetry-breaking order parameter, we will instead refer to the phases according to their energy hierarchy at the critical temperature, 
\(T_c\).}

\begin{figure}[tp]
    \centering
    \includegraphics[width=1.0\linewidth]{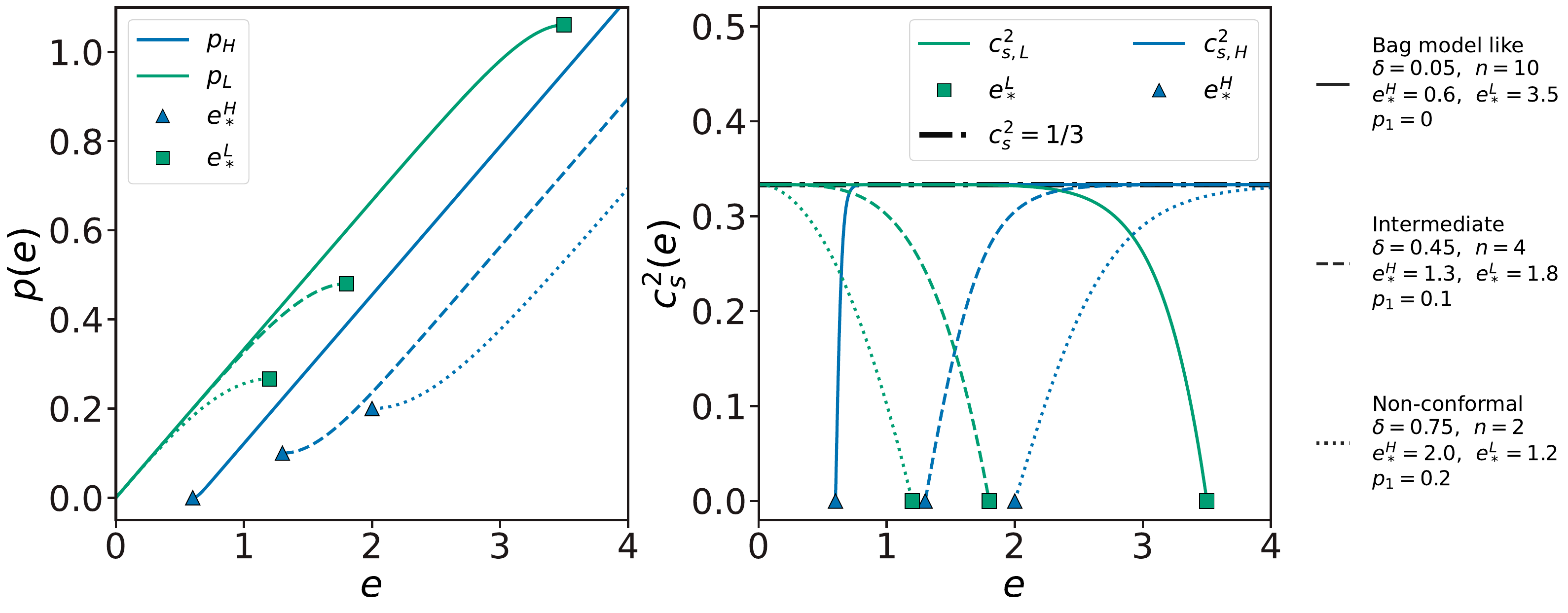}
    \caption{Pressure (left) and speed of sound squared (right) as functions of the energy density for different choices of parameters in \eqref{eq: EoS-p}. The turning points $\etL$ and $\etH$ are indicated for each model. We see that this parametrization of the EoS is flexible enough to reproduce both bag-model-like equations of state and QCD-like non-conformal ones.}
    \label{fig: EoS}
\end{figure}

Our EoS ansatz satisfying all the requirements above is
\begin{equation}
\begin{aligned}
        p_{\LL}(e) & = \frac{1}{3} \left( e - \frac{e^{\,n+1}}{(n+1)\,(\etL)^{\,n}} 
        \right),\\[2mm]
        p_{\HH}(e) \, & =\, \frac{1}{3}\,\delta \ \log\left[\cosh\left(\frac{e-\etH}{\delta}\right)\right] + p_1 \,.
\end{aligned}
\label{eq: EoS-p}
\end{equation}
In this equation, $p_{\LL}, p_{\HH}, p_1, e, \etL, \etH$ and $\delta$ are dimensionful parameters that we make effectively dimensionless by measuring them in units of an intrinsic scale that we set to unity, as in Eq.~\eqref{lamda}. The parameter $n$ is directly dimensionless. The intuition for the role of all these  parameters is as follows. The quantities \(\etL\) and \(\etH\) determine the location of the turning points of the low- and high-energy branches, respectively. At these points the speed of sound vanishes: 
\begin{equation}
    \left.\frac{d p_{\LL}}{de}\right|_{e=\etL}
    =
    \left.\frac{d p_{\HH}}{de}\right|_{e=\etH}
    =
    0\,.
\end{equation}
The parameter \(p_1\) fixes the value of the pressure at the high-energy turning point,
\begin{equation}
    p_{\HH}(\etH) = p_1,
\end{equation}
and therefore corresponds to a vertical shift of the high-energy branch in the $p-e$ plane. 
By construction, the speed of sound  asymptotically approaches  its conformal value at low and high energies:
\begin{equation}
    \lim_{e\to 0}\frac{d p_{\LL}}{de}
    =
    \lim_{e\to \infty}\frac{d p_{\HH}}{de}
    =
    \frac{1}{3}.
\end{equation}
 Finally, the parameters \(\delta\) and \(n\) control how rapidly the speed of sound varies with the energy density. Increasing \(n\) makes the variation of the speed of sound in the low-energy phase sharper, while an analogous effect in the high-energy phase is obtained by decreasing \(\delta\). In \fig{fig: EoS} we show the resulting forms of \(p(e)\) and \(c_s^2(e)\) for representative values of the parameters \(\etH\), \(\etL\), and \(p_1\). For the values corresponding to the solid line, the EoS shares the qualitative features of that in  \fig{fig: EoS-holography}: fairly limited amounts of supercooling and superheating, single-valuedness as a function of the energy density, and  a speed of sound that interpolates between the conformal value, \(c_s^2 = 1/3\), and zero in a manner qualitatively similar to that shown in \fig{fig: EoS-holography}.

The bag-model limit is approached 
by making $\etL$ and $\etH$ sufficiently large and small, respectively. Under these circumstances,  the EoS \eqref{eq: EoS-p} resembles that of a bag model in the large range $\etL \gg e \gg \etH$, with bag constant such that 
\begin{equation}
    p_1 = -\frac{4\epsilon}{3} + \frac{e_*^H}{3} + \frac{\delta \log 2}{3}\,.
\end{equation}
In this scenario, deviations from a constant speed of sound are localized to the regions $e \sim \etH$ and $e \sim \etL$. This is qualitatively similar to what occurs in \eqref{eq: Radiation + potential EoS} for
$N \gg 1$, 
where the EoS also departs from the conformal value $c_s^2 = 1/3$ when zooming into the vicinity of the turning points (see \fig{fig: g_eff_figure}).

Before turning to the study of bubble expansion, it is useful to define an entropy function in order to determine which hydrodynamic constraints are genuinely more restrictive than the second law of thermodynamics. Since in \eqref{eq: EoS-p} we have only specified the reduced EoS, $p(e)$, the entropy is not uniquely determined but depends on two additional integration constants, one for each branch. Indeed, we can obtain the entropy density by integrating a first-order differential equation that follows from the definition of the enthalpy,
\begin{equation}
e + p(e) = Ts
\quad \Rightarrow \quad
\frac{1}{s}\frac{ds}{de} = \frac{1}{e+p(e)} \,,
\label{eq: s(e)-ODE}
\end{equation}
where we used the thermodynamic identity $T = de/ds$. This equation can be integrated independently for each branch of the EoS, $\HH$ and $\LL$, once appropriate boundary conditions are specified. We choose these to be the values of the entropy at the turning points:
\begin{equation}
\stH \equiv s_{\HH}(\etH) \,,
\qquad
\stL \equiv s_{\LL}(\etL) \,.
\end{equation} 
The entropy density is therefore characterized by the parameters already appearing in \eqref{eq: EoS-p}, together with the additional constants $\stH$ and $\stL$. These encode the information that is lost when specifying only the reduced EoS, namely the effective number of degrees of freedom associated with each branch, which is proportional to $\stH$ and $\stL$, respectively.

\section{Hydrodynamics of 
expanding bubbles}
\label{sec: section 3}
At late times, the expansion of a bubble is well described by a perfect fluid, since the large size of the bubble suppresses gradient corrections. The equations of motion are given by the conservation of the stress-energy tensor,
\begin{equation}
    \nabla_{\mu}T^{\mu\nu} = 0,\quad  T^{\mu\nu} = w u^{\mu}u^{\nu} + p g^{\mu\nu}\,,
    \label{eq:ideal-hydro}
\end{equation}
where $u^{\mu} = \gamma(1,\vec v)$ is the fluid 4-velocity,  $\gamma = 1/\sqrt{1-v^2}$, and $g_{\mu\nu}$ is the metric of the background geometry, which we set to be flat with signature $(-,+,+,+)$. At these scales, the only relevant microscopic information is the reduced EoS, relating the pressure and the energy density $p = p(e)$, which closes the system of equations. Furthermore, the absence of intrinsic macroscopic scales implies that bubble growth is self-similar at late times, that is, all quantities depend solely on the ratio $\xi = r/t$, where $r$ is the distance to the bubble center. This property simplifies the problem substantially.

Solving \eqref{eq:ideal-hydro} has been the subject of extensive research---see e.g.~\cite{Espinosa2010, Barni:2024lkj, Bea:2024bxu}. This analysis typically begins by deriving a system of equations for the enthalpy and fluid velocity by projecting the stress-tensor conservation along, and perpendicular to, the fluid flow, leading to
\begin{equation}
    \begin{aligned}
        \gamma^2(1-\xi v)\left(\frac{\mu(\xi,v)^2}{c_s^2}-1\right)\partial_{\xi}v = (d-1)\frac{v}{\xi}
        \,,\\[2mm]
        \frac{\partial_{\xi}w}{w} = \gamma^2\left(1+\frac{1}{c_s^2}\right)\mu(\xi,v)\partial_{\xi}v \,,
    \end{aligned}
    \label{eq:self-similar-eoms}
\end{equation}
where $v$ is the radial fluid velocity, 
\begin{equation}
    \mu(\xi,v) =- \frac{v- \xi}{1-v\xi}
\end{equation}
is the Lorentz-boosted fluid velocity in a frame moving with velocity $\xi$ (note the overall minus sign convention), and $d$ is the number of spatial dimensions, which we set to $d=3$ henceforth. For notational brevity, we write $\mu(\xi,v)$ as $\mu(\xi)$ from this point onward, with the understanding that $v$ is always evaluated at $v(\xi)$.

A crucial feature of Eqs.~\eqref{eq:self-similar-eoms} is that, when the speed of sound is constant---as in the bag-model 
EoS---the system decouples into a nested set of equations. In contrast, for a non-constant speed of sound the two equations remain fully coupled through $c_s(e)$. This underlies many of the qualitative differences that we will find between non-conformal models and the bag model.

As mentioned, at late times bubbles are large and gradients are suppressed. This approximation only fails at the bubble wall and at possible shocks formed in the evolution, where gradients remain large.  However, given their negligible width in the self-similar variable $\xi$, they can be described as discontinuities. Specifically, without invoking further microscopic information, one can still impose so-called ``junction conditions'' arising from integrating the stress-tensor conservation equation in the radial direction, leading to
\begin{equation}
    w_+v_+^2\gamma_+^2+p_+ = w_-v_-^2\gamma_-^2 +  p_- \,, \quad w_+v_+\gamma_+^2 = w_-v_-\gamma_-^2,
    \label{eq:matching-original}
\end{equation}
which are often rewritten in the more familiar form
\begin{equation}
    v_+v_- = \frac{p_+-p_-}{e_+-e_-}\,, \quad \frac{v_+}{v_-} = \frac{e_-+p_+}{e_++p_-}.
    \label{eq:matching-usual}
\end{equation}
Here, the `$\pm$' subscripts refer to the fluid quantities immediately ahead of and behind the discontinuity, respectively, and all velocities are measured in the rest frame of the discontinuity. In this frame, both fluid velocities are negative. Since they share the same sign, however, we follow the customary convention in which $v_\pm$ denotes the absolute value of the fluid velocity on either side of the wall, so that both quantities are strictly positive. Note that this choice of convention leaves \eqref{eq:matching-usual} unchanged. 

Given an EoS, one can solve Eqs.~\eqref{eq:self-similar-eoms} together with \eqref{eq:matching-usual} in order to obtain the different types of bubble expansions. These are characterized by the  values of two parameters,  $(\xi_w,\alpha_N)$, with $\xi_w$ the wall velocity and $\alpha_N = \alpha(T_N)$ the phase transition strength at the nucleation temperature.\footnote{Although we have found that $\alpha_N$ need not be single-valued for some QCD-like EoS, for the family of EoS considered in Sec.~\ref{space_of_sols}, $\alpha_N(T)$ remains single-valued over the temperature range relevant for bubble solutions, so the transition strength is unambiguously fixed by the nucleation temperature.} This is defined as \cite{Ares:2020lbt}
\begin{equation}\label{eq: alphadef}
    \alpha(T) = \left.\frac{4}{3}\frac{\theta_{\HH}-\theta_{\LL}}{w_{\HH}}\right\vert_{T} \,,
\end{equation}
where the subscripts $H$ and $L$ refer to the high- and low-energy phases, and $\theta$ is the trace of the stress tensor
\begin{equation}
    \theta = \frac{1}{4}\left(e-3p\right).
    \label{eq:trace-theta}
\end{equation}
In the case of a bag equation of state \eqref{eq: Bag-Model-pressure}, the transition strength reduces to the familiar expression
\begin{equation}\label{eq: transition strength}
    \alpha(T) = \frac{\epsilon}{a_{\HH} T^4}.
\end{equation}

From a microscopic perspective, the wall velocity and the transition strength are not independent parameters. Hydrodynamics alone, however, does not determine the relation between them; rather, it constrains the region of the $(\xi_w,\alpha_N)$ plane in which consistent bubble solutions can exist. This yields non-trivial constraints on the possible bubble configurations without requiring additional microscopic input. In the next section, we illustrate these constraints in the regime where they become most restrictive, namely for a QCD-like EoS. Then, in Sec.~\ref{space_of_sols}, we perform a systematic analysis for the general class of EoS defined in \eqref{eq: EoS-p}, with the goal of understanding how the space of allowed bubble solutions interpolates between the bag-model limit and the QCD-like regime.

\section{Bubble solutions for a QCD-like EoS}
\label{qcdeos}
Here we will analyze the bubble solutions in the regime where the difference with a bag model is most dramatic. To this end, we will consider the EoS introduced in \fig{fig: EoS-holography}. We will see that the EoS alone excludes the existence of detonations completely,   both in supercooled and in superheated transitions. 

\begin{figure}[tp]
    \centering
    \includegraphics[width=0.9\textwidth]{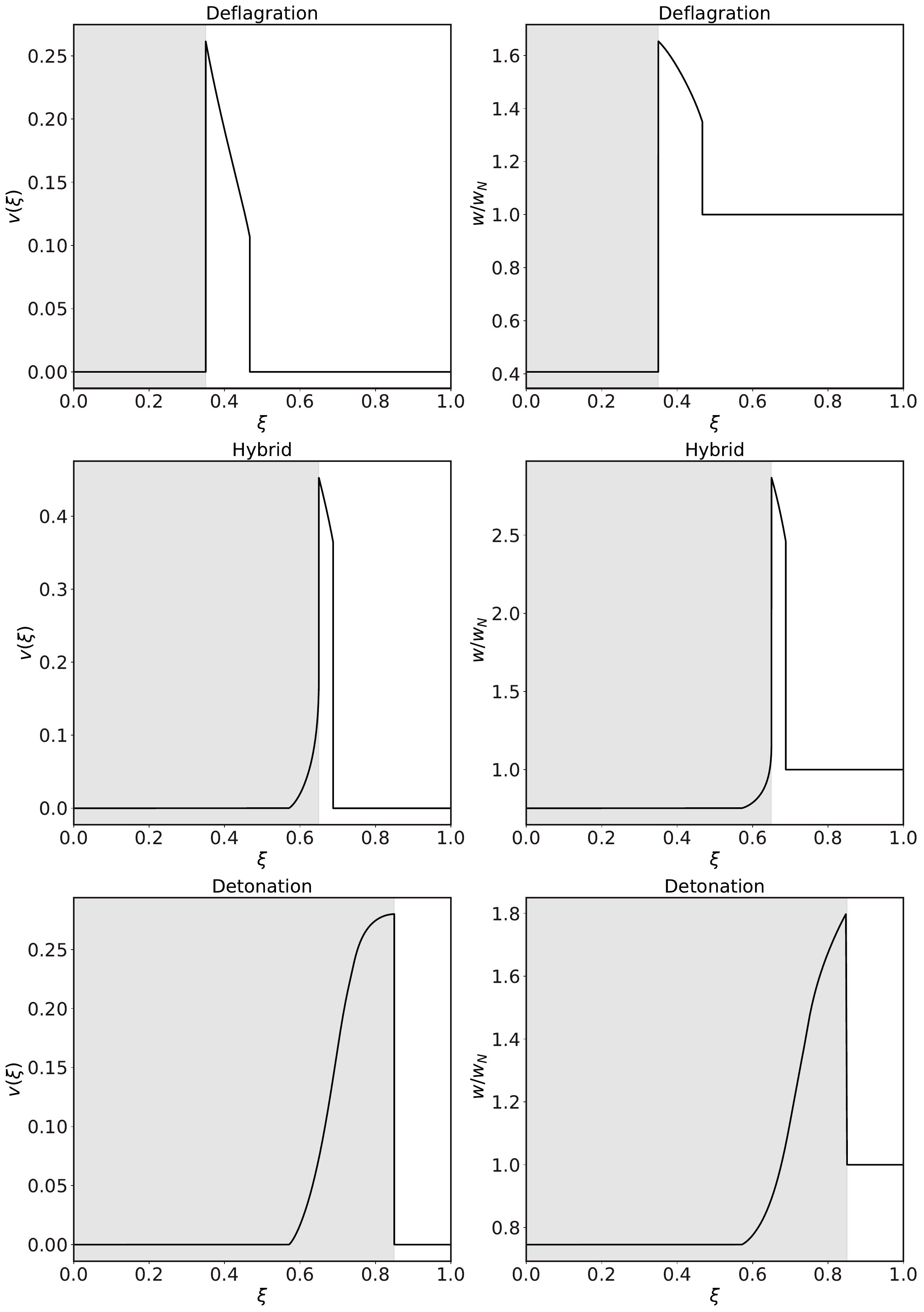}
    \caption{Velocity in the bubble rest frame (left column) and enthalpy, normalized to the nucleation enthalpy (right column), as functions of the self-similar variable for deflagration, hybrid, and detonation solutions. The region behind the wall is shaded in gray. The EoS used is given by $p_1=0$, $n=4$, $\delta=1$, $\etH=0.5$, and $\etL=2.0$ in \eqref{eq: EoS-p}.}
    \label{fig: three flows}
\end{figure}

Consider the supercooled case first. In this case, detonations are characterized by the condition 
\begin{equation}
\label{vv}
    v_+ \geq v_- \,,
\end{equation}
since the fluid immediately in front of the wall is at rest, whereas the fluid immediately behind the wall is moving in the same direction as the wall---see \fig{fig: three flows}. Recall that, in our conventions, $v_\pm$ denote the absolute values of the corresponding velocities. 
The two equations in \eqref{eq:matching-original} then  lead to two necessary conditions for detonations to exist: 
\begin{align}
    p_- &\geq p_+ \,, \label{pp}\\
    w_- &\geq w_+ \,. \label{ww}
\end{align}
Since $v_+v_- > 0$, substituting \eqref{pp} in the first equation in \eqref{eq:matching-usual} implies that 
\begin{equation}
\label{ee}
    e_- \geq e_+ \,.
\end{equation}
Since $v_+v_- < 1$, the above equations together with \eqref{eq:matching-usual} imply that 
\begin{equation}
\label{delta_e}
    e_- - e_+ \geq p_- - p_+ \,,
\end{equation}
meaning that the jump in energy density across the wall is larger than the jump in pressure. 

\begin{figure}[tp]
    \centering
    \includegraphics[width=0.7\linewidth]{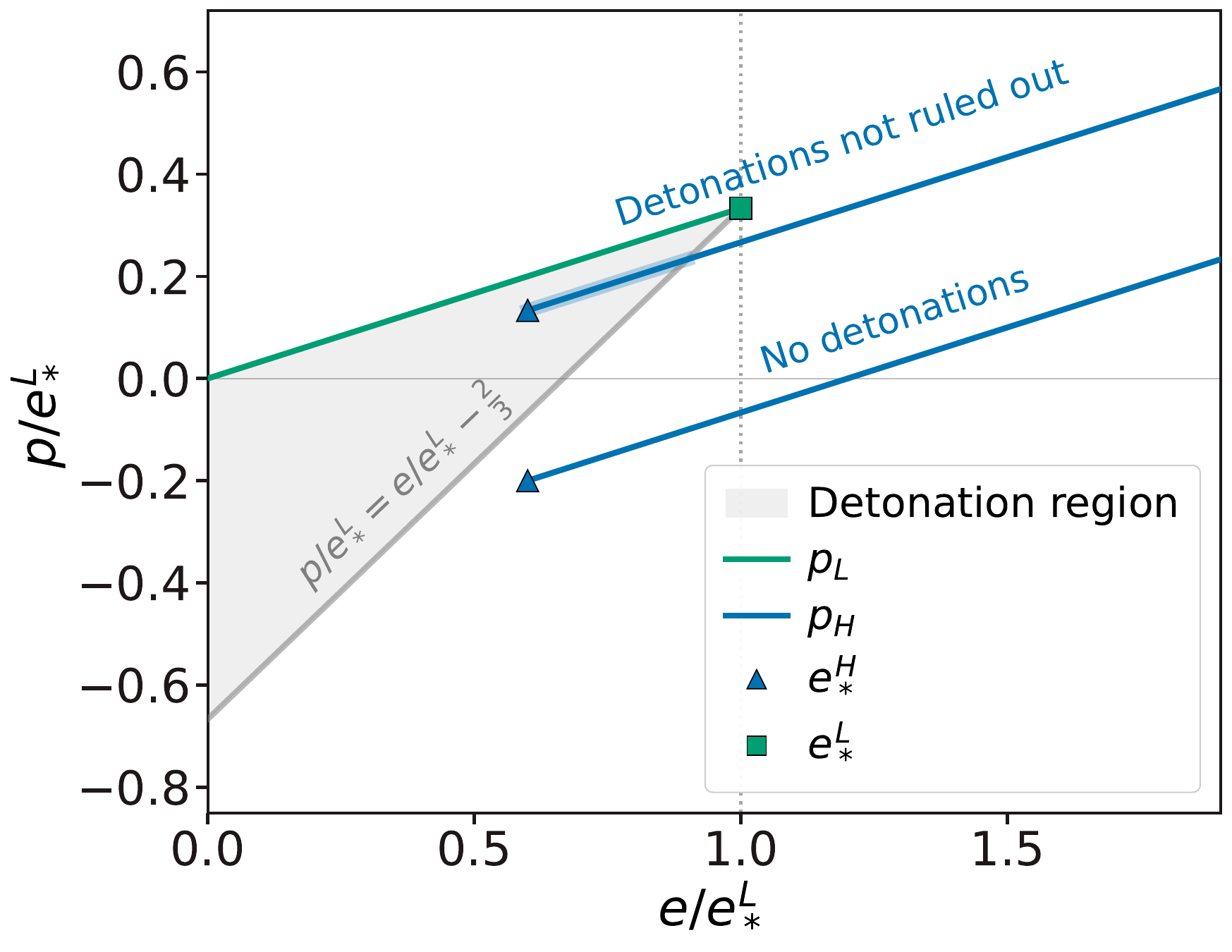}
    \caption{Two bag models in which detonations are permitted and excluded, respectively. The reduced EoS is given by \eqref{redu}. The low-energy branch, shown as a solid green line, is common to both models and exists only up to a maximum energy density $\etL$, marked by the green square. The high-energy branches are shown as solid blue lines. Both extend down to the same minimum energy density, $\etH=0.6\etL$, marked by the blue triangle, but differ in their bag constants. Equation~\eqref{delta_e} is a necessary, but not sufficient, condition for detonations to exist. For a bag model, this requires the high-energy branch to enter the shaded gray region. The upper EoS, with $\epsilon=0.05 \, \etL$, satisfies this condition. The lower one, with $\epsilon=0.3 \,\etL$, violates \eqref{delta_e}, and detonations are therefore excluded in this model.
    }
    \label{fig:det_region_bag_model}
\end{figure}

For supercooled transitions, the phases behind and ahead of the wall are the low- and the high-energy phases, respectively. Therefore, as previously pointed out in \cite{Sanchez-Garitaonandia:2023zqz}, Eqs.~\eqref{pp}--\eqref{ee} mean that there must exist points on the low-energy branch with higher pressure, enthalpy and energy density than some points on the high-energy branch. This condition cannot be satisfied for QCD-like EoS: as shown in \fig{fig: EoS-holography} and in the top-left panel of \fig{fig: g_eff_figure}, in this case any state of the low-energy phase has lower energy density than any state in the high-energy phase, that is, 
\begin{equation}
\label{ccc}
    \max(e_{\LL}) < \min(e_{\HH}) \,.
\end{equation}
Note that this condition is automatically implied if thermodynamic quantities are single-valued functions of the energy density, since this prevents the low-energy branch from overlapping with the high-energy branch. By contrast, the conditions \eqref{pp}--\eqref{ee} can be satisfied in a bag-model EoS, as illustrated by the right-most column of \fig{fig: g_eff_figure}. However, even in some of these cases detonations may still be excluded because \eqref{delta_e} cannot be satisfied. This is illustrated in \fig{fig:det_region_bag_model} for a bag model with finite amounts of supercooling and superheating. In this case, the existence of detonations implies an upper bound on the bag constant:
\begin{equation}
\label{upperbag}
2\epsilon< \etL - \etH \,.
\end{equation}
This bound follows by saturating \eqref{delta_e} under the most favorable conditions, namely for the pair of states that maximizes the left-hand side of the equation. This is achieved for $e_-=\etL$ and $e_+=\etH$. We also note that violating any of the conditions \eqref{pp}--\eqref{delta_e} is sufficient to exclude detonations, but it is not necessary, since detonations may also be ruled out by global constraints on the flow---see Sec.~\ref{sec: section 4}.

A similar argument shows that detonations are also excluded in superheated transitions.\footnote{For superheated transitions, we follow the terminology in \cite{Bea:2024bxu}, where the term ``detonation'' refers to supersonic walls in which the plasma ahead of the wall is at rest. This is opposite to the terminology in \cite{Barni:2024lkj}, where  superheated detonation are solutions with subsonic walls.} In this situation, the fluid ahead of the wall is at rest in the laboratory frame, while the fluid behind the wall moves with negative velocity, since the bubble needs to absorb energy from outside in order to grow---see the middle row of Fig.~5 in \cite{Bea:2024bxu}. Consequently, in the  rest frame of the wall, where $v_\pm$ are defined, one has
\begin{equation}
\label{vv2}
v_+ \leq v_- \,.
\end{equation}
Repeating the previous analysis then reverses the inequalities in Eqs.~\eqref{pp}--\eqref{ee}. At the same time, however, the roles of the low- and high-energy phases are interchanged relative to the supercooled case: the fluid behind the wall is now in the high-energy phase, whereas the fluid ahead of the wall is in the low-energy phase. These two reversals compensate each other, and therefore the condition \eqref{ccc} remains unchanged. As a consequence, detonations are also excluded in superheated transitions with a QCD-like EoS. 

The existence or otherwise of hybrid solutions requires a more involved analysis,  because the possible obstructions arise from the non-existence of global solutions to Eqs.~\eqref{eq:self-similar-eoms} and \eqref{eq:matching-usual} that  allow for the  construction of a complete hybrid flow. Interestingly, this analysis reveals that, as mathematical solutions,  hybrid flows do not exist for the QCD-like model of \fig{fig: EoS-holography}, but they do exist for the QCD-like model of \fig{fig: g_eff_figure}. However, in the latter case these mathematical solutions are not physically acceptable solutions  because they violate the second law of thermodynamics. 

\begin{figure}[t!]
    \centering
    \includegraphics[width=1.0\linewidth]{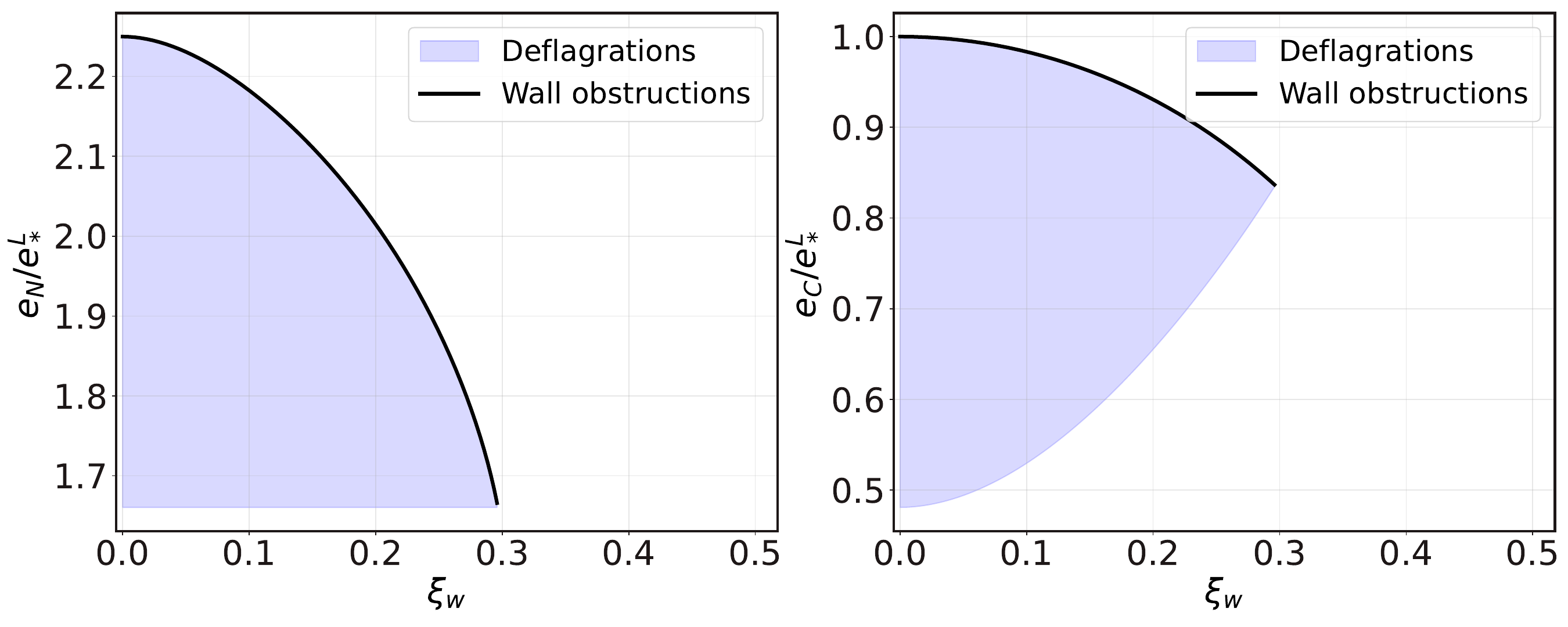}
    \\[4mm]
    \includegraphics[width=1.0\linewidth]{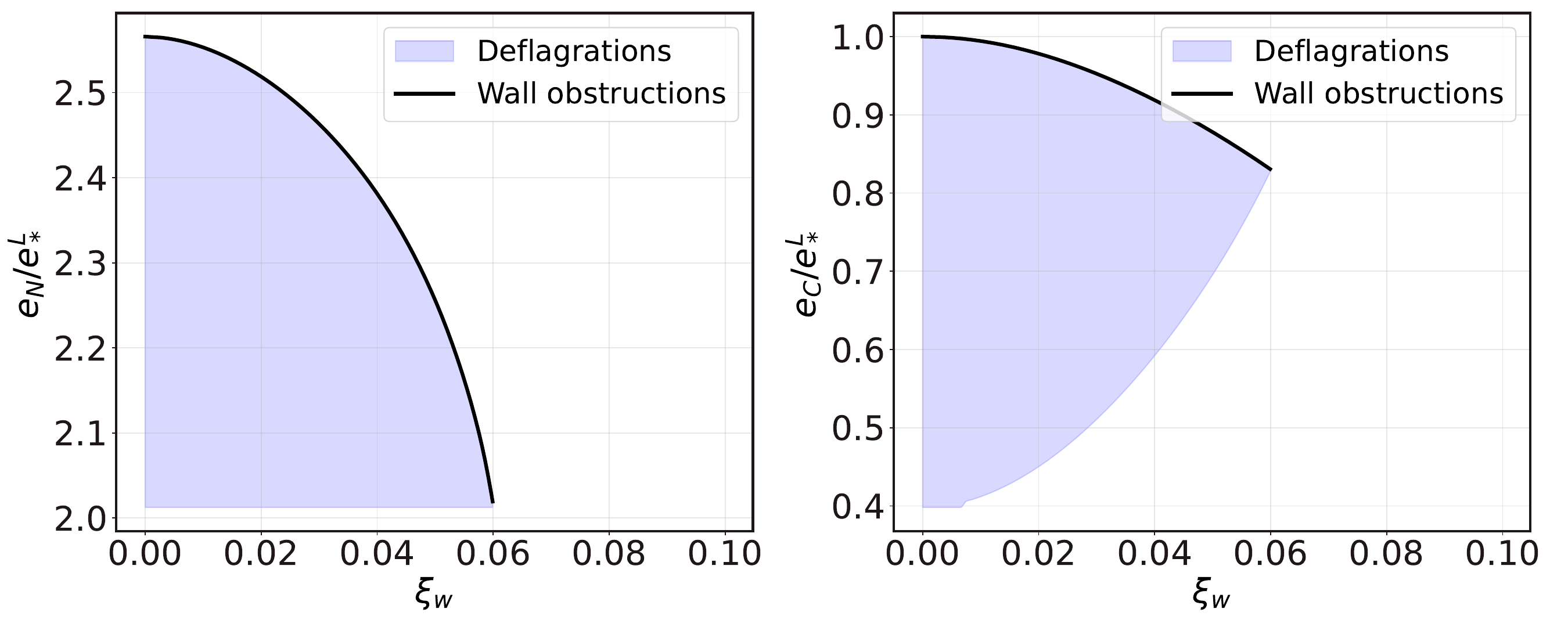}
    \caption{The shaded regions show the allowed nucleation energy, $e_N$, (left column) and the energy density at the center of the bubble, $e_C$, (right column) as functions of the bubble wall velocity, $\xi_w$, for the model of \fig{fig: EoS-holography} (top row) and for pure $SU(3)$ YM theory (bottom row). We observe an obstruction, shown by the solid black line, that bounds the deflagration solutions from above: beyond this line, no such solutions exist. We emphasize that the shaded regions represent the hydrodynamically allowed solutions; imposing positive entropy production at the wall further reduces these regions. }
    \label{fig: bubbles-holography}
\end{figure}

Deflagrations do occur in QCD-like models, but they are restricted to relatively low wall velocities by the limited amount of allowed supercooling and superheating. This is illustrated in the top row of \fig{fig: bubbles-holography} for the model of \fig{fig: EoS-holography}, and in the bottom row  for pure $SU(3)$ YM theory. To produce the plots in the latter case, only the EoS over the range of energy densities shown in the inset on the right-hand side of \fig{fig: cs2_holography}  is required. We see that the much smaller size of the phase-transition region in YM, illustrated in the speed-of-sound plot in \fig{fig: cs2_holography}, translates here into a much smaller range of allowed bubble
wall velocities.

From the analysis in this section, it is clear that non-conformality introduces obstructions to bubble expansion that are entirely absent in the bag model. To understand how these obstructions arise and what their origin is, below we will study bubble solutions across a family of EoS that interpolate between a bag-model-like regime and one in which non-conformal features dominate, similarly to \fig{fig: EoS-holography}.

\section{New obstructions to bubble expansion}
\label{sec: section 4}
As the EoS departs from a bag-model form and becomes qualitatively similar to that in \fig{fig: EoS-holography}, two properties have a major impact on the space of allowed bubble solutions in the $(\xi_w,\alpha_N)$ plane. These are the non-constant speed of sound along the flow, and the existence of maximum superheating and supercooling for the low- and high-energy phases, respectively. In what follows, we examine the types of obstructions that arise as a consequence of these features. We divide them into two classes: those that arise at the bubble wall, which we call ``wall obstructions'' (WO), and those that arise along the flow, which we call ``flow obstructions'' (FO).

Their distinction is not only related to where the obstruction originates, but also to how it depends on the EoS. WO depend only on the local values of $p(e)$ and $c_s(e)$, whereas FO are sensitive to how these quantities vary with energy, namely to $\partial_e c_s$. Physically, wall obstructions arise from local constraints at the discontinuity, while flow obstructions are associated with the global structure of the solutions.

\subsection{Wall obstructions}

Let us first focus on WO, which affect only deflagrations and detonations. As pointed out in \cite{Espinosa2010}, deflagrations are bounded by the subsonic wall condition
\begin{equation}
\xi_w \leq c_s(e_-).
\label{eq: deflagration limit}
\end{equation}
In the bag model, this condition simply marks the boundary between deflagrations and hybrids and therefore does not represent a genuine obstruction. This is illustrated in \fig{fig:alpha-a}. By contrast, when the speed of sound is not constant, this boundary can become a true obstruction for deflagrations, since the solution may no longer be continuously extendable into a hybrid branch.
This behaviour is most clearly visible in the middle and bottom rows of \fig{fig:alpha_N vs xi_w}, where the deflagration and hybrid regions are shown in purple and green, respectively, and the WO is indicated by a solid black curve. For sufficiently small values of $\alpha_N$, the deflagration branch terminates at the black curve, corresponding to the saturation of \eqref{eq: deflagration limit}, yet no hybrid solutions exist beyond it. As the model departs further from the bag-model limit---moving from left to right and from top to bottom in \fig{fig:alpha_N vs xi_w}---an increasing portion of the deflagration branch fails to connect continuously to a hybrid configuration upon saturating \eqref{eq: deflagration limit}.

For EoS for which the necessary condition $w_-\geq w_+$ can be satisfied, the precise limiting condition for detonations is obtained by observing that these solutions cannot superheat the fluid behind the wall beyond the maximum temperature allowed within the low-energy branch. Equivalently,
\begin{equation}
e_- \leq \etL \,.
\label{eq:maximum-superheating}
\end{equation}
Saturating \eqref{eq:maximum-superheating} at the wall and solving the matching conditions \eqref{eq:matching-usual} then yields a relation between $\xi_w$ and $\alpha_N$ that defines the boundary of the allowed detonation region. This obstruction is visible in \fig{fig:alpha-c} as the solid black curve delimiting the red region corresponding to detonations.

\subsection{Flow obstructions}
\begin{figure}[tp]
    \centering
    \includegraphics[width=\linewidth]{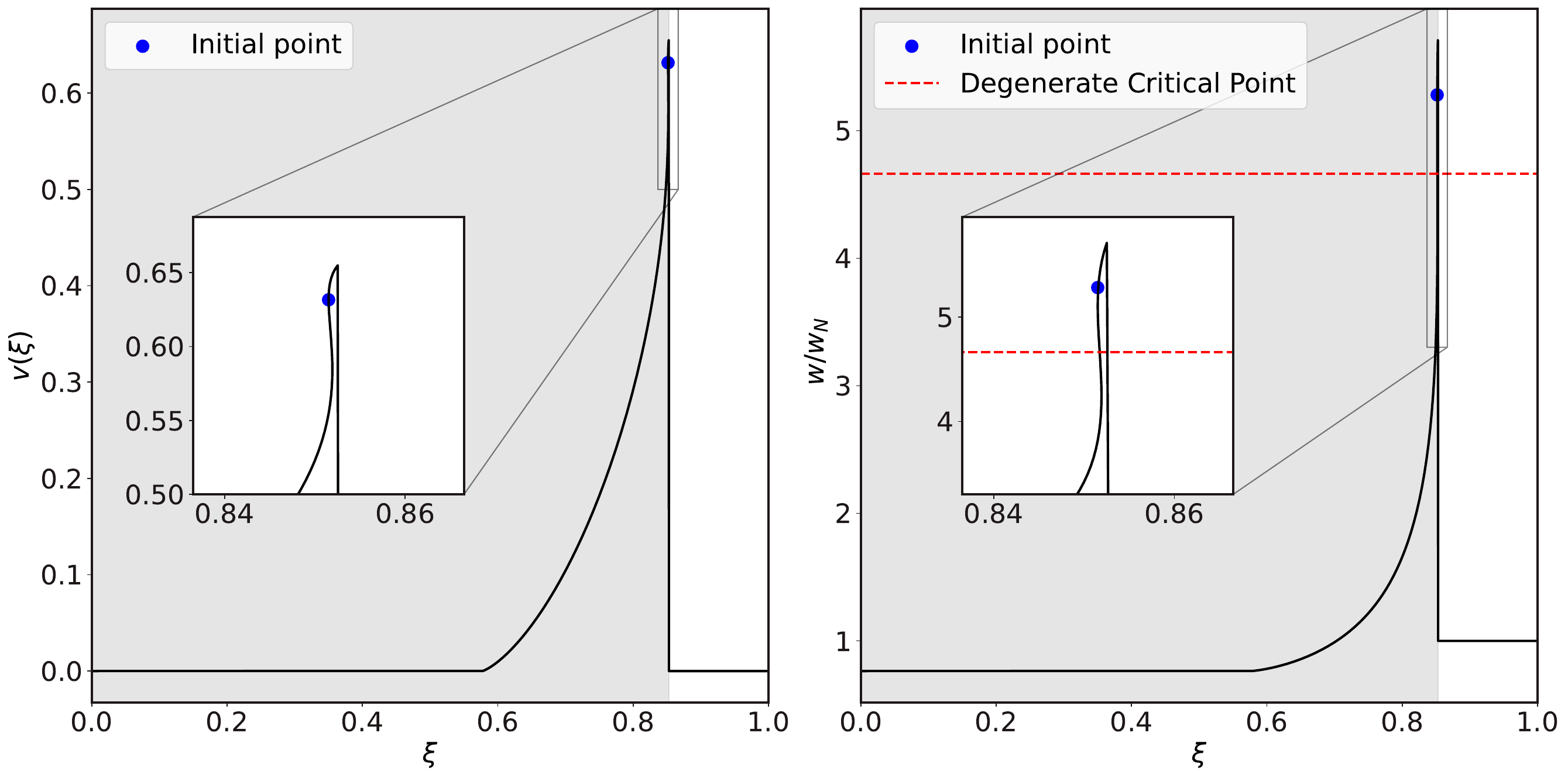}\\
    \centering
    \includegraphics[width = 0.8\linewidth]{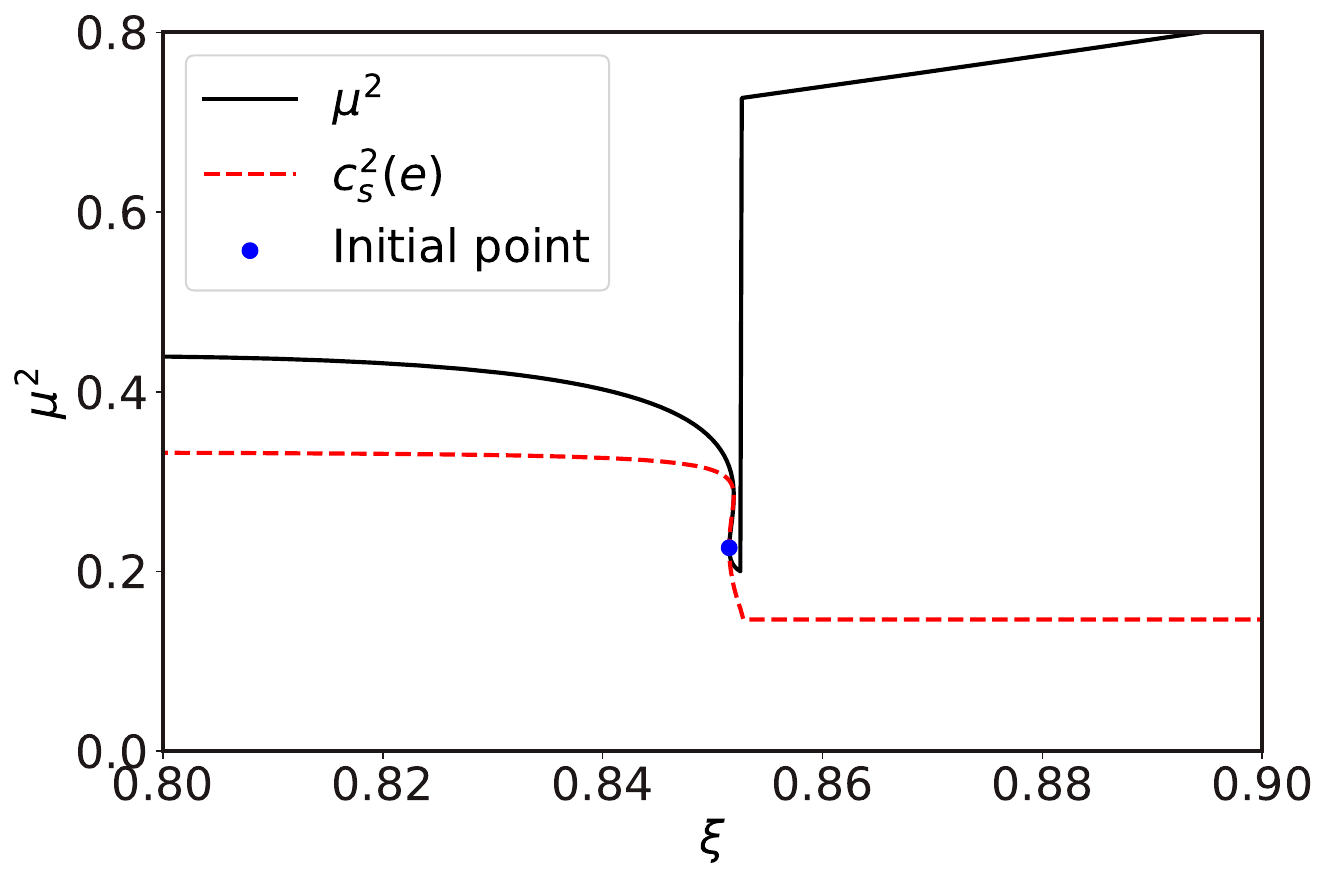}
    \caption{Forbidden detonation beyond the DCP (top), together with $\mu^2$ and $c_s^2$ as functions of the self-similar variable $\xi$ (bottom). We observe that the flow becomes multivalued along the rarefaction wave, preventing the construction of the corresponding detonation solution. Starting from an initial point (blue dot) with $\mu(\xi,v)=c_s(e)$ and energy density above the DCP, we integrate the flow backwards and forwards. The wall appears as a single discontinuity on the forward branch at $\xi=\xi_w$, where the energy jumps to $e_N$.}
    \label{fig:mu and cs vs xi}
\end{figure}

\begin{figure}[tp]
    \centering
    \includegraphics[width = 0.85\textwidth]{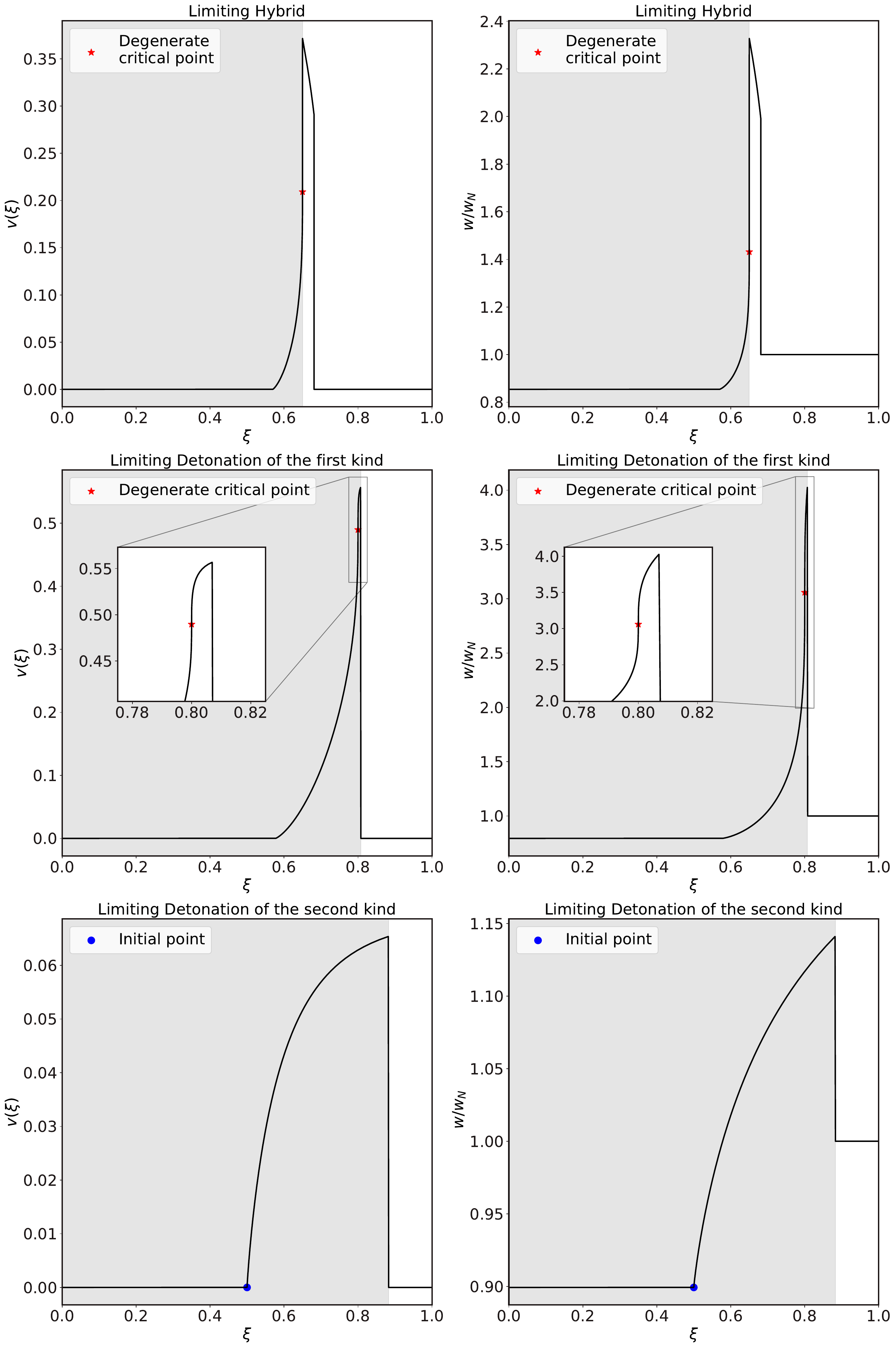}
    \caption{Examples of limiting solutions constrained by flow obstructions. The first and second rows correspond to a limiting hybrid and detonation, respectively. 
    The bottom row corresponds to a limiting detonation of the second kind, in which case $\partial_\xi v\to \infty$ as $\xi = c_s(e)$ and $v = 0$. The blue marker denotes the initial point from which these solutions are constructed, located at $v \to 0$ and energy density fixed by $\mu(\xi, v) = c_s(e)$, where the flow equation degenerates.}
    \label{fig: limiting_flows}
\end{figure}

FO are associated with rarefaction waves and therefore affect only hybrids and detonations.\footnote{As we will show, these obstructions can be avoided by introducing shocks. This is presumably the reason why deflagrations seem unaffected by these obstructions, as their flow already contains a shock.} Rarefaction waves are characterized by $\mu(\xi_w) = v_- \geq c_s(e_-)$ (see e.g.~\cite{Espinosa2010}). For an EoS with constant speed of sound, this condition is enough to construct a continuous flow connecting with the state at rest inside the bubble. This is because $\mu(\xi)$ grows, while $c_s$ is fixed, with decreasing $\xi$, ensuring that $\mu(\xi) > c_s$ and that $\partial_{\xi}v$ remains finite. In contrast, for a non-constant speed of sound, even if $\mu(\xi_w) = v_- \geq  c_s(e_-)$ is fulfilled, both $\mu(\xi)$ and $c_s$ grow as $\xi$ decreases, which can cause $\partial_{\xi}v$ to diverge further along the flow, as shown in \fig{fig:mu and cs vs xi}. As a consequence, the flow becomes multivalued, making it impossible to connect continuously to the inside of the bubble at rest. The limiting condition corresponds to the case in which $\partial_{\xi}v$ diverges at a single point without the flow becoming multivalued, as seen in the top and middle panels in \fig{fig: limiting_flows}.

If we parametrize the flow in terms of $\xi(v)$, the aforementioned limiting condition corresponds to a degenerate critical point (DCP) at which both the first and second derivatives of $\xi$ with respect to $v$ vanish. Using \eqref{eq:self-similar-eoms} we see that, for this to happen, the following two conditions must be satisfied:
\begin{equation}
    \mu(\xi) = c_s(e) \,, \quad \frac{d c_s^2}{d e} = -\frac{2(1-c_s^2(e))c_s^2(e)}{e+p(e)} \,.
    \label{eq:critical sound}
\end{equation}
The first condition is the usual consequence of having a flow with diverging $\partial_{\xi}v$. This is also the condition imposed to obtain Jouguet detonations or hybrids. The second condition, which obviously cannot be satisfied if the speed of sound is constant,  depends only on the EoS and not on the specific flow solution. Given a specific EoS, if the second equation in \eqref{eq:critical sound} is satisfied for some energy $e_\star$, we say that the model has a DCP and further obstructions will arise for both hybrids and detonations.

Unfortunately, studying the obstruction from a DCP implies solving for a large portion of the self-similar flow, in contrast to the simpler conditions obtained in the previous subsection. The way we proceed in practice is by building the limiting detonations, i.e.,~those that possess a DCP characterized by \eqref{eq:critical sound} at some point in their rarefaction wave.

For a given wall speed, $\xi_w$, the DCP occurs at the point $\xi_\star$ along the flow. Let $e_\star=e(\xi_\star)$.\footnote{This should not be confused with the turning points $\etL, \etH$.} Since this point lies within a rarefaction wave, $\xi_\star$ can be chosen anywhere between the wall, $\xi_\star=\xi_w$, and the point where the flow reaches the bubble interior, $\xi_\star=c_s(e_\star)$. Integrating the flow with this initial condition both outwards and inwards, and then imposing the matching conditions, we obtain a relation between $\alpha_N$ and $\xi_w$ for which detonations and hybrids exhibit a DCP somewhere along the rarefaction wave. These solutions are shown as dashed curves in
\fig{fig:alpha_N vs xi_w}, and delimit the green and red regions corresponding to hybrids and detonations, respectively. The would-be detonations and hybrids outside these regions are inconsistent, because they have multivalued self-similar flows.

It is important to note that, when the DCP is placed immediately behind the wall, $\xi_\star=\xi_w$, the solution takes the form of a Jouguet detonation or hybrid. This is indeed what happens, as can be seen in \fig{fig:alpha_N vs xi_w}: the dashed green and purple lines meet at a single point, from which the Jouguet-velocity condition emerges, shown as the dashed blue line. At this point, hybrids and detonations are continuously connected.

\begin{figure}[t]
    \centering
    \includegraphics[width=\linewidth]{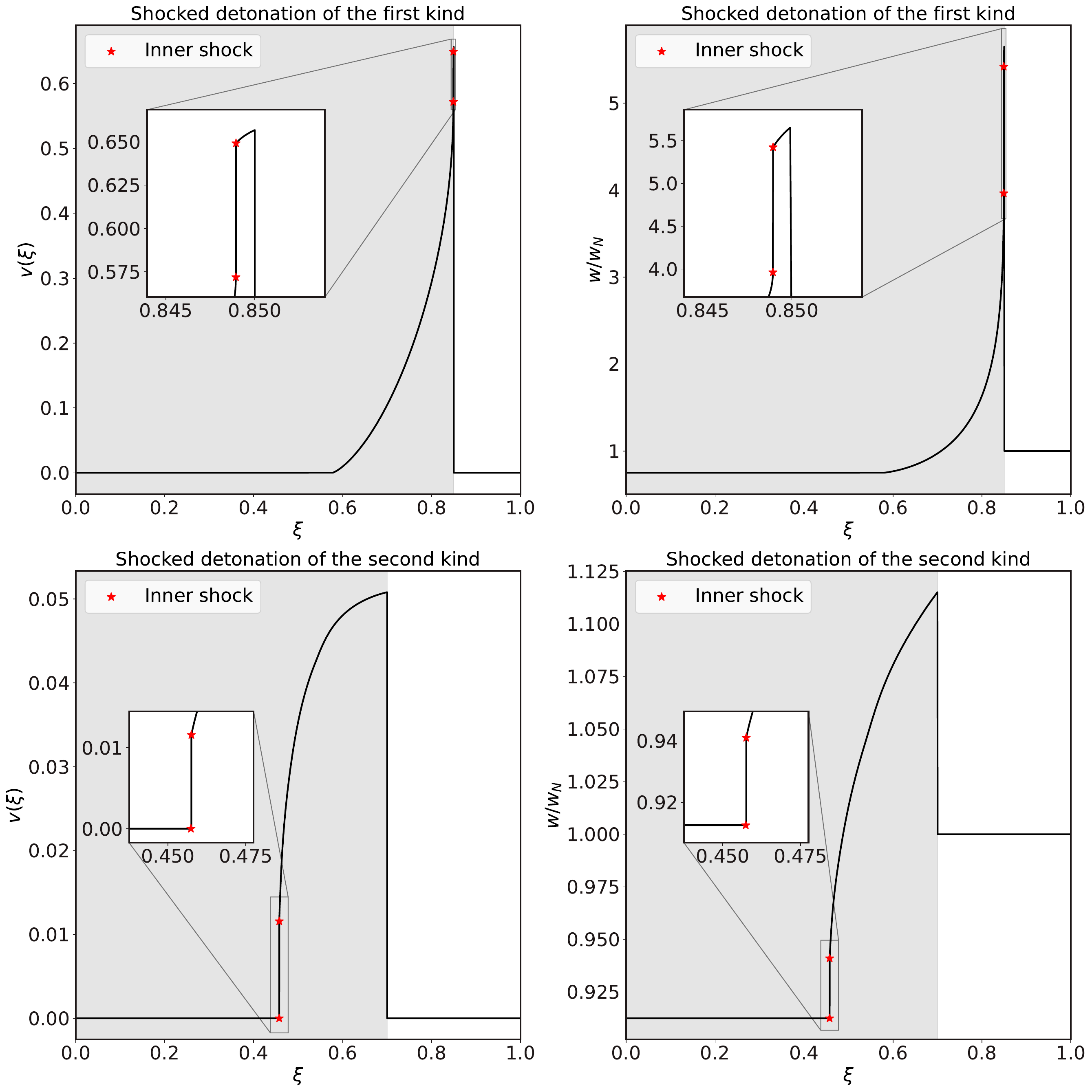}
    \caption{Fluid profiles for two classes of shocked detonations, obtained by introducing an additional internal shock into a limiting detonation configuration---see Fig.~\ref{fig: limiting_flows}. The top row shows a first-kind shocked detonation, in which the internal shock connects two states on the low-energy branch. The bottom row shows a second-kind shocked detonation, in which the internal shock brings the fluid directly to rest. The left and right panels show, respectively, the velocity in the bubble rest frame, $v(\xi)$, and the enthalpy normalized to its nucleation value, as functions of the self-similar coordinate $\xi$. The shaded region denotes the bubble interior, $\xi<\xi_w$, with the bubble wall located at its boundary. Red stars mark the pre- and post-shock states across the internal discontinuity.}
    \label{fig:hyper_detonations flows}
\end{figure}

Before moving on, let us note that the exclusion of solutions beyond the dashed curves in \fig{fig:alpha_N vs xi_w} assumes that the flow remains continuous. If shocks are allowed inside the rarefaction wave, additional detonation solutions can be constructed. This new type of solutions consists of connecting the upper and lower branches of the multivalued flow in \fig{fig:mu and cs vs xi}, allowing us to construct a full rarefaction wave connecting with the interior of the bubble. These shocks connect supersonic states at each side, and are allowed by the second law precisely when the flow immediately behind the wall has greater energy than the DCP one, $e_\star$. As a consequence, detonations can be extended beyond the apparent obstruction introduced by a DCP. An example of these detonations can be found in the top row of \fig{fig:hyper_detonations flows}.

In principle, for a given nucleation energy and bubble wall velocity, one can construct a one-parameter family of shocked detonations, each with the shock at a different position $\xi_{sh}$. To determine which solution dominates, we select the shock position that maximizes the entropy production at the discontinuity. This turns out to be equivalent to imposing the Jouguet condition $v_- = c_s$ at the shock, similarly to hybrid solutions.

Finally, we identify a different type of obstruction associated with the fixed point of the flow at $\xi=c_s(e)$ and $v=0$. This is the point at which the rarefaction wave connects continuously to the bubble interior. In the bag model, all rarefaction waves approach this fixed point from larger to smaller values of $\xi$---see e.g.~\cite{Espinosa2010}. Equivalently, any flow emerging from the fixed point can be integrated outwards to construct a detonation. This is no longer true for more general EoS: rarefaction-wave solutions can approach the fixed point from either larger or smaller values of $\xi$. The flows that can be used to construct a full detonation are those that emerge from the fixed point with positive slope. The boundary of the allowed flows is determined by the condition $\partial_{\xi}v\to\infty$ at $\xi=c_s(e_C)$, where $e_C$ is the energy density inside the bubble.

By integrating outwards from $\xi_0=c_s(e_C)$, with a small initial velocity $v(\xi_0)\ll 1$, for different choices of $e_C$, we obtain the corresponding limiting detonations, as shown in the bottom row of 
\fig{fig: limiting_flows}. In the space of solutions, these are represented by the lower segments of the purple dot-dashed lines in 
\fig{fig:alpha_N vs xi_w}, which connect smoothly to the limiting detonations discussed above at $e_C=e_\star$. Any detonation lying on the other side of the lower purple dot-dashed lines in 
\fig{fig:alpha_N vs xi_w} would start with an inward-going flow, $\partial_{\xi}v<0$. These solutions cannot be built with a single continuous rarefaction wave, but solutions with a shock exist and are allowed. The way to build them is by exploiting the fact that inward-going flows eventually turn around and go towards greater $\xi$. These flows are analogous to those in \fig{fig:mu and cs vs xi}, but with the lower branch lying in the $v<0$ half plane. A shock can then be introduced, connecting a point of the upper branch of the flow with the state at rest inside the bubble. An example of this kind of solutions can be found in the bottom row of \fig{fig:hyper_detonations flows}.

\begin{figure}[tp]
    \centering
    \begin{subfigure}{0.49\linewidth}
        \includegraphics[width=\linewidth]{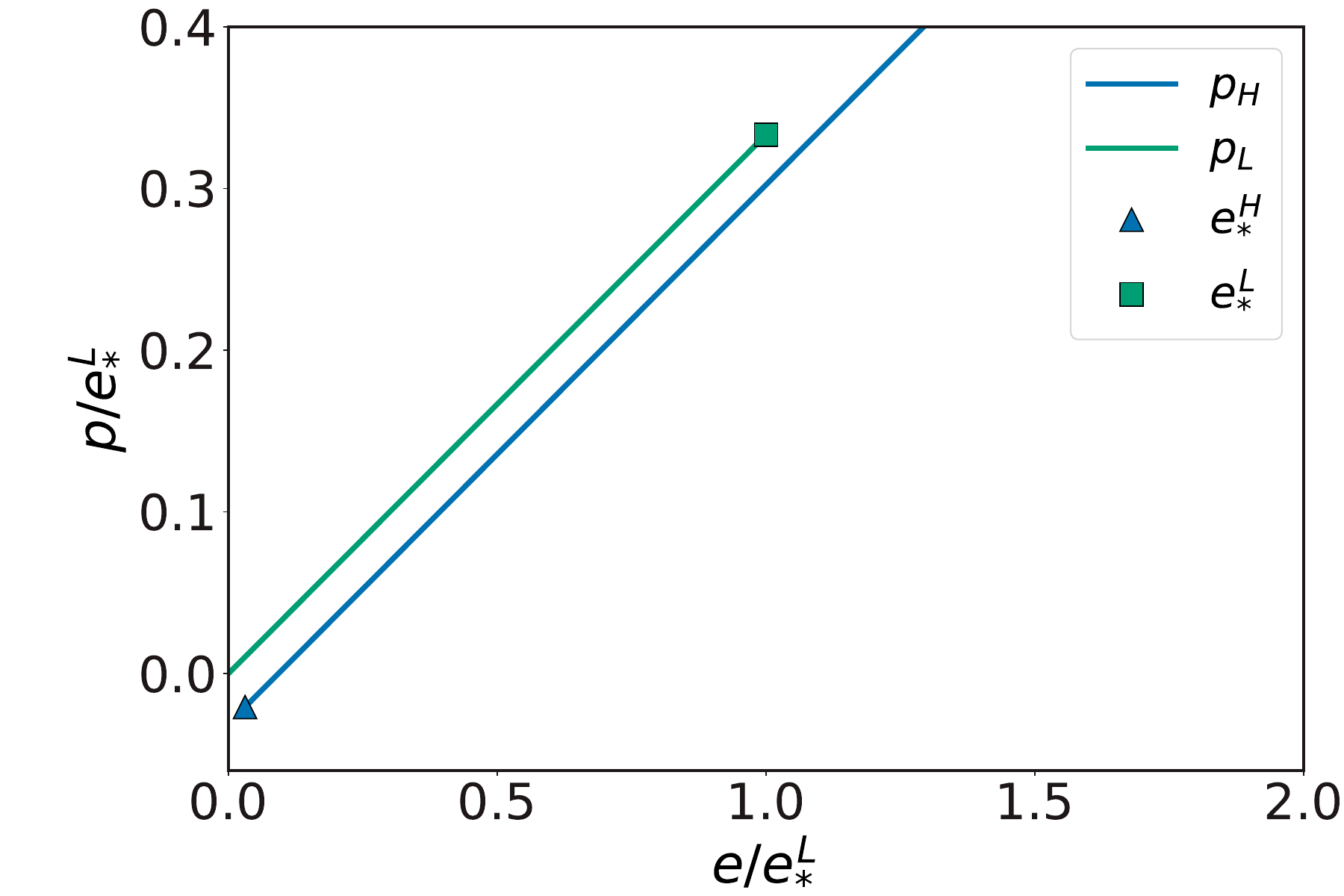}
        \caption{}
        \label{fig:EoS-a}
    \end{subfigure}\hfill
    \begin{subfigure}{0.49\linewidth}
        \includegraphics[width=\linewidth]{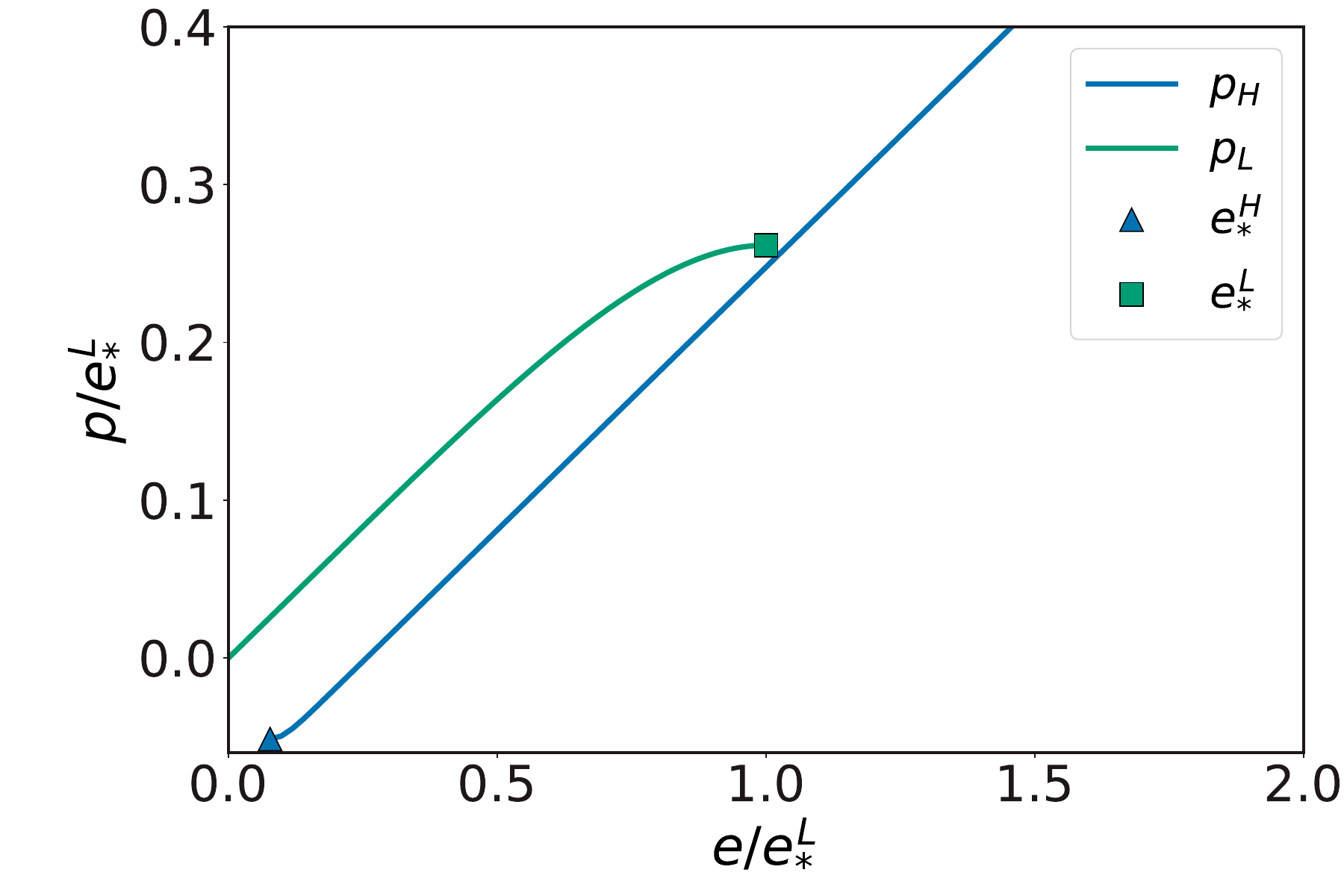}
        \caption{}\label{fig:EoS-b}
    \end{subfigure}\\[15pt]
    \begin{subfigure}{0.49\linewidth}
        \includegraphics[width=\linewidth]{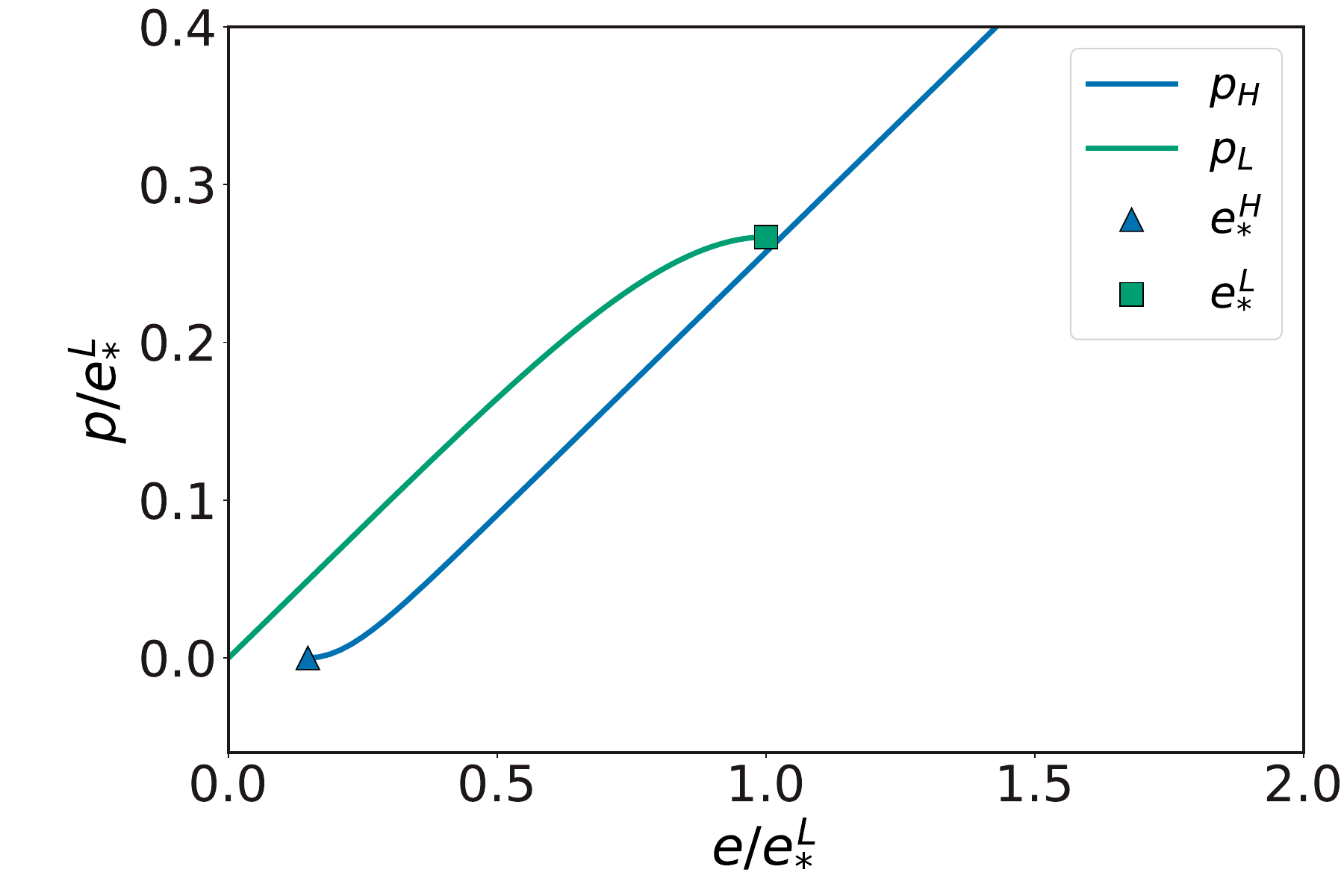}
        \caption{}\label{fig:EoS-c}
    \end{subfigure}\hfill
    \begin{subfigure}{0.49\linewidth}
        \includegraphics[width=\linewidth]{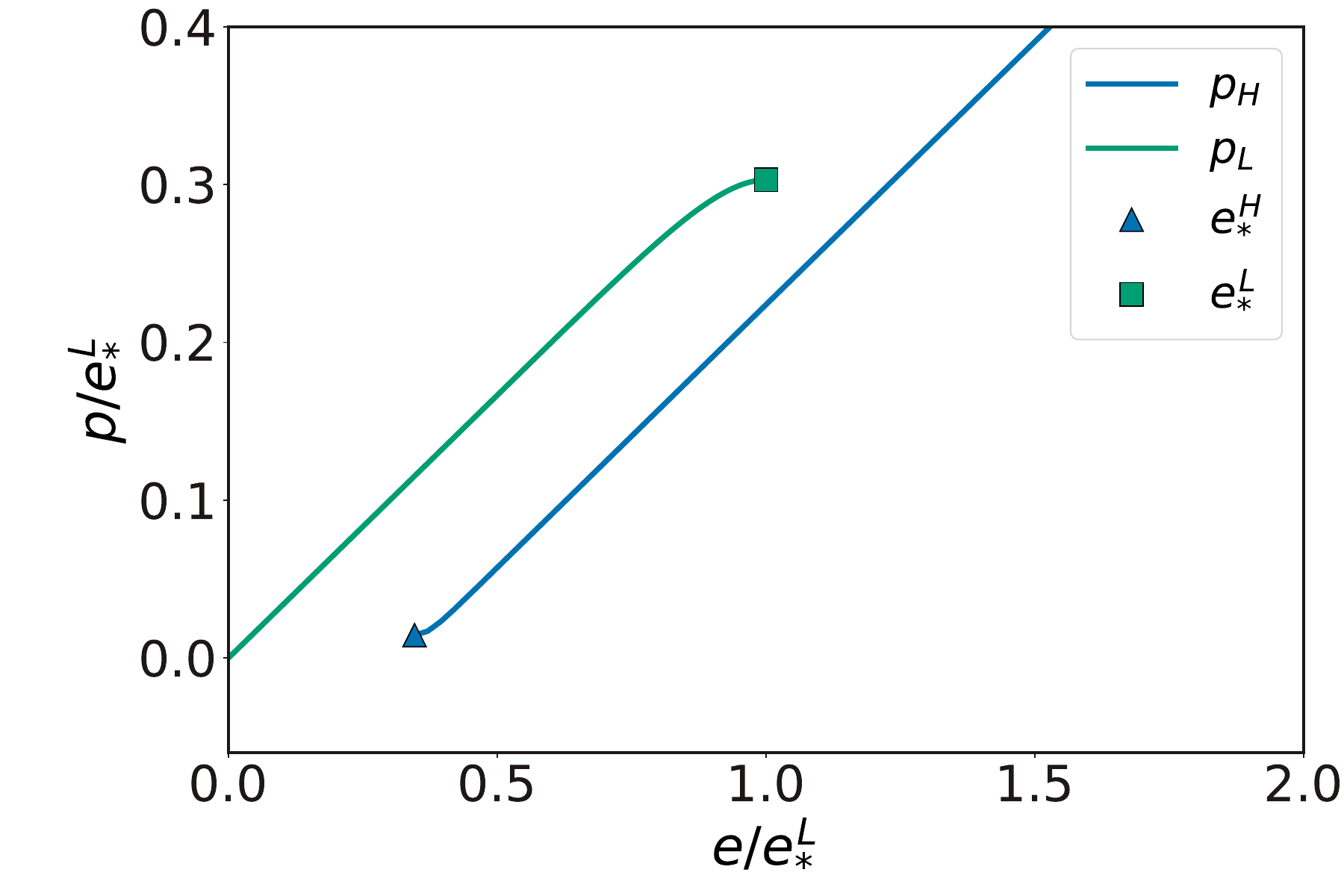}
        \caption{}\label{fig:EoS-d}
    \end{subfigure}\\[15pt]
    \begin{subfigure}{0.49\linewidth}
        \includegraphics[width=\linewidth]{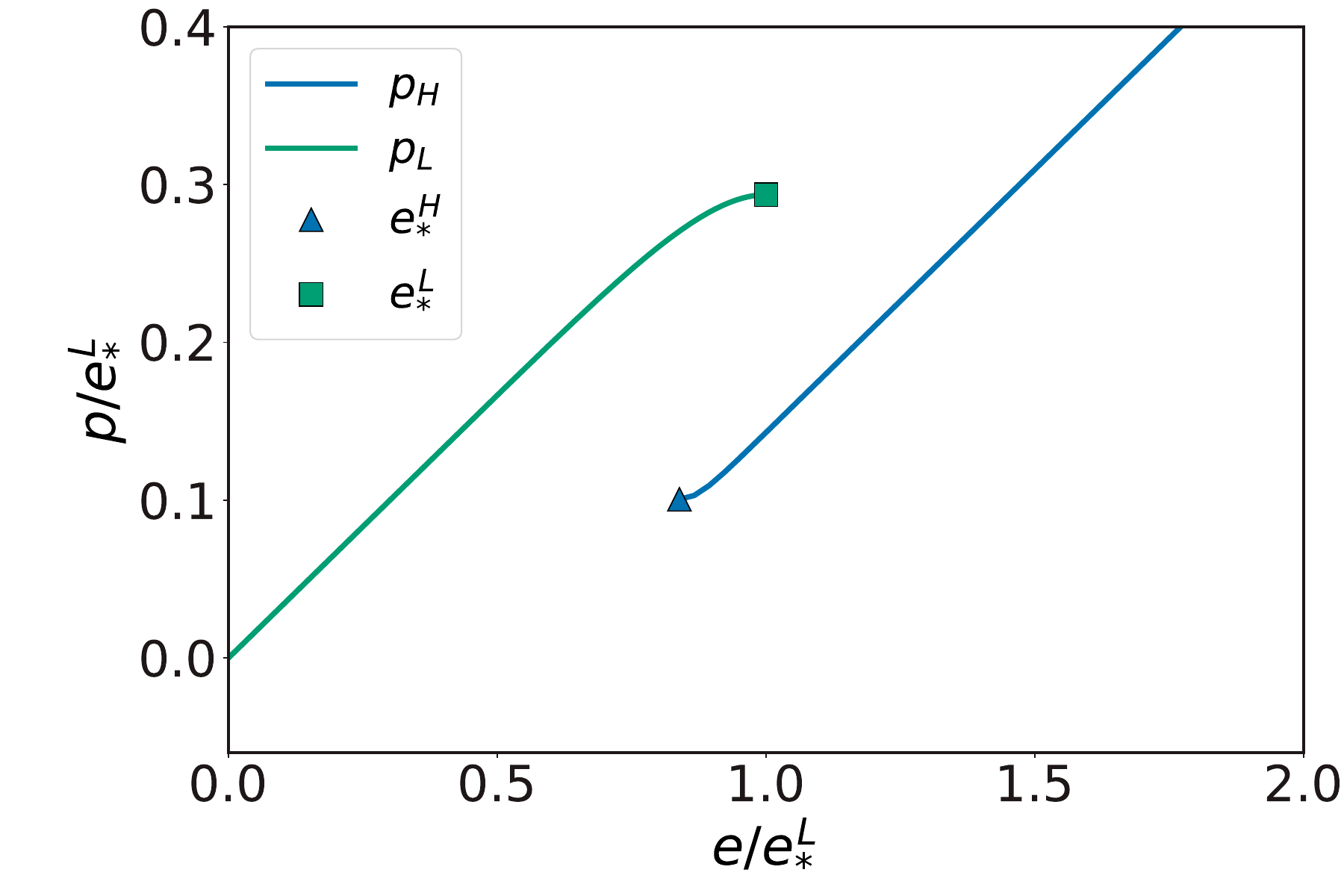}
        \caption{}\label{fig:EoS-e}
    \end{subfigure}\hfill
    \begin{subfigure}{0.49\linewidth}
        \includegraphics[width=\linewidth]{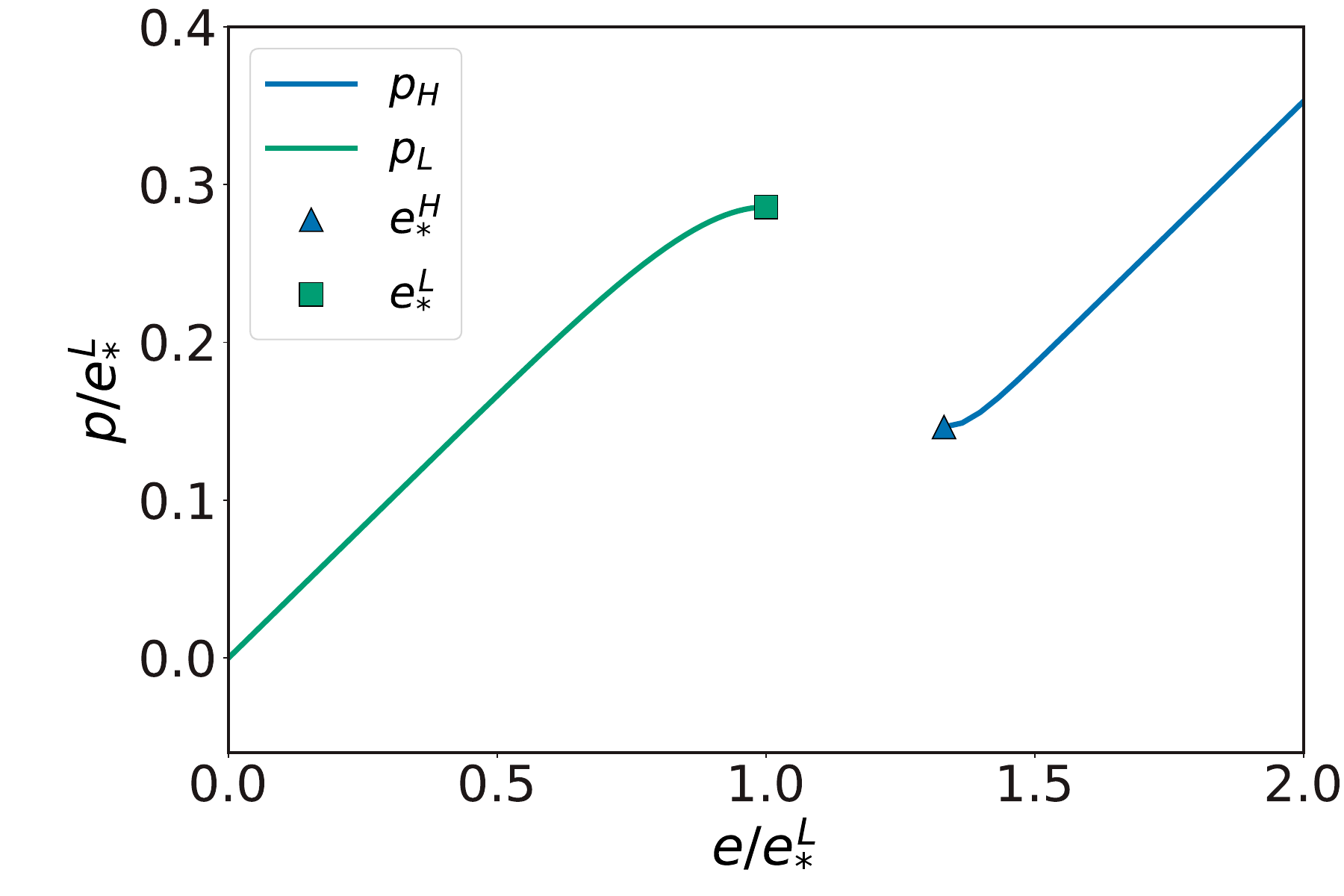}
        \caption{}\label{fig:EoS-f}
    \end{subfigure}
    \caption{Six representative EoS from the family \eqref{eq: EoS-p} which, from top to bottom and from left to right, interpolate from a bag-model to a QCD-like EoS. Panel~(\subref{fig:EoS-a}) is a bag-model EoS with a finite amount of supercooling and superheating; the remaining panels progressively modify it, ending in the strongly-coupled, holography-motivated EoS of panel~(\subref{fig:EoS-f}).}
    \label{fig:EoS-family}
\end{figure}

\begin{figure}[tp]
    \centering
    \begin{subfigure}{0.49\linewidth}
        \includegraphics[width=\linewidth]{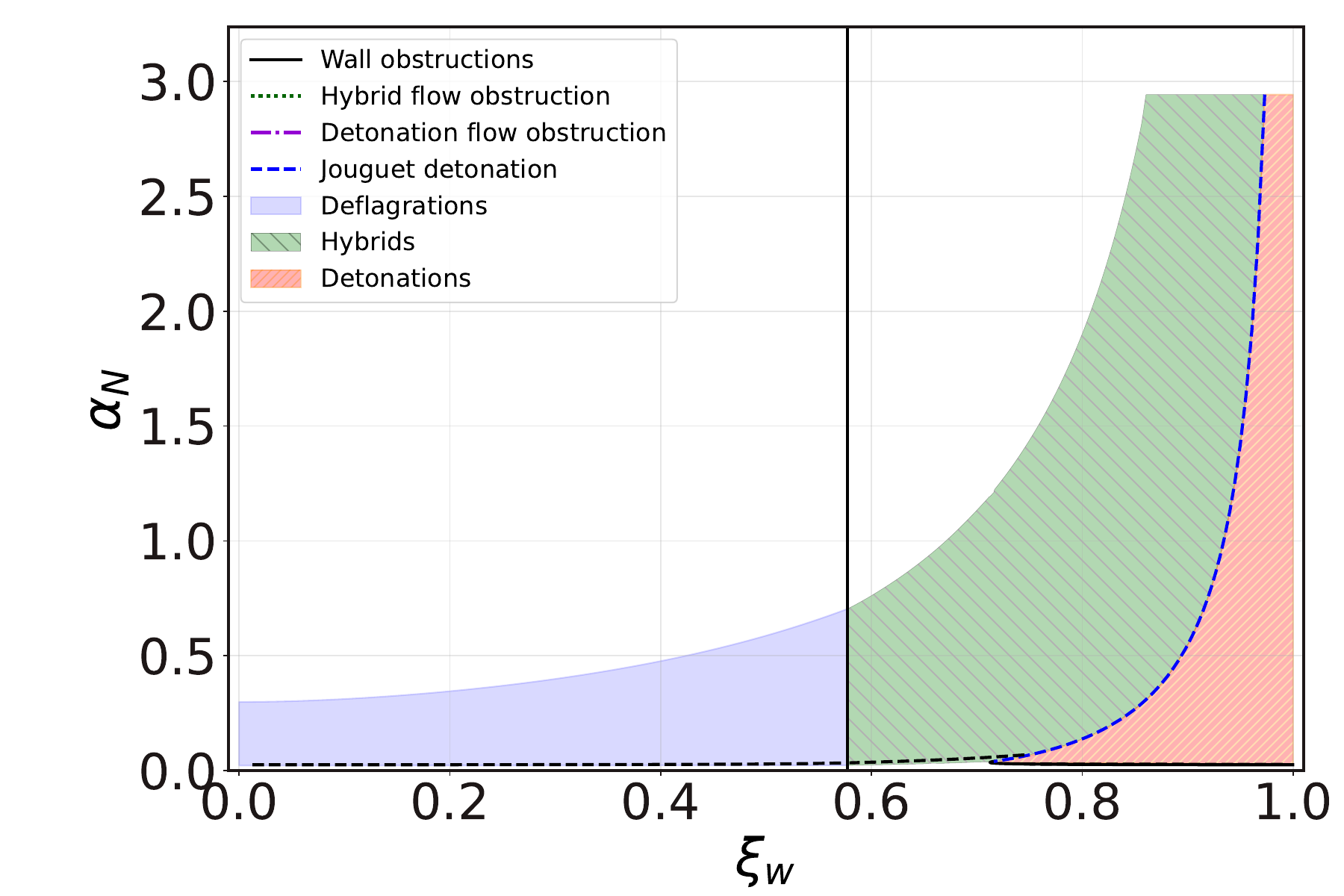}
        \caption{}\label{fig:alpha-a}
    \end{subfigure}\hfill
    \begin{subfigure}{0.49\linewidth}
        \includegraphics[width=\linewidth]{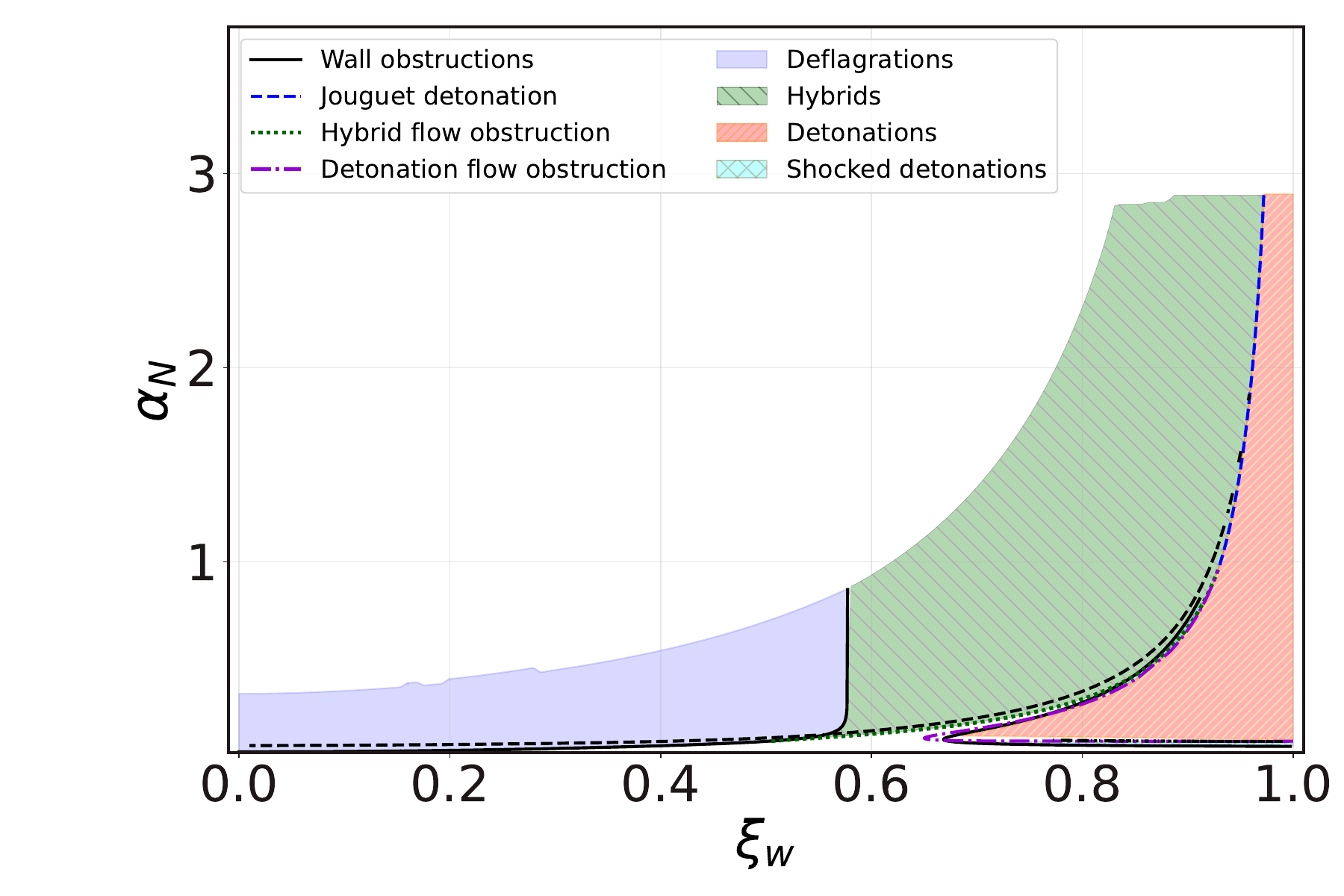}
        \caption{}\label{fig:alpha-b}
    \end{subfigure}\\[15pt]
    \begin{subfigure}{0.49\linewidth}
        \includegraphics[width=\linewidth]{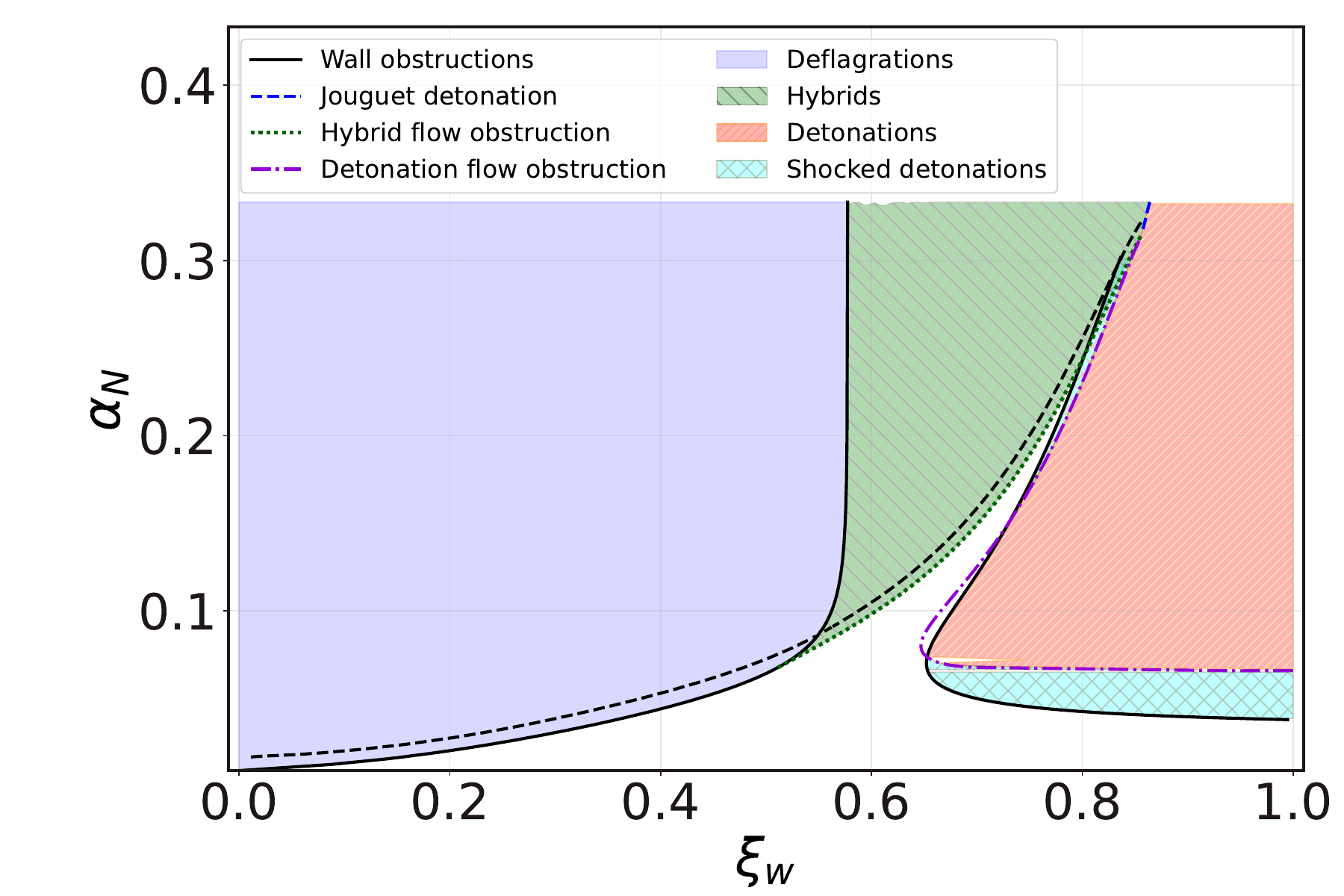}
        \caption{}\label{fig:alpha-c}
    \end{subfigure}\hfill
    \begin{subfigure}{0.49\linewidth}
        \includegraphics[width=\linewidth]{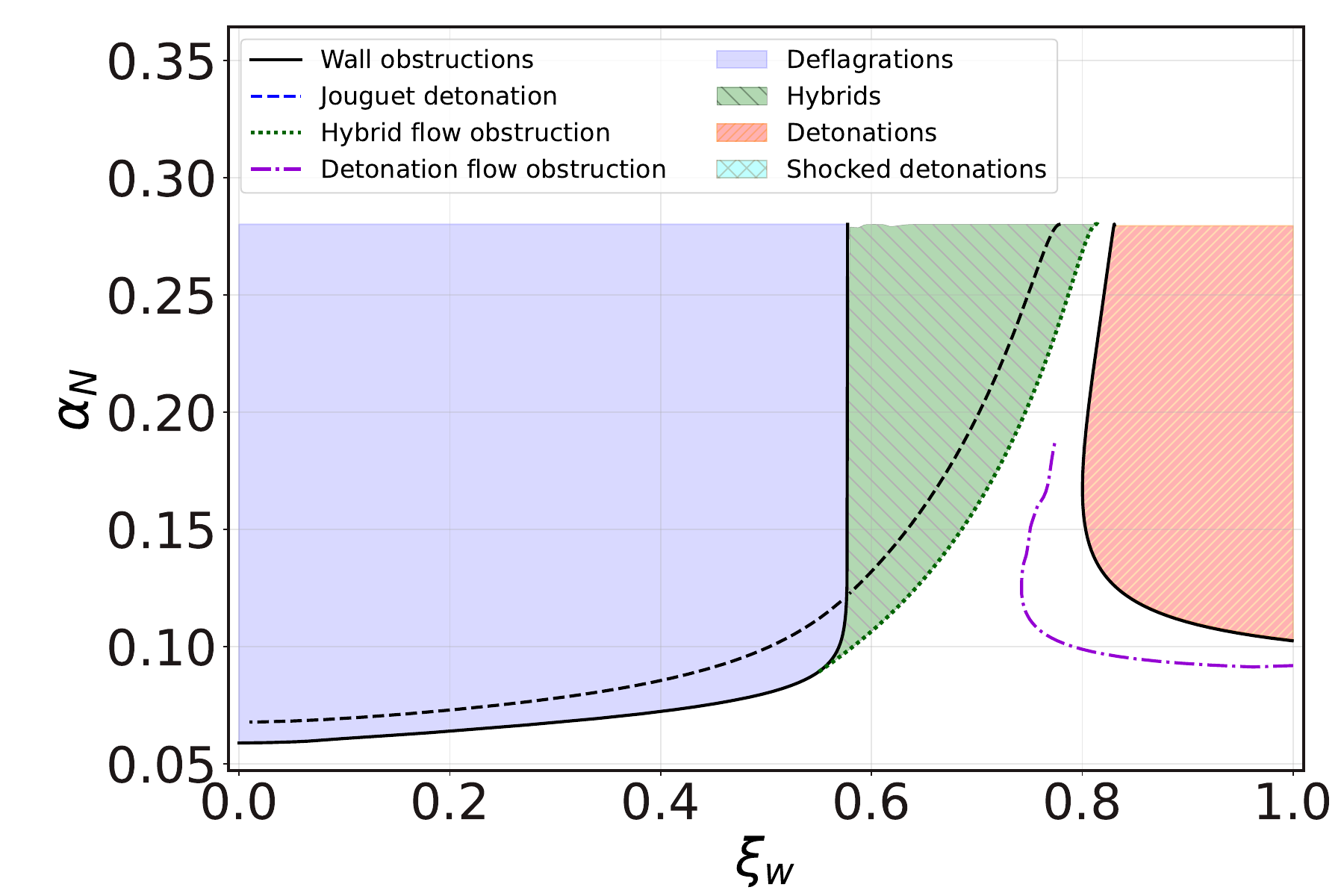}
        \caption{}\label{fig:alpha-d}
    \end{subfigure}\\[15pt]
    \begin{subfigure}{0.49\linewidth}
        \includegraphics[width=\linewidth]{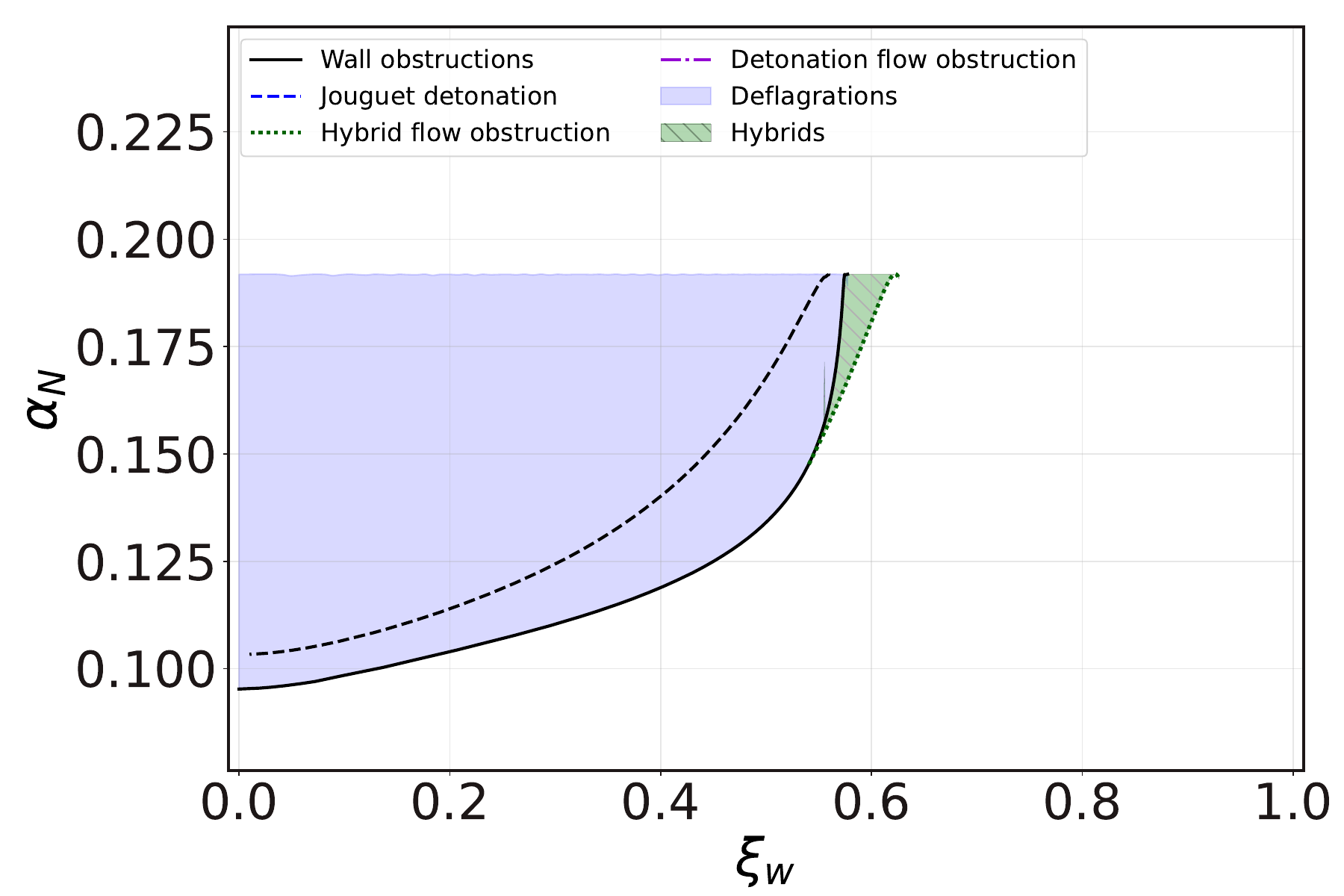}
        \caption{}\label{fig:alpha-e}
    \end{subfigure}\hfill
    \begin{subfigure}{0.49\linewidth}
        \includegraphics[width=\linewidth]{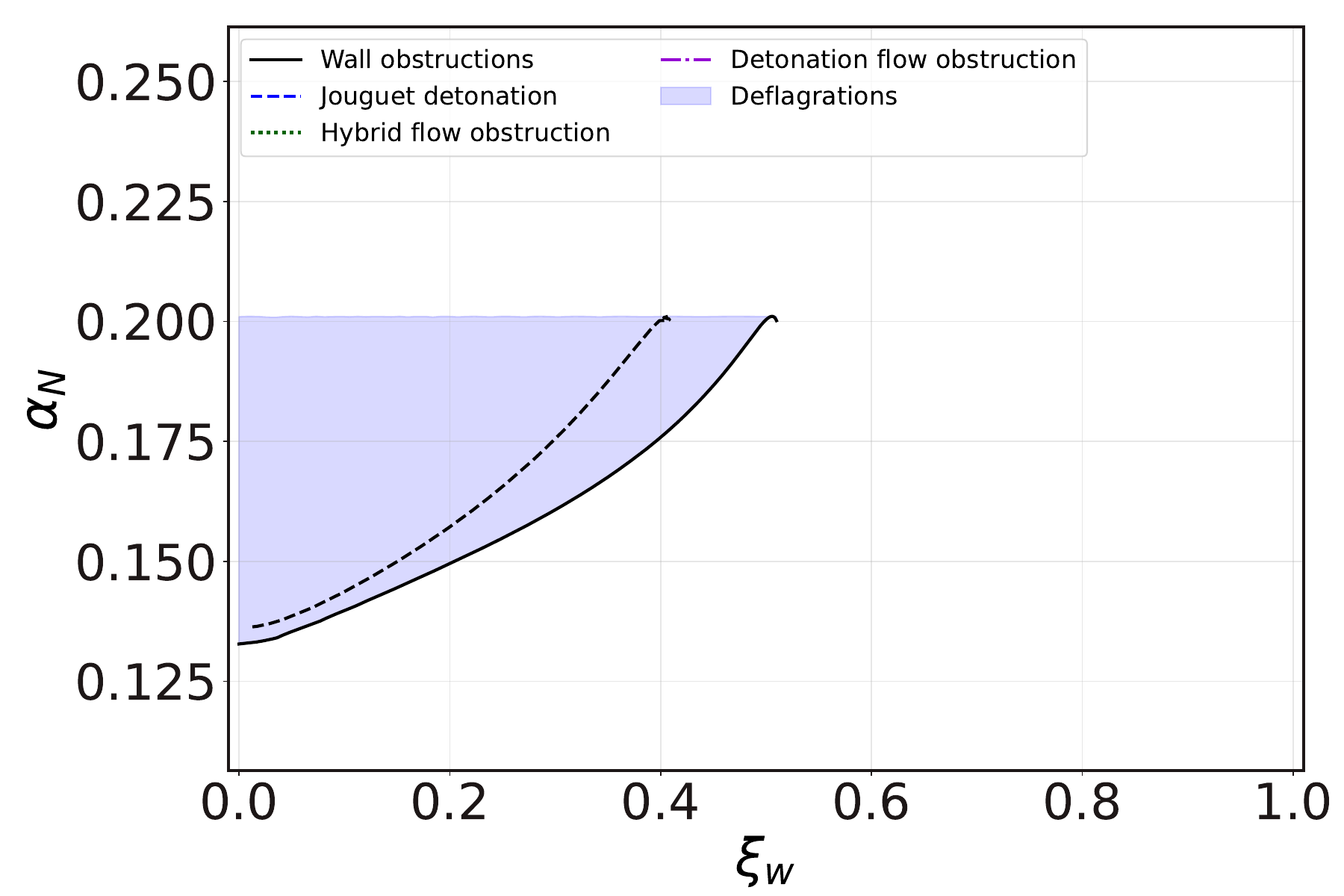}
        \caption{}\label{fig:alpha-f}
    \end{subfigure}
    \caption{Space of allowed solutions (colored regions) in the $(\xi_w, \alpha_N)$ plane, where the transition strength is computed according to \eqref{eq: alphadef}. This requires knowledge not only of the reduced EoS, but of the full EoS---see the main text for the choice made for the latter. Solutions above and below the black dashed curves have positive and negative entropy production at the bubble wall, respectively. The latter must therefore be discarded as physical expanding solutions, although their time reversals can be interpreted as physical collapsing solutions. 
    }
    \label{fig:alpha_N vs xi_w}
\end{figure}

\begin{figure}[tp]
    \centering
    \begin{subfigure}{0.49\linewidth}
        \includegraphics[width=\linewidth]{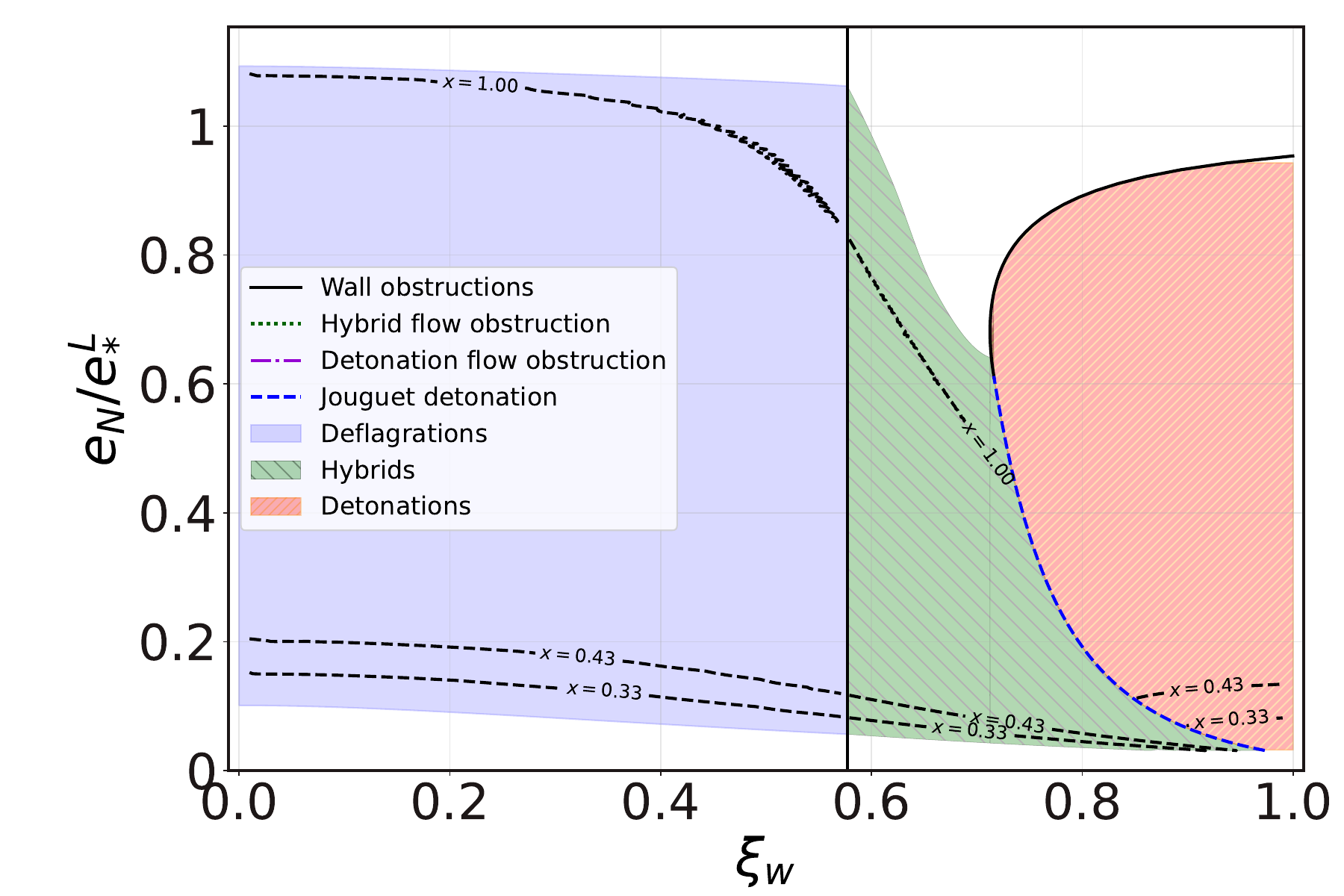}
        \caption{}\label{fig:eN-a}
    \end{subfigure}\hfill
    \begin{subfigure}{0.49\linewidth}
        \includegraphics[width=\linewidth]{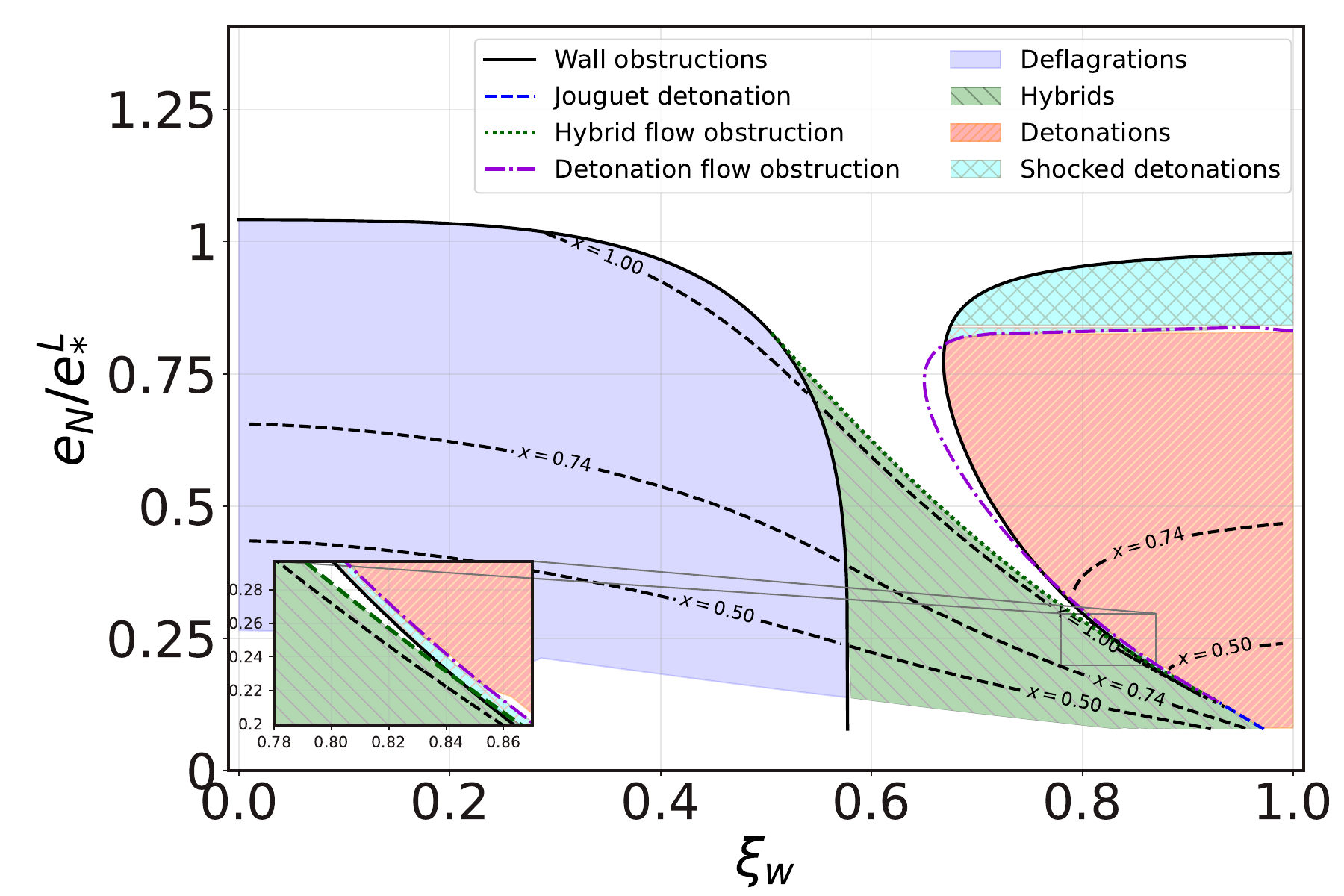}
        \caption{}\label{fig:eN-b}
    \end{subfigure}\\[15pt]
    \begin{subfigure}{0.49\linewidth}
        \includegraphics[width=\linewidth]{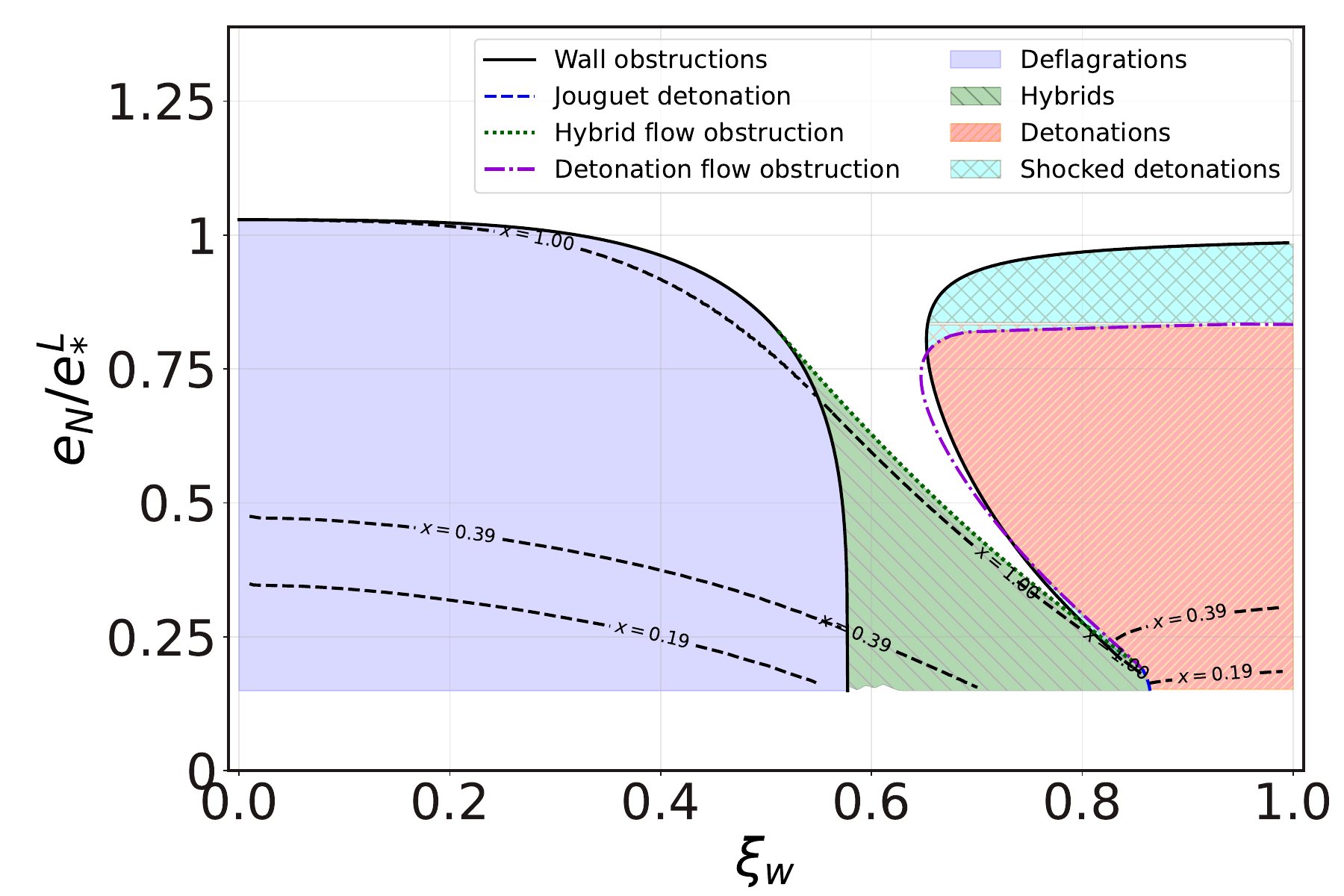}
        \caption{}\label{fig:eN-c}
    \end{subfigure}\hfill
    \begin{subfigure}{0.49\linewidth}
        \includegraphics[width=\linewidth]{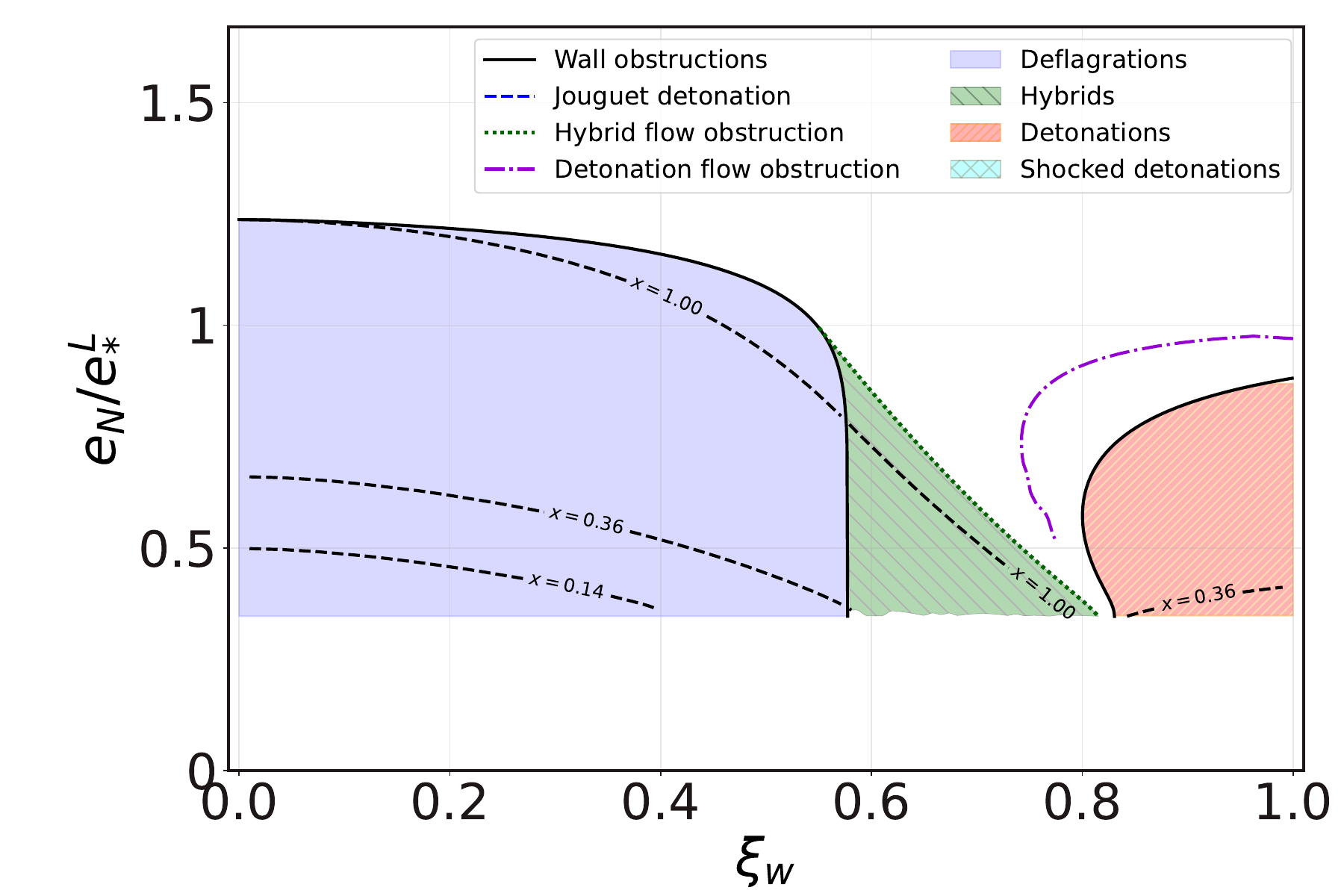}
        \caption{}\label{fig:eN-d}
    \end{subfigure}\\[15pt]
    \begin{subfigure}{0.49\linewidth}
        \includegraphics[width=\linewidth]{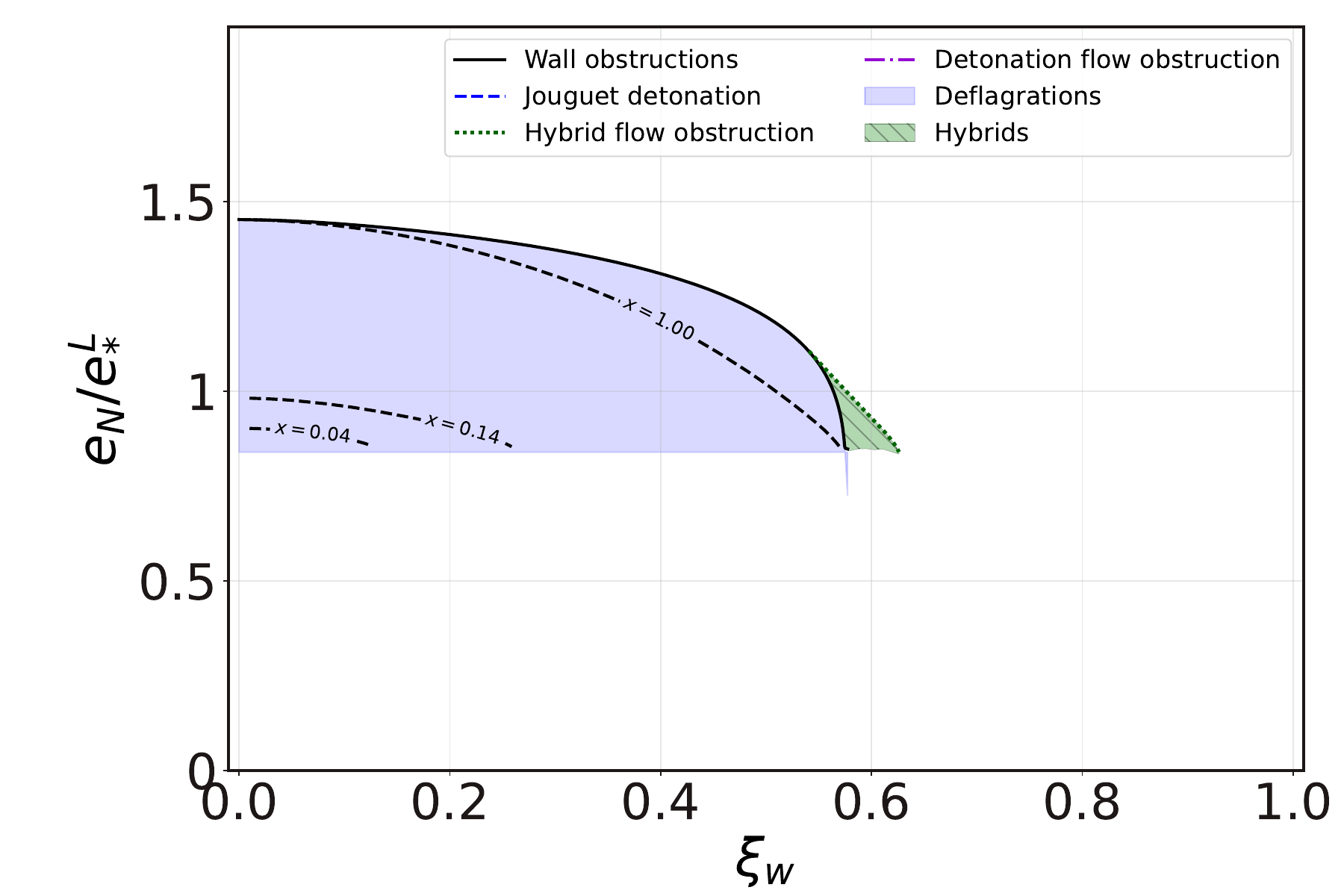}
        \caption{}\label{fig:eN-e}
    \end{subfigure}\hfill
    \begin{subfigure}{0.49\linewidth}
        \includegraphics[width=\linewidth]{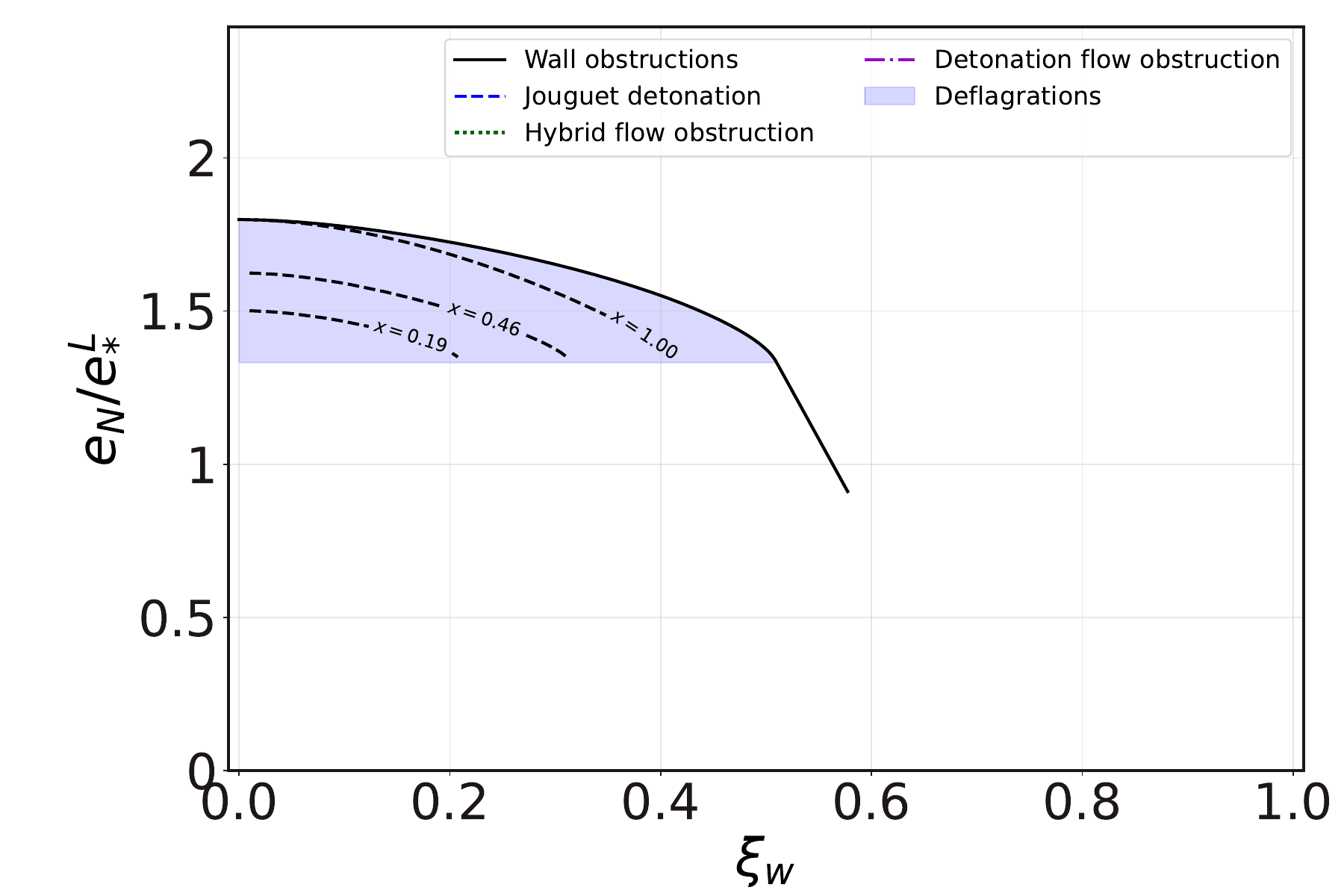}
        \caption{}\label{fig:eN-f}
    \end{subfigure}
    \caption{Space of allowed solutions (colored regions) in the $(\xi_w, e_N)$ plane. The different panels correspond to the EoS shown in \fig{fig:EoS-family}. The black dashed curves mark zero entropy production at the bubble wall for different choices of the full EoS, parametrized as in \eqref{eq: x entropy}.  Solutions below and above these curves have positive and negative entropy production at the wall, respectively. The latter must therefore be discarded as physical expanding solutions, although their time reversals can be interpreted as physical collapsing solutions.
    }
    \label{fig:eN vs xi_w}
\end{figure}

\begin{figure}[tp]
    \centering
    \begin{subfigure}{0.49\linewidth}
        \includegraphics[width=\linewidth]{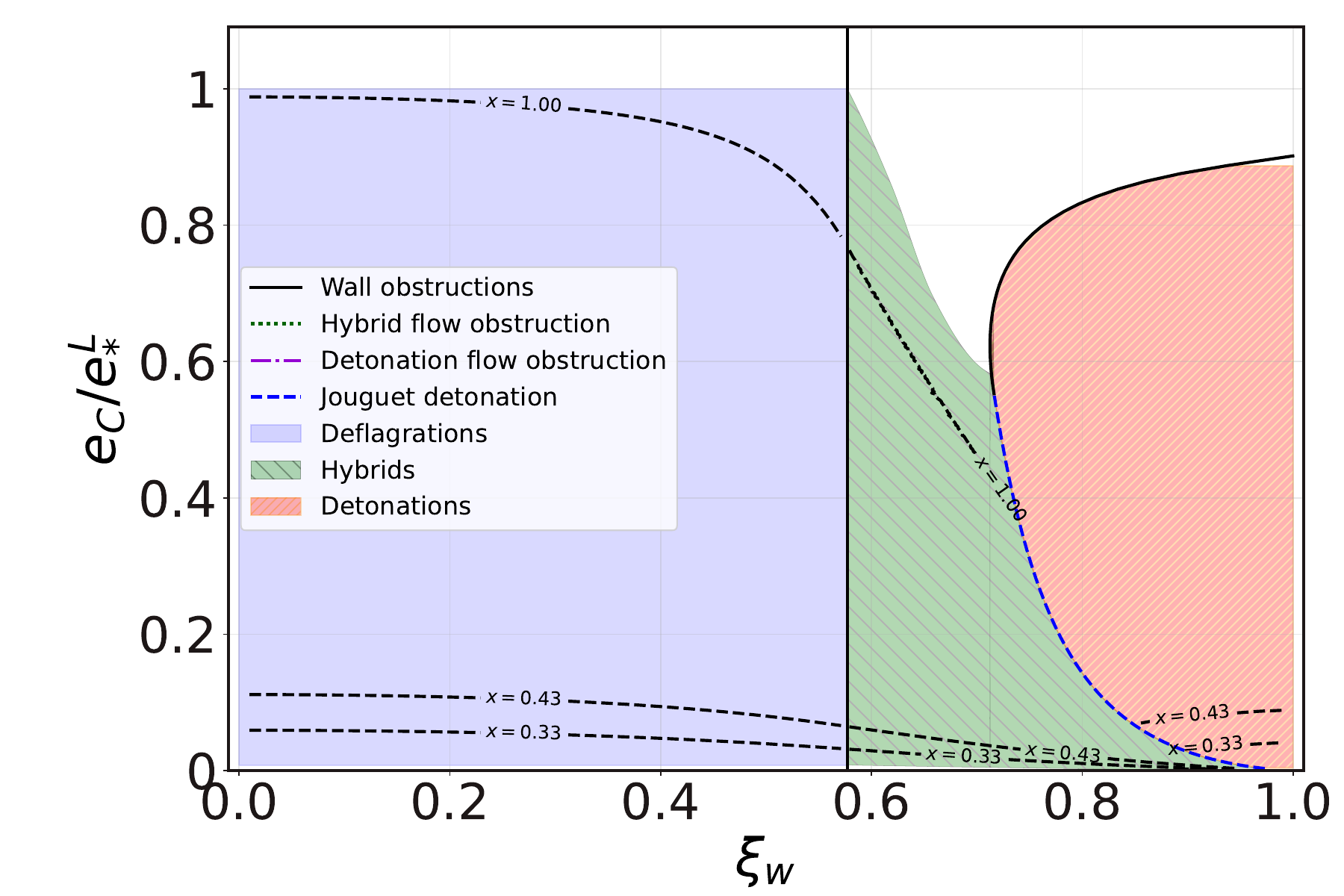}
        \caption{}\label{fig:eC-a}
    \end{subfigure}\hfill
    \begin{subfigure}{0.49\linewidth}
        \includegraphics[width=\linewidth]{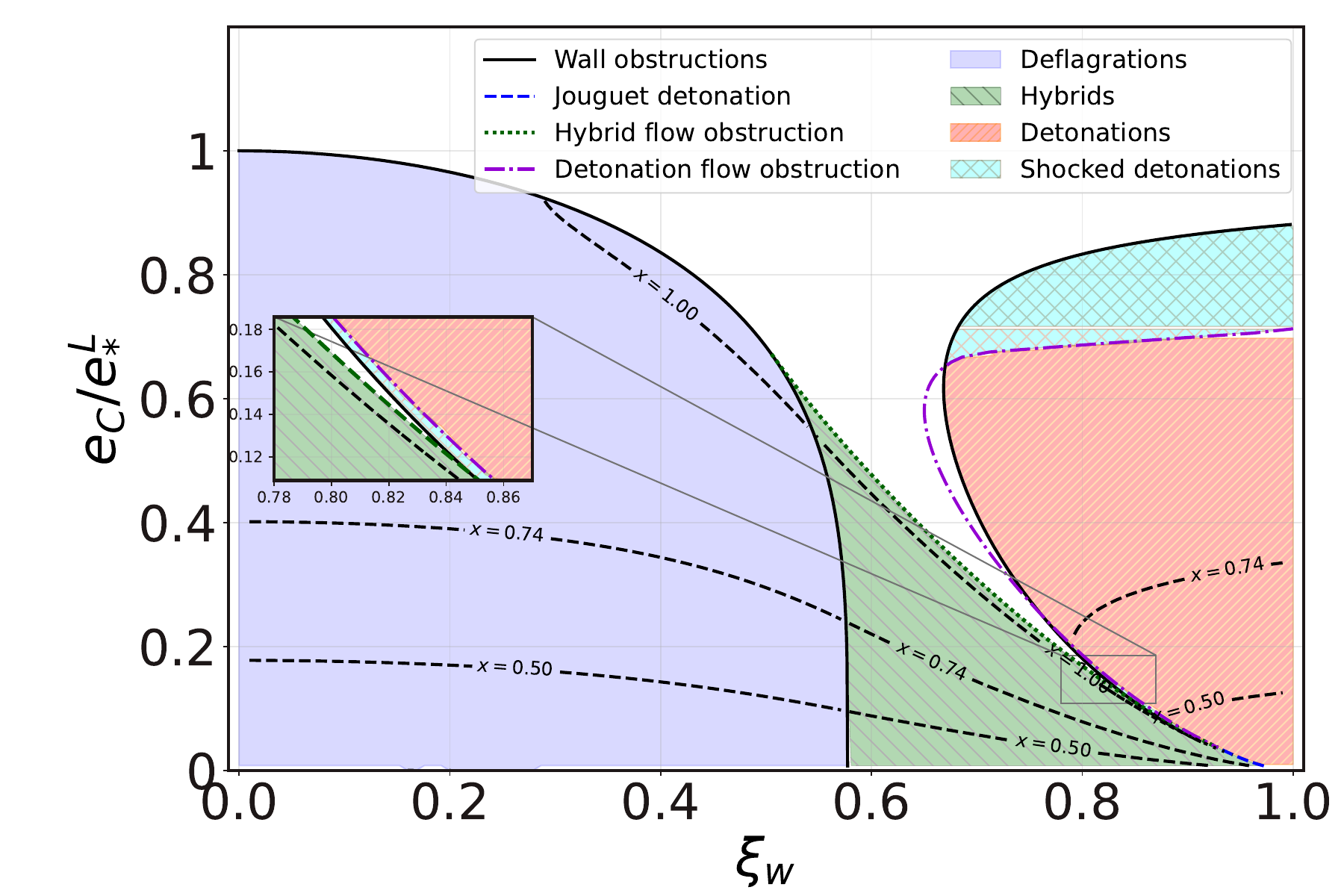}
        \caption{}\label{fig:eC-b}
    \end{subfigure}\\[15pt]
    \begin{subfigure}{0.49\linewidth}
        \includegraphics[width=\linewidth]{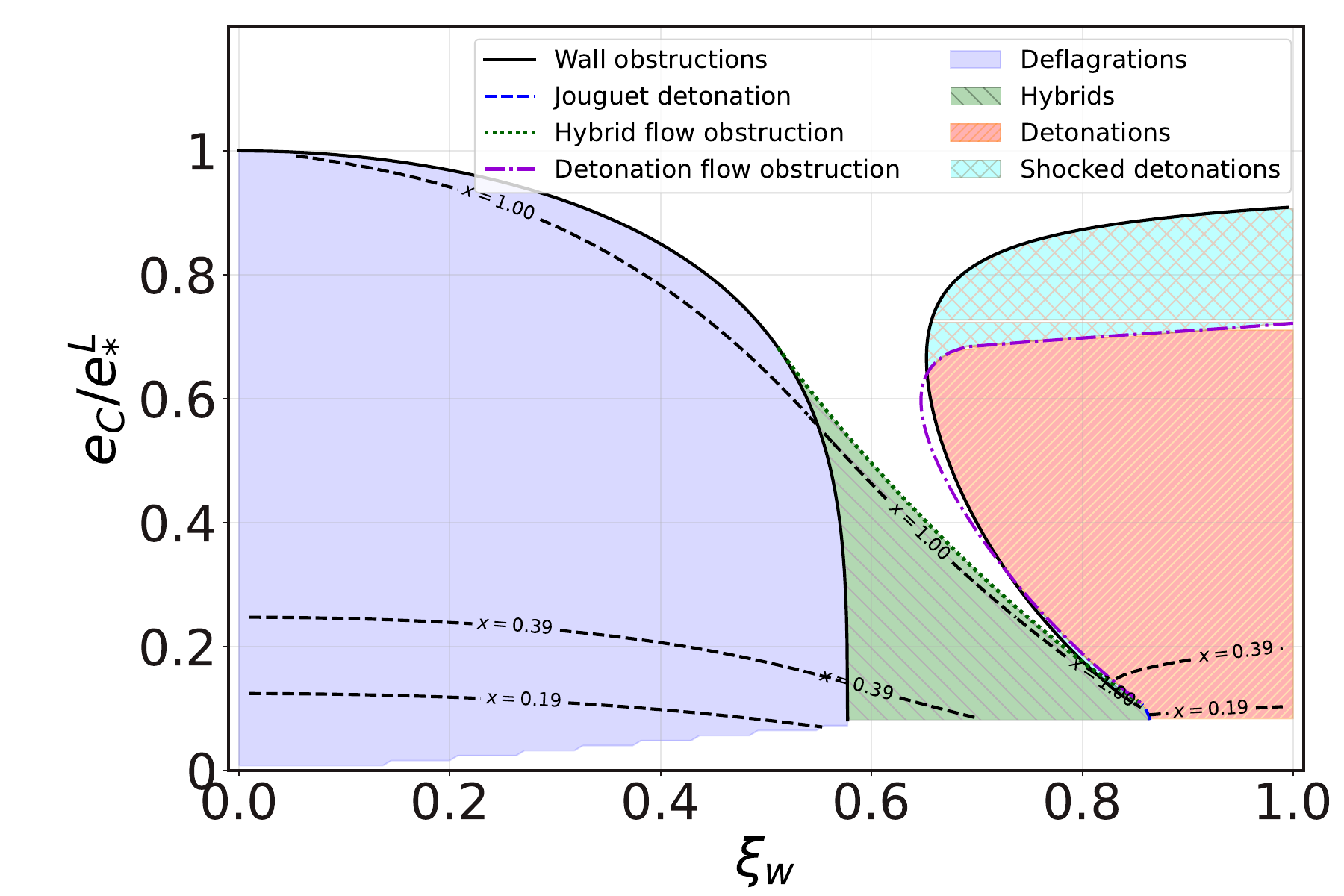}
        \caption{}\label{fig:eC-c}
    \end{subfigure}\hfill
    \begin{subfigure}{0.49\linewidth}
        \includegraphics[width=\linewidth]{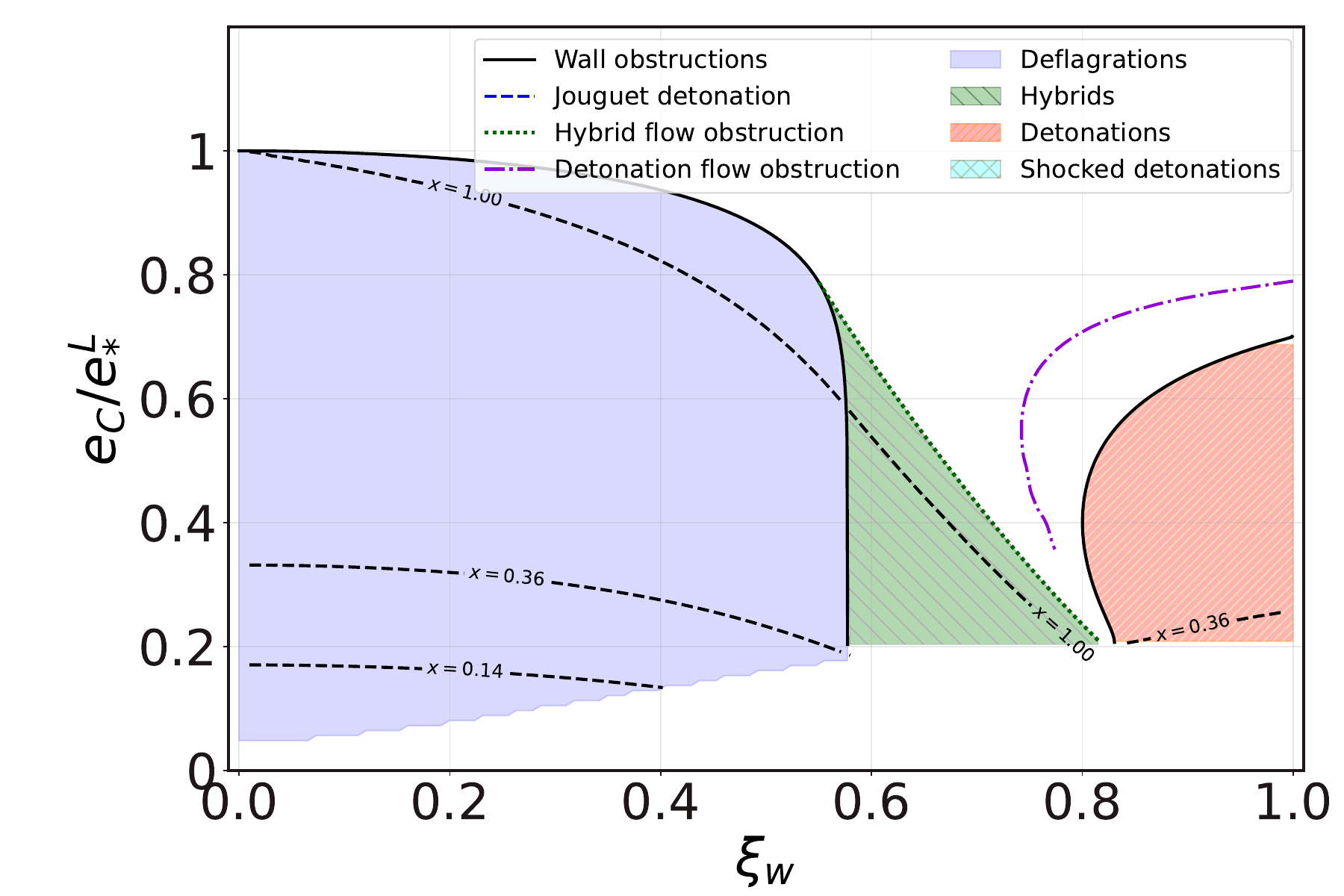}
        \caption{}\label{fig:eC-d}
    \end{subfigure}\\[15pt]
    \begin{subfigure}{0.49\linewidth}
        \includegraphics[width=\linewidth]{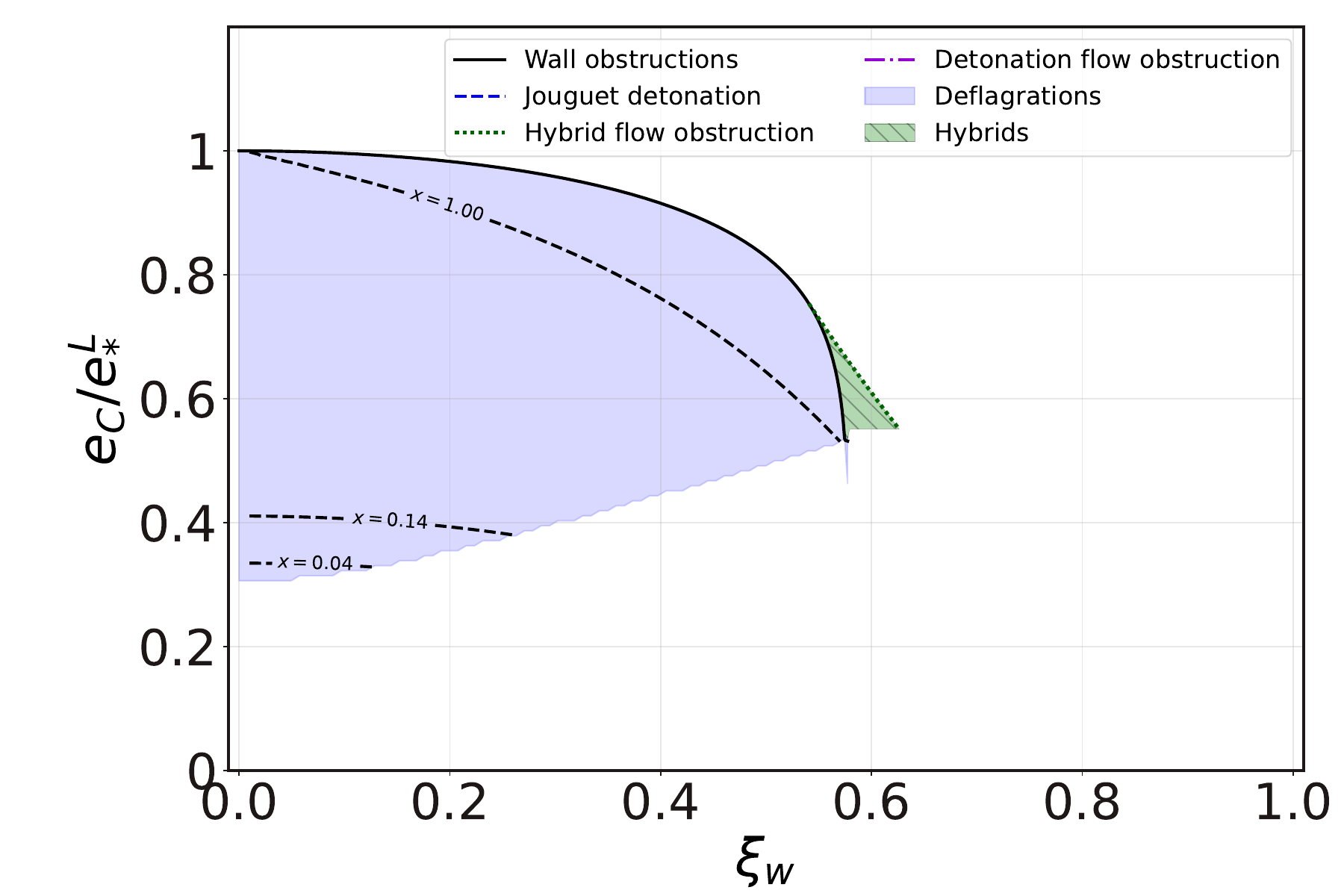}
        \caption{}\label{fig:eC-e}
    \end{subfigure}\hfill
    \begin{subfigure}{0.49\linewidth}
        \includegraphics[width=\linewidth]{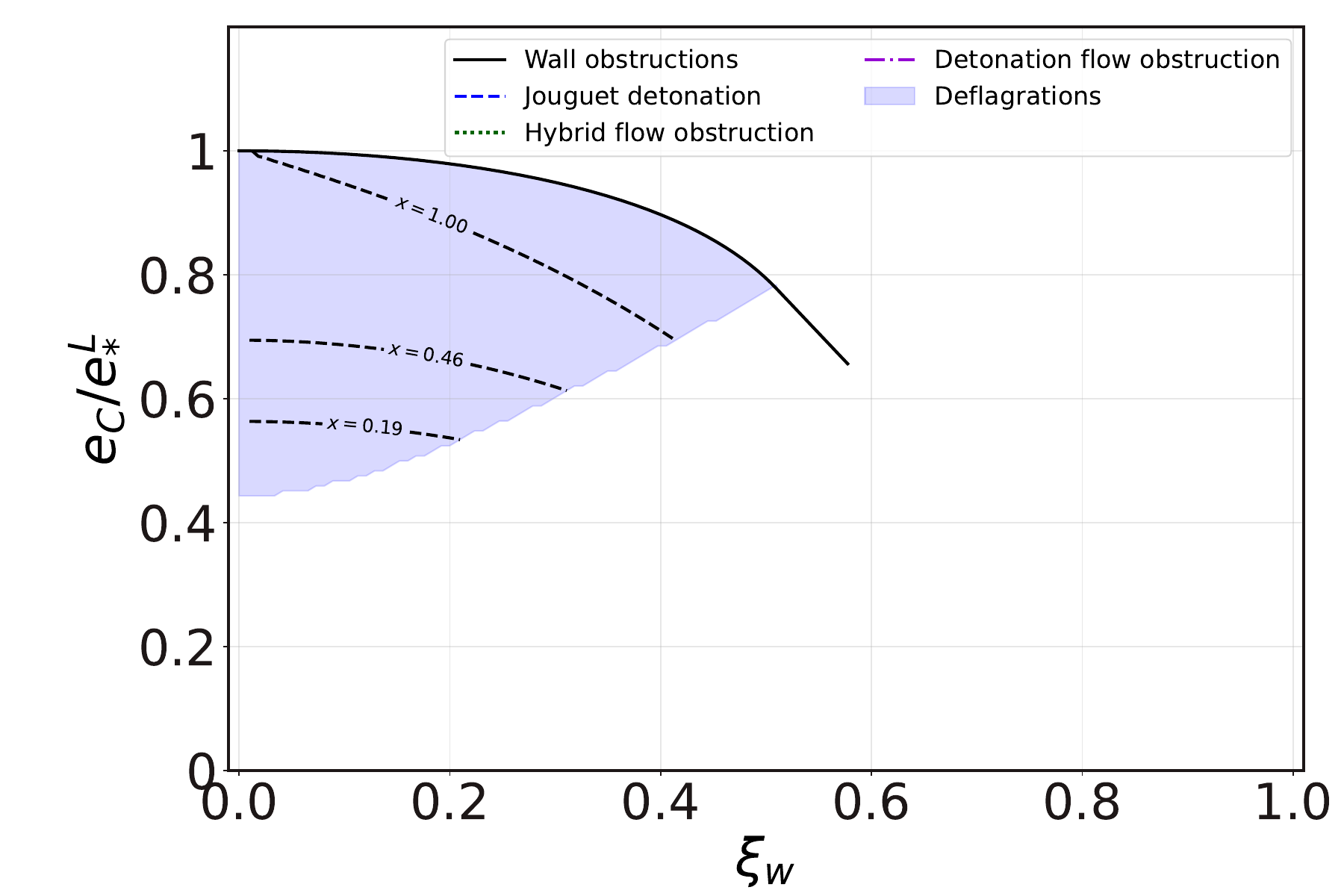}
        \caption{}\label{fig:eC-f}
    \end{subfigure}
    \caption{Space of allowed solutions (colored regions) in the $(\xi_w, e_C)$ plane. The different panels correspond to the EoS shown in \fig{fig:EoS-family}. The black dashed curves mark zero entropy production at the bubble wall for different choices of the full EoS, parametrized as in \eqref{eq: x entropy}.  Solutions below and above these curves have positive and negative entropy production at the wall, respectively. The latter must therefore be discarded as physical expanding solutions, although their time reversals can be interpreted as physical collapsing solutions.
    } 
    \label{fig:ec vs xi_w}
\end{figure}

\section{Space of solutions for different EoS}
\label{space_of_sols}
Having identified the new obstructions present in these EoS, we now examine their impact on the space of allowed bubble solutions. The structure of this space is controlled by the interplay between wall and flow obstructions, whose relative importance depends strongly on the underlying EoS. To illustrate this, we select from the family \eqref{eq: EoS-p} the six representative EoS shown in \fig{fig:EoS-family}, which, from top to bottom and from left to right, interpolate from a bag-model-to a QCD-like EoS. Roughly speaking, this interpolation is achieved by moving the endpoint of the high-energy branch, marked by the blue triangle, upwards and to the right relative to the endpoint of the low-energy branch, marked by the green square. The EoS in the top-left panel is a bag model, with $c_s^2=1/3$ on each branch, but with finite ranges of supercooling and superheating. The EoS in the bottom-right panel is QCD-like, with thermodynamic observables that are single-valued functions of the energy density.

The resulting spaces of allowed solutions are shown in Figs.~\ref{fig:alpha_N vs xi_w}, \ref{fig:eN vs xi_w} and \ref{fig:ec vs xi_w}, where we display the allowed regions in the $(\xi_w,\alpha_N)$, $(\xi_w,e_N)$ and $(\xi_w,e_C)$ planes, respectively. Here $e_N$ denotes the nucleation energy density, while $e_C$ denotes the energy density at the center of the bubble. These three sets of plots contain the same information, presented in different parametrizations, except for one important distinction: the reduced EoS is sufficient to produce Figs.~\ref{fig:eN vs xi_w} and \ref{fig:ec vs xi_w}, whereas Fig.~\ref{fig:alpha_N vs xi_w} requires knowledge of the full EoS. We will return below to the specific choice of this additional information made in order to construct Fig.~\ref{fig:alpha_N vs xi_w}. The deflagration, hybrid and detonation regions in Figs.~\ref{fig:alpha_N vs xi_w}--\ref{fig:ec vs xi_w} are shown in purple, green and red, respectively. Solid curves denote WOs, while dashed colored curves denote FOs. The dashed black curves indicate solutions with zero entropy production (see below).

To understand how the space of solutions changes with the EoS, we start with~\fig{fig:alpha-a}. As mentioned above, this EoS corresponds to a bag model with finite amounts of supercooling and superheating. We see that deflagrations transition into hybrids at \mbox{$\xi_w=c_s=1/\sqrt{3}$}, indicated by the vertical solid black line, while hybrids connect continuously to detonations at the Jouguet velocity, shown by the dashed blue curve. As the transition strength increases, solutions with smaller wall velocities disappear, until only very fast detonations remain. Any deviations from the bag-model behaviour at small transition strength arise from the fact that we have restricted  the range of energies over which each phase is defined.

As the EoS departs from the bag-model limit, new features begin to appear. In \fig{fig:alpha-b}, 
we show the space of allowed solutions for EoS of 
\fig{fig:EoS-b}, whose speed of sound is approximately constant except near the endpoints of the superheating and supercooling ranges, where it vanishes. At large transition strength, the flow does not probe states close to $\etH$ or $\etL$, and the solution space therefore resembles that of the bag model. Deviations from the constant-speed-of-sound behaviour become important only at small transition strength. Deflagrations are bounded by the WO, shown as a solid black line, and hybrids by the DCP, shown as a dashed green line. Detonations are delimited both by the DCP, shown as a purple dot-dashed line, beyond which they transition to shocked detonations, and by the maximum allowed superheating of the low-energy phase, shown as a solid black line.

Restricting the range of supercooling, corresponding to the EoS in \fig{fig:EoS-c}, reduces the maximum accessible transition strength. As a result, the part of the solution space that resembled the bag-model case is removed. This can be seen by comparing \fig{fig:alpha-b} and \fig{fig:alpha-c}: the latter closely resembles the low-$\alpha_N$ region of the former.

Moving to the EoS in \fig{fig:EoS-d}, we see for the first time the appearance of a gap between hybrids and detonations in the space of solutions. Further reducing the maximum amounts of supercooling and superheating leads to the two plots in \fig{fig:alpha-e} and \fig{fig:alpha-f}. In \fig{fig:alpha-e}, detonations are absent, while in \fig{fig:alpha-f} both hybrids and detonations are excluded. 

Finally, we must determine which of the hydrodynamically allowed solutions remain admissible once the second law of thermodynamics is imposed. To do so, we draw curves of zero entropy production at the wall.\footnote{The sign of entropy production at shocks is unaffected by the choice of $s_*^L$ and $s_*^H$, as they take place within a single phase of the EoS, and we have verified that all shocks found are allowed.} These curves depend not only on the reduced EoS, $p(e)$, but also on the choice of entropy function $s(T)$, and therefore on the full EoS. At fixed reduced EoS, this freedom can be parametrized by the location of the critical temperature within its allowed range. We define
\begin{equation}
    x \equiv \frac{T_c - T(e_*^H)}{T(\etL) - T(e_*^H)}\,,
    \label{eq: x entropy}
\end{equation}
so that $x=0$ places $T_c$ at its lowest possible value, corresponding to the point of maximal supercooling, while $x=1$ places it at its highest possible value, corresponding to the point of maximal superheating. For a fixed choice of parameters in \eqref{eq: EoS-p}, different choices of $\stL$ and $\stH$ generate a family of entropy functions $s(T)$, and hence a family of zero-entropy-production curves labeled by $x$. Larger values of $x$ correspond to a larger metastable region on the high-energy branch, or equivalently to a larger range of nucleation temperatures, and  lead to less restrictive second-law bounds.

There is, however, an important distinction between the different parametrizations of the solution space. In the $(\xi_w,e_N)$ and $(\xi_w, e_C)$ planes, the axes are energy densities and therefore depend only on the reduced EoS. In these plots, changing $s(T)$ leaves the hydrodynamic obstruction structure unchanged and merely moves the zero-entropy-production curve on top of the same background. By contrast, in the $(\xi_w, \alpha_N)$ plane the transition strength $\alpha_N$ is defined by comparing the two phases at a common temperature. Since the temperature assignment follows from the full EoS, changing $s(T)$ relabels the states in temperature and changes which pair of states enters the definition of $\alpha_N$. Thus the obstruction contours themselves are deformed, rather than only the zero-entropy-production curve being displaced. For this reason, in the $\alpha_N$ plots we show the zero-entropy-production contour only for a single representative value of $x$ close to unity, corresponding to the least restrictive second-law bound.

In Fig.~\ref{fig:alpha_N vs xi_w}, the solutions below the dashed black curves must be discarded for the model labeled by the corresponding value of $x$. In Figs.~\ref{fig:eN vs xi_w} and \ref{fig:ec vs xi_w}, instead, the discarded solutions lie above the dashed black curves.  We see that, for deflagrations and hybrids, the second law is always more restrictive than the hydrodynamic obstructions: the WO for deflagrations and the FO for hybrids are reached only after the entropy-production condition has already been violated. By contrast, for detonations, the hydrodynamic obstructions can be the most restrictive constraint for suitable choices of $s(T)$.

Nevertheless, whenever the hydrodynamic obstructions produce a gap between different classes of solutions, this gap necessarily persists after imposing the second law---see Figs.~\ref{fig:alpha-d} and \ref{fig:alpha-e}. If no hydrodynamic gap is present, however, the entropy-production condition may or may not generate one, depending on the choice of $s(T)$, as illustrated by Figs.~\ref{fig:alpha-a}, \ref{fig:alpha-b} and \ref{fig:alpha-c}.

Figures~\ref{fig:eN vs xi_w} and~\ref{fig:ec vs xi_w} also make it possible to characterize the phase-separated states in the static limit, $\xi_w\to 0$. The main lesson is that the largest range of allowed central energy densities, $e_C$, is obtained in this limit. As the wall velocity increases, this range shrinks. This behaviour is not obvious a priori, but emerges from the numerical analysis of the different EoS considered here.

For QCD-like EoS, shown in the middle and bottom rows, the lower edge of the allowed $e_C$ region moves upward as $\xi_w$ increases. This reflects the fact that the amount of supercooling available on the high-energy branch is finite: the nucleation energy cannot be lowered beyond the endpoint $e_*^H$. Thus the configuration with the smallest allowed $e_C$ is tied to the maximally supercooled state, for which $e_N=e_*^H$. Increasing the wall velocity then forces the central energy density to increase, reducing the allowed range of $e_C$.

For bag-model-like EoS, shown in the top row, the limiting behaviour is different. In this case the lower bound on the central energy density is simply $e_C\geq 0$. The widest range of $e_C$ is still reached in the static limit, but the corresponding statement does not hold for the nucleation energy. Instead, as $\xi_w$ increases, the lower boundary remains pinned at $e_C=0$, while the associated nucleation energy $e_N$ decreases.

A second conclusion concerns the relation between the phase-separated states and the critical temperature. If one works only with the reduced EoS, $p(e)$, then any pair of states with the same pressure, one on each stable branch, can be regarded as a possible coexistence configuration in the static limit $\xi_w\to 0$. This is because the reduced EoS contains no information about the temperature assignment of these states, and therefore does not determine which equal-pressure pair corresponds to the true critical temperature.

Once the two integration constants of the entropy density are fixed and Eq.~\eqref{eq: s(e)-ODE} is solved, this degeneracy is removed. The full EoS assigns a temperature to each state, and only one equal-pressure pair also has a common temperature. This pair is the true phase-coexistence state at $T_c$. Equivalently, it is the only static configuration with vanishing entropy production at the wall, as required in the limit $\xi_w\to 0$. In Figs.~\ref{fig:eN vs xi_w} and~\ref{fig:ec vs xi_w}, this state is identified by the intersection of the zero-entropy-production curves, shown as black dashed lines, with the vertical line $\xi_w=0$.

In the bag model, this last point leads to a lower bound on the critical temperature.  Mechanical equilibrium, $p_H=p_L$, does not by itself select a unique transition. Instead, it is satisfied along the entire line
\begin{equation}
    e_L = e_H - 4\epsilon \,,
\end{equation}
so any equal-pressure pair on this line is admissible from the point of view of the reduced EoS alone. In the static limit, however, the central energy density is the energy density of the low-energy phase, $e_C=e_L$. Imposing the endpoint value $e_C=e_L=0$ therefore selects the endpoint of the coexistence line, for which $e_N=e_H=4\epsilon$. Using
\begin{equation}
    e_N = a_H T_c^4 + \epsilon
\end{equation}
then fixes the critical temperature to be
\begin{equation}
    T_c = \left(\frac{3\epsilon}{a_H}\right)^{1/4}.
\end{equation}
The lower bound $e_C\geq 0$ thus forbids nucleation energies below $e_N=4\epsilon$  in the static limit. Since lowering $e_N$ would correspond to lowering $T_c$ at fixed high-energy EoS, this means that the critical temperature is bounded from below,
\begin{equation}
    T_c \geq \left(\frac{3\epsilon}{a_H}\right)^{1/4}.
\end{equation}
Thus, in the bag model, the absence of static coexistence endpoints below $e_N=4\epsilon$ and the existence of a minimum critical temperature are two equivalent ways of stating the same constraint.

\section{Kinetic energy fraction}
\label{sec: section 6}

The kinetic energy fraction is a key parameter in estimates of the gravitational-wave spectrum \cite{Caprini:2019egz}. Nonlinear effects during bubble collisions mean that the true kinetic energy fraction cannot always be accurately approximated by that of a single bubble \cite{Cutting:2019zws}. Nevertheless, we will use the single-bubble kinetic energy fraction as a proxy to identify which bubble solutions are expected to produce the strongest gravitational-wave signal.

Let us emphasize that our focus here is not the direct modification of the kinetic energy fraction by a non-constant speed of sound for a given bubble solution. Rather, we study an indirect effect: the non-constant speed of sound gives rise to hydrodynamic obstructions, which restrict the space of admissible solutions and thereby limit the kinetic energy fractions that can actually be realized. Direct effects of a non-constant speed of sound on the fluid energy budget, including comparisons with the constant-speed-of-sound case, have been studied in \cite{Giese:2020rtr,Giese:2020znk}.

The kinetic energy fraction is defined as the ratio between the kinetic energy and the total energy of the fluid in motion. To make this definition precise, we first compute the energy density of the fluid in the lab frame from \eqref{eq:ideal-hydro}. Using the enthalpy, $w=e+p$, and the trace variable defined in \eqref{eq:trace-theta}, this energy density can be decomposed as
\begin{equation}
    T^{tt} \,\,= \,\, w\gamma^2-p
    \,\,=\,\, \frac{3}{4}w\gamma^2
    + \frac{1}{4}w\gamma^2 v^2
    + \theta .
\end{equation}
The first term in the last expression is usually identified with the thermal energy, the second with the kinetic energy, and the last with the vacuum-energy contribution, since $\theta$ reduces to the bag constant in the bag-model. Motivated by this splitting, we define the kinetic energy fraction as
\begin{equation}
    K =
    \frac{
    \int_{V_{\text{bubble}}} \mathrm{d}^3x\, w\gamma^2 v^2
    }{
    \int_{V_{\text{bubble}}} \mathrm{d}^3x\, (w\gamma^2-p)
    } .
    \label{eq: kinetic-energy-fraction}
\end{equation}
Here the integrals are taken over the sphere, centered on the bubble, that contains all the fluid in motion. Thus
\begin{equation}
    V_{\text{bubble}} = \frac{4\pi}{3}\,\xi_{\max}^3 t^3,
    \qquad
    \xi_{\max}=\max(\xi_w,\xi_{\text{sh}}),
\end{equation}
where $\xi_{\text{sh}}$ is the position of the shock ahead of the wall, if present.

This definition is also motivated by the structure of the stress tensor. The numerator is the part of the fluid energy associated with the non-vanishing spatial components of the stress-energy tensor, and it is these components that source gravitational radiation \cite{Kamionkowski:1993fg}. Of course, an isolated spherical bubble does not radiate gravitational waves by itself; rather, this quantity provides the relevant energy budget once spherical symmetry is broken, for example during bubble collisions.

The denominator can be simplified considerably by using energy conservation. The region $V_{\text{bubble}}$ is chosen to enclose all the fluid in motion. Therefore, the total energy contained in this region at time $t$ must equal the energy that would have been contained in the same spatial volume in the unperturbed configuration, namely the high-energy phase at the nucleation temperature and at rest. Hence
\begin{equation}
    \int_{V_{\text{bubble}}} \mathrm{d}^3x\, (w\gamma^2-p)
    =
    \int_{V_{\text{bubble}}} \mathrm{d}^3x\, (w_N-p_N)
    =
    e_N V_{\text{bubble}}
    =
    \frac{4\pi}{3}\,\xi_{\max}^3 t^3 e_N .
\end{equation}

Substituting this result into the denominator of \eqref{eq: kinetic-energy-fraction}, and changing to the self-similar variable $r=\xi t$, we obtain
\begin{equation}
     K =
     \frac{3}{e_N \xi_{\max}^3}
     \int_{0}^{\xi_{\max}} \mathrm{d}\xi\, \xi^2 w\gamma^2 v^2 .
\end{equation}
Equivalently, in terms of the transition strength, this can be written as
\begin{equation}
     K =
     \frac{3\alpha_N}{
     \left[(1+\alpha_N)\theta_{H,N}-\theta_{L,N}\right]\xi_{\max}^3}
     \int_{0}^{\xi_{\max}} \mathrm{d}\xi\, \xi^2 w\gamma^2 v^2 ,
\end{equation}
where
\begin{equation}
    \theta_{H,N}=\theta_H(T_N),
    \qquad
    \theta_{L,N}=\theta_L(T_N).
\end{equation}

For the bag model \eqref{eq: Bag-Model-pressure}, for which $\theta_{\HH}=\epsilon$ and $\theta_{\LL}=0$, the previous expression reduces to
\begin{equation}
    K =
    \frac{3\alpha_N}{(1+\alpha_N)\epsilon\,\xi_{\max}^3}
    \int_{0}^{\xi_{\max}} \mathrm{d}\xi\, \xi^2 w\gamma^2 v^2 .
    \label{eq:kinetic-energy-fraction-bag-model}
\end{equation}
This differs slightly from the form often used in the literature, see e.g.~\cite{Caprini:2019egz,Giese:2020znk,Barni:2024lkj}, where the denominator contains $\xi_w^3$ rather than $\xi_{\max}^3$. We use \eqref{eq:kinetic-energy-fraction-bag-model} because $\xi_{\max}$ follows directly from the energy-balance argument above: the normalization volume should enclose all the fluid in motion, whose outer edge is located at $\xi_{\max}=\max(\xi_w,\xi_{\text{sh}})$.

\begin{figure}[tp]
    \centering
    \begin{subfigure}{0.49\linewidth}
        \includegraphics[width=\linewidth]{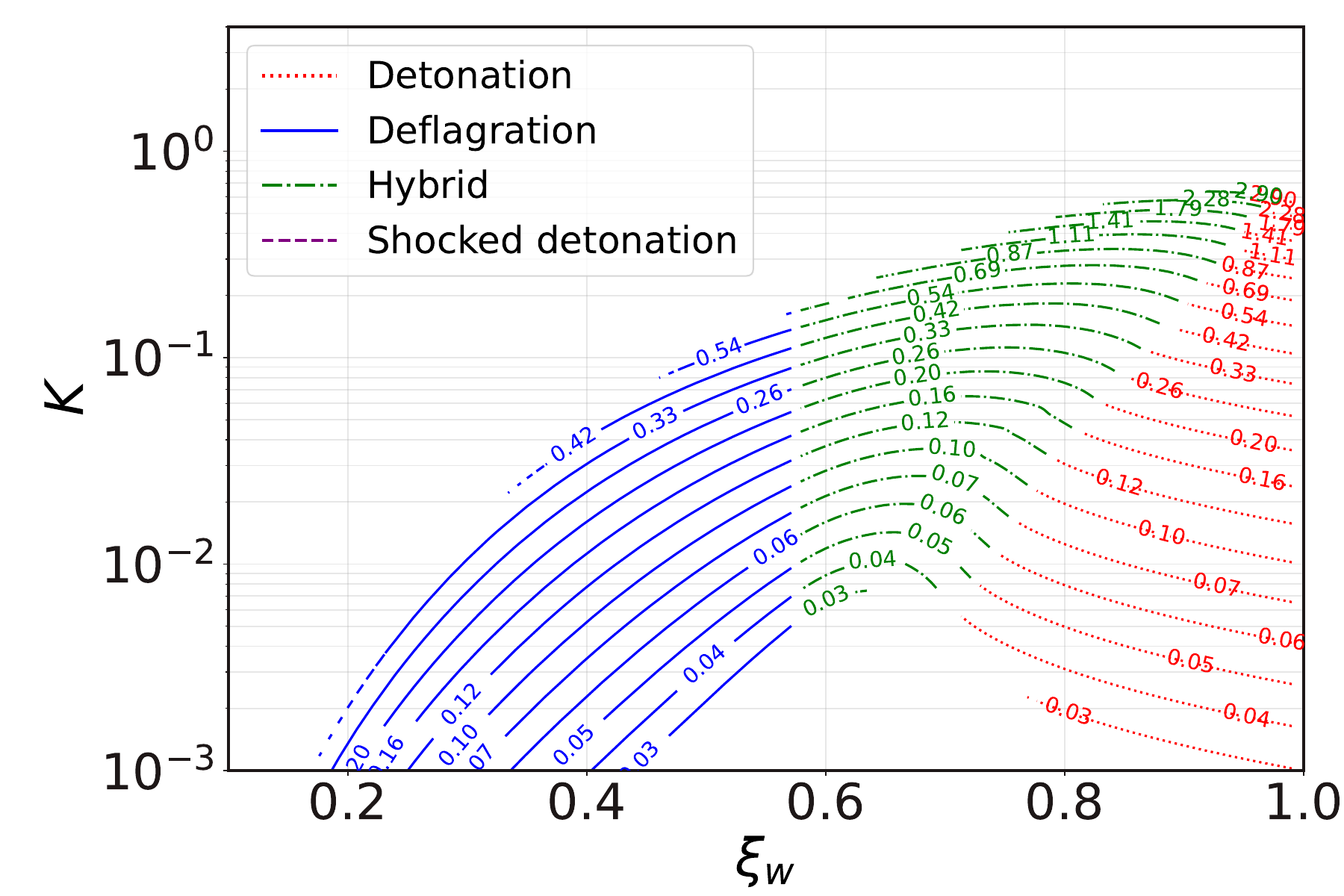}
        \caption{}\label{fig:eff-a}
    \end{subfigure}\hfill
    \begin{subfigure}{0.49\linewidth}
        \includegraphics[width=\linewidth]{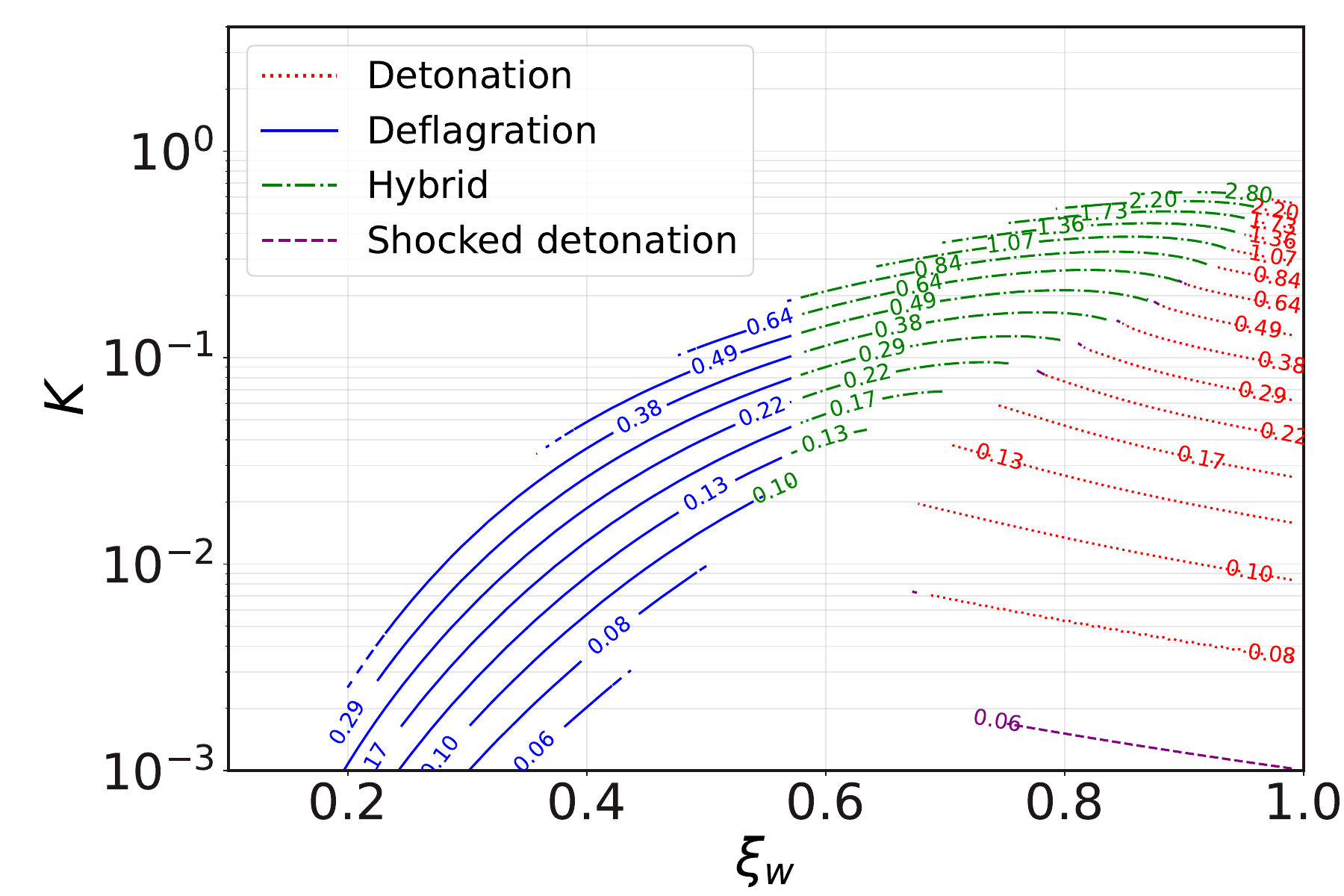}
        \caption{}\label{fig:eff-b}
    \end{subfigure}\\[15pt]
    \begin{subfigure}{0.49\linewidth}
        \includegraphics[width=\linewidth]{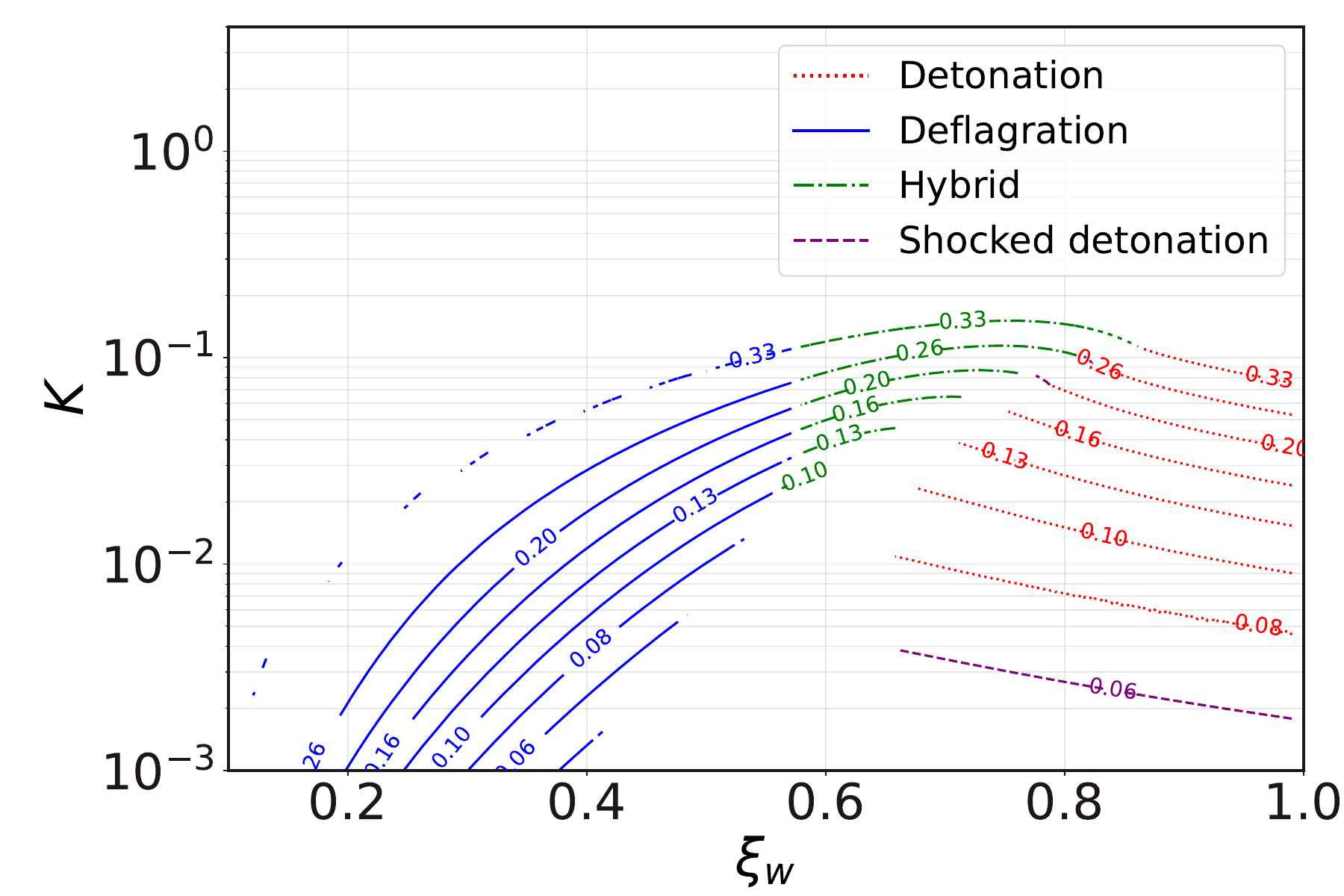}
        \caption{}\label{fig:eff-c}
    \end{subfigure}\hfill
    \begin{subfigure}{0.49\linewidth}
        \includegraphics[width=\linewidth]{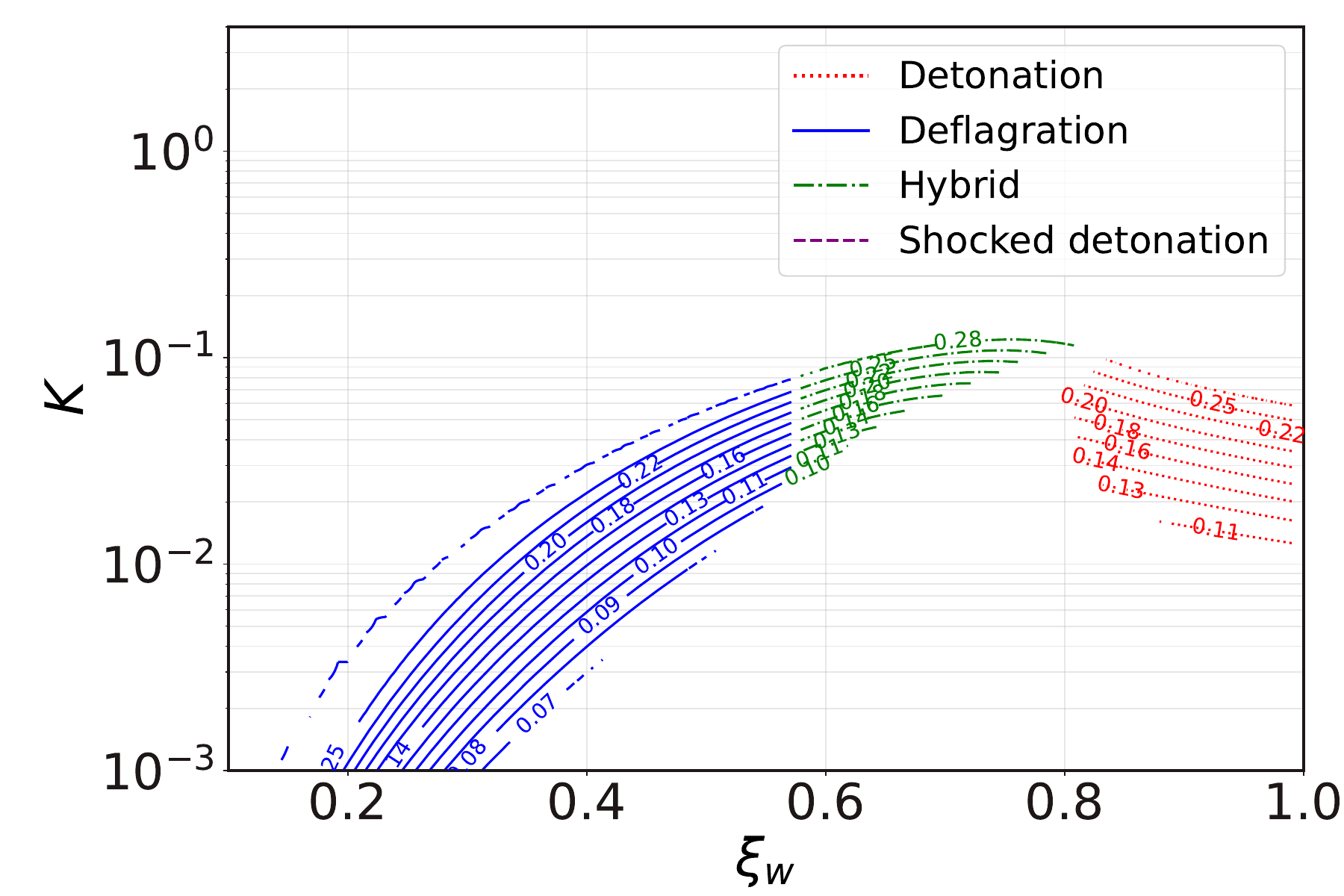}
        \caption{}\label{fig:eff-d}
    \end{subfigure}\\[15pt]
    \begin{subfigure}{0.49\linewidth}
        \includegraphics[width=\linewidth]{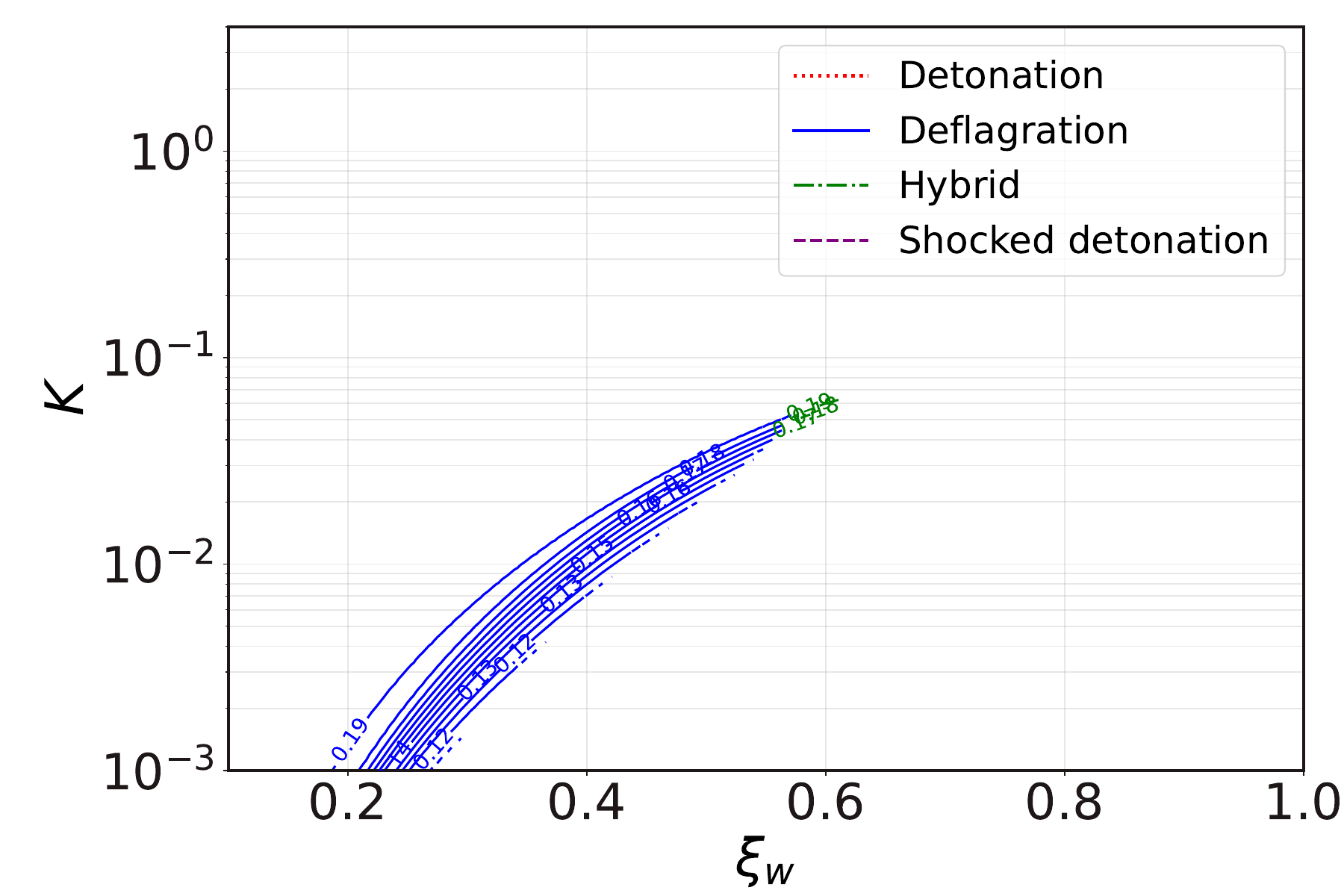}
        \caption{}\label{fig:eff-e}
    \end{subfigure}\hfill
    \begin{subfigure}{0.49\linewidth}
        \includegraphics[width=\linewidth]{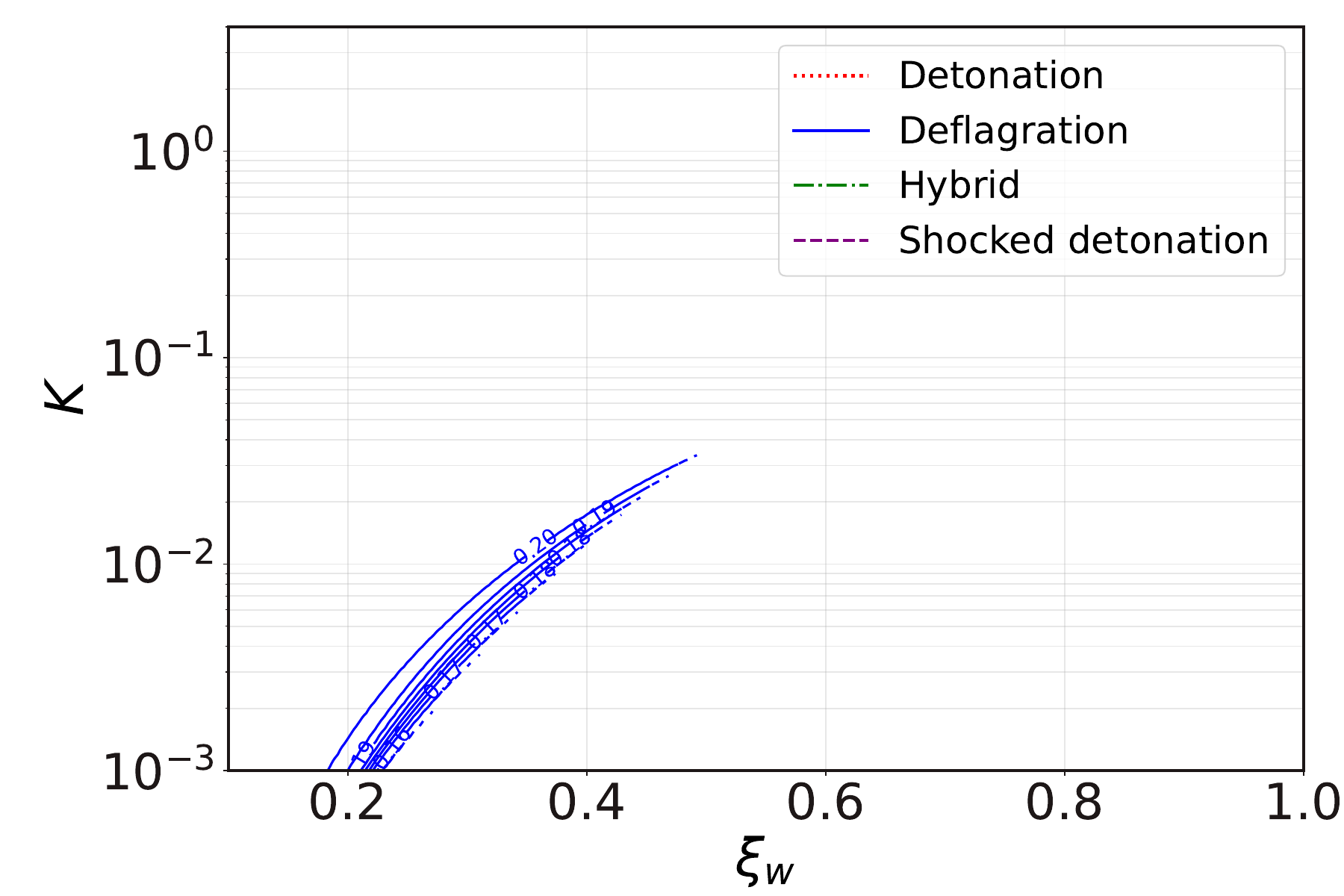}
        \caption{}\label{fig:eff-f}
    \end{subfigure}
    \caption{Kinetic energy fraction, $K$, defined in \eqref{eq: kinetic-energy-fraction}, as a function of the bubble wall velocity. Each panel corresponds to one of the EoS shown in \fig{fig:EoS-family}. The different colors denote different types of self-similar solutions, and the labels on the curves indicate the corresponding value of the transition strength, $\alpha_N$.
    }
    \label{fig: kinetic energy fraction}
\end{figure}

We show the kinetic energy fraction for the bubble solutions of \fig{fig:alpha_N vs xi_w} in \fig{fig: kinetic energy fraction}. Qualitatively, the curves resemble those obtained in the bag model \cite{Espinosa2010}, shown in the top-left panel. The main difference is that, in the present case, hydrodynamic obstructions can remove some of the solutions that would otherwise be the most efficient at converting energy into bulk fluid motion.

For example, consider a theory characterized by an EoS of the type shown in Figs.~\ref{fig:EoS-c} or \ref{fig:EoS-d}, and by a transition strength $\alpha_N=0.1$. In this case, the gap between hybrids and detonations visible in Figs.~\ref{fig:alpha-c} and \ref{fig:alpha-d} translates into a reduced kinetic energy fraction. Indeed, if faster hybrids were available at this value of $\alpha_N$, the kinetic energy fraction would continue to grow towards the would-be maximum of the green curve with $\alpha_N=0.1$ in \fig{fig:eff-f}. Since these solutions are obstructed, this maximum cannot be reached.

The suppression is even more pronounced in cases where both hybrids and detonations are completely excluded, as in the case of the EoS of 
\fig{fig:EoS-f}. Then the largest achievable kinetic energy fraction is set by deflagrations alone and is therefore significantly reduced. 

Overall, hydrodynamic obstructions affect the kinetic energy fraction primarily in an indirect way: they remove regions of parameter space that would otherwise contain the most efficient gravitational-wave sources. We also emphasize that the plots in \fig{fig: kinetic energy fraction} do not impose the second law of thermodynamics. Including this constraint would further restrict the allowed solutions and typically enlarge the gap between the hybrid and detonation regimes.

\section{Discussion}
\label{sec: section 7}

In this work we have explored the implications of non-conformal equations of state (EoS) for the expansion of isolated bubbles in first-order phase transitions. The EoS considered here interpolate between two physically well-motivated limits: the bag-model limit, commonly used to describe electroweak-like transitions, and the QCD-like limit, characterized by a strongly varying speed of sound and finite ranges of supercooling and superheating. QCD-like theories are relevant for phase transitions in neutron star mergers and core-collapse supernovae, and may also arise in the early universe in beyond-the-Standard-Model scenarios, for example in dark sectors. Since the gravitational-wave signal from first-order phase transitions may provide one of the few observational probes of such theories, understanding bubble dynamics beyond the bag-model approximation is essential.

Our main result is the identification of a new set of hydrodynamic obstructions to single-bubble expansion that arise when the EoS departs from the bag-model form \cite{Espinosa2010}. These obstructions can be divided into two classes. The first are \textit{wall obstructions}, which arise from the matching conditions at the bubble wall. They depend on whether the states required on the two sides of the wall exist within the allowed branches of the EoS, and are therefore sensitive to features such as the finite range of supercooling or superheating and the values of the speed of sound along those branches. The second are \textit{flow obstructions}, which arise when the rarefaction wave cannot be continued smoothly through the fluid. Although these are global obstructions to constructing a complete flow profile, their onset is controlled by a local property of the EoS: the behaviour of the speed of sound near the point where the rarefaction becomes critical, in particular the value of $c_s(e)$ and its derivative with respect to the energy density.

By considering a family of EoS interpolating between the bag-model limit, holographic models such as \cite{Bea2018}, and QCD-like theories such as pure $SU(3)$ YM theory \cite{Lucini:2023irm}, we have shown how these obstructions reshape the space of admissible solutions in the $(\xi_w,\alpha_N)$ plane. This is illustrated in Fig.~\ref{fig:alpha_N vs xi_w}. As the EoS departs from the bag model, the regions corresponding to deflagrations, hybrids and detonations no longer connect in the standard way. Instead, gaps can open between different branches of solutions, most notably between hybrids and detonations.

An important implication is that the wall velocity cannot be treated as a freely adjustable parameter. Although determining the precise relation between $\xi_w$ and $\alpha_N$ requires microscopic input, hydrodynamics alone already constrains the range of admissible wall velocities. Combining these hydrodynamic constraints with existing methods for estimating the wall velocity \cite{Ai:2023see,Bea:2021zsu,Sanchez-Garitaonandia:2023zqz,AiLaurentVanDeVis:bounds,AiNagelsVanvlasselaer,Krajewski:2023clt,LiWangYuwen:strongcoupling} may therefore lead to robust bounds on $\xi_w$ without requiring a complete microscopic computation. This is particularly relevant for QCD-like theories, for which the allowed wall velocities are found to be rather small, in qualitative agreement with estimates based on other approaches \cite{BeaGilibertiMateos:QCDlike,Kang:2024xqk,ClineLaurent:QCD}.

These obstructions also affect the expected gravitational-wave signal. Using the kinetic energy fraction as a proxy for the emission strength, while keeping in mind the limitations of this approximation for deflagrations and hybrids \cite{Cutting:2019zws}, we find that hydrodynamic gaps can suppress the maximum kinetic energy fraction attainable at fixed transition strength. In particular, when the most efficient hybrid solutions are removed by a flow obstruction, the system cannot access the configurations that would otherwise dominate the kinetic energy budget. In more extreme cases, where both hybrids and detonations are excluded, the maximum kinetic energy fraction is set by deflagrations alone and is significantly reduced.

There are several natural extensions of this work. First, it would be interesting to apply the same analysis to superheated, or inverse, transitions \cite{Barni:2024lkj,Bea:2024bxu}. Similar obstructions are expected to arise, although their impact on the space of solutions may differ. Second, one should include a conserved number density, which would bring the setup closer to the conditions relevant for matter in neutron stars. For a constant speed of sound, the dynamics of such a conserved current decouples and no new qualitative features appear \cite{Bea:2024bxu}. When the speed of sound varies with temperature, however, the dynamics of the stress tensor and of the charge current become fully coupled, potentially leading to new phenomena relevant for first-order phase transitions in neutron star mergers \cite{Casalderrey-Solana:2022rrn}.

Finally, it would be important to understand how the relation between the kinetic energy fraction and the gravitational-wave amplitude is modified for non-conformal EoS \cite{Caprini:2019egz}. For EoS that differ substantially from the bag model, the changes may be not only quantitative but also qualitative. This could affect predictions for gravitational waves from phase transitions in neutron star mergers and core-collapse supernovae \cite{Casalderrey-Solana:2022rrn,Bleau:2026ala}.

\section*{Acknowledgments}
We thank Carlos Hoyos, Benoit Laurent, Miguel Vanvlasselaer and Jorinde van de Vis for discussions. We are especially grateful to Giulio Barni for extensive discussions and suggesting the possibility that more solutions may exist between hybrids and detonations, which led us to the discovery of shocked detonations. We thank  Ed Bennett, Biagio Lucini, Maurizio Piai and Davide Vadacchino for valuable discussions concerning the lattice data employed here, and Ed Bennett for his assistance in accessing and reading the data. We acknowledge financial support from the ``Center of Excellence Maria de Maeztu 2025--2029'' award to the \mbox{ICCUB}, grant CEX2024-001451-M, funded by AEI/10.13039/501100011033. PT is supported by the project ``Dark Energy and the Origin of the Universe'' (PRE2022-102220), funded by MCIN/AEI/10.13039/501100011033, by grant 
No.~00017375 from the Simons Foundation, and by project PID2022-141125NB-I00 from the Spanish Ministry of Science, Innovation and Universities. DM and PT acknowledge financial support from Grant No.~PID2022-136224NB-C22 from the Spanish Ministry of Science, Innovation and Universities, and from Grant No.~2021-SGR-872 funded by the Catalan Government.
This research is also funded by the European Union (ERC, HoloGW, Grant Agreement 
No.~101141909). Views and opinions expressed are, however, those of the authors only and do not necessarily reflect those of the European Union or the European Research Council. Neither the European Union nor the granting authority can be held responsible for them.
MSG acknowledges support from a Royal Society - Research Ireland University Research Fellowship via grants URF/R1/211027 and 
RF/ERE/231191.

\section*{AI and LLMs usage disclaimer}
Standard AI programming tools were used during the development of the code that yielded the results shown in this paper. These were used to increase efficiency in tasks such as debugging and modifying existing code. Large language models were used to assist in editing and polishing the manuscript, but they did not contribute to the scientific content or results.

\appendix

\section{From Electroweak- to QCD-like theories in holography}\label{app: backreaction}
In this appendix we explain how to interpolate continuously between an electroweak-like and a QCD-like EoS, the latter being similar to those obtained in strongly coupled holographic models. As discussed around Eq.~\eqref{eq: Radiation + potential EoS}, this interpolation can be achieved by varying the effective number of degrees of freedom, $\geff$, that do not participate in the phase transition.

When $\geff\gg 1$, most of the degrees of freedom remain spectators during the transition. Their radiation contribution dominates the thermodynamics, and the resulting EoS is bag-model-like: the two stable branches have an approximately constant speed of sound, $c_s^2\simeq 1/3$. By contrast, when $\geff\sim 1$, the number of spectator degrees of freedom is comparable to the number of degrees of freedom involved in the transition. The transition dynamics then has a much stronger impact on the thermodynamics, and the resulting EoS exhibits a speed of sound that can deviate significantly from its conformal value as the temperature is varied.

In holography, this interpolation can be realized by varying the backreaction of the matter fields on the geometry. We quantify the amount of backreaction by a parameter $\chi$, defined so that the five-dimensional Einstein equations take the schematic form
\begin{equation*}
    G_{\mu\nu} = \chi\, T_{\mu\nu} \,,
\end{equation*}
where, for convenience, we have fixed $\kappa \equiv 8\pi G_5=\sqrt{2}$.

A prototypical realization of this idea is provided by gauge theories with a large number of colours, $N_c$, and a much smaller number of flavours, $N_f$. In the dual description, the flavours are introduced by $N_f$ D-branes embedded in a gravitational background whose action is of order $N_c^2$ \cite{Karch:2002sh}. The contribution of the branes is instead of order $N_cN_f$, so their relative backreaction is controlled by the ratio $\chi\sim N_f/N_c$. In the limit $\chi=0$, the branes are probes of the geometry. For small but non-zero $\chi$, their backreaction is weak and can be treated perturbatively \cite{Mateos:2006yd}.

In this appendix, we analyze how varying the backreaction parameter $\chi$ affects the thermodynamics of the dual gauge theory in a particular holographic model. We consider classical five-dimensional Einstein gravity coupled to a scalar field:
\begin{equation}
\label{eq: EFE and KG action}
    S
    =
    \int d^5x \sqrt{-g}
    \left[
        \frac{1}{4}R
        - \frac{\chi}{2}\partial_\mu\phi\,\partial^\mu\phi
        - \chi V(\phi)
        - \frac{1}{2}\Lambda
    \right],
\end{equation}
where $\Lambda$ is the cosmological constant, related to the AdS radius by $\Lambda=-6/L^2$. From now on we set $L=1$. For the scalar potential we use the model of \cite{Bea2018}, defined through the superpotential
\begin{equation}
\label{eq: superpotential}
    W(\phi)
    =
    -\frac{3}{2}
    -\frac{\phi^2}{2}
    -\frac{\phi^4}{4\phi_M^2}
    +\frac{\phi^6}{\phi_Q}\,,
\end{equation}
in terms of which
\begin{equation}
\label{eq: holo potential}
    V(\phi)
    =
    -\frac{4}{3}W(\phi)^2
    +\frac{1}{2}W'(\phi)^2
    -\frac{\Lambda}{2}\,.
\end{equation}
The first two terms already contain the vacuum energy supporting the AdS$_5$ geometry, which is why we subtract it explicitly. In
\eqref{eq: EFE and KG action} the cosmological constant is kept as a separate term, so that $\chi$ multiplies only the genuinely scalar part of the potential and the limit $\chi\to0$ recovers pure AdS$_5$. Here $\phi_M$ and $\phi_Q$ are free parameters that can be tuned to obtain a crossover, or a first- or second-order phase transition. In the following we set $\phi_Q=10$ and $\phi_M=0.88$.

Following \cite{DeWolfe2010}, we take the metric and scalar field to have the form
\begin{equation*}
    ds^2
    =
    \frac{1}{r^2}
    \left(
        -f(r)g(r)\,dt^2
        + \frac{dr^2}{f(r)}
        + g(r)\,d\bm{x}^2
    \right),
    \qquad
    \phi=\phi(r) .
\end{equation*}
In these coordinates, the Einstein--Klein--Gordon equations reduce to the following system of ordinary differential equations:
\begin{equation*}
    \begin{split}
        &2 r f'(r) g'(r)
        + g(r)\left(r f''(r)-3 f'(r)\right)=0,\\
        &3 r g'(r)^2
        -3 g(r)\left(r g''(r)+g'(r)\right)
        -4 r \chi g(r)^2 \phi'(r)^2=0,\\
        &r^2 f(r) \phi''(r)
        + r^2 f'(r) \phi'(r)
        + \frac{2 r^2 f(r) g'(r)\phi'(r)}{g(r)}
        -3 r f(r)\phi'(r)
        -V'(\phi)=0,\\
        &\frac{3 f'(r) g'(r)}{4 f(r)g(r)}
        -\frac{3 f'(r)}{2 r f(r)}
        +\frac{2\chi V(\phi(r))}{r^2 f(r)}
        -\frac{6}{r^2 f(r)}
        +\frac{3 g'(r)^2}{2 g(r)^2}
        -\frac{6 g'(r)}{r g(r)}
        +\frac{6}{r^2}
        -\chi \phi'(r)^2=0 .
    \end{split}
\end{equation*}

We solve this system by fixing the value of the scalar field at the horizon, $\phi_h$. The remaining near-horizon data for $f$, $g$ and $\phi$ are then fixed by regularity. Varying $\phi_h$ generates a family of static black-brane solutions. For each solution, the entropy density and temperature are given by
\begin{equation*}
    s
    =
    \frac{\pi g(r_H)^{3/2}}{r_H^{3/2}\phi_A^3},
    \qquad
    T
    =
    \frac{|f'(r_H)|\sqrt{g(r_H)}}{4\pi\phi_A},
\end{equation*}
where $r_H$ is the position of the horizon, which we set to $r_H=1$.

The quantity $\phi_A$ is the source of the operator dual to the scalar field in the boundary theory. It is extracted from the near-boundary expansion of the bulk fields. Solving the equations order by order as $r\to 0$, one finds
\begin{equation*}
    \begin{split}
        g(r)
        &=
        1-\frac{1}{3}r^2\chi\phi_A^2
        +\mathcal{O}(r^4),\\
        f(r)
        &=
        1+\mathcal{O}(r^4),\\
        \phi(r)
        &=
        r\phi_A
        +\frac{1}{3}r^3
        \left[
            3\phi_B
            +2(\chi-1)\phi_A^3\log r
        \right]
        +\mathcal{O}(r^4).
    \end{split}
\end{equation*}
Here $\phi_A$ and $\phi_B$ are related, respectively, to the source and expectation value of the operator dual to the scalar field.

\begin{figure}[ht!]
    \centering
    \includegraphics[width=0.49\textwidth]{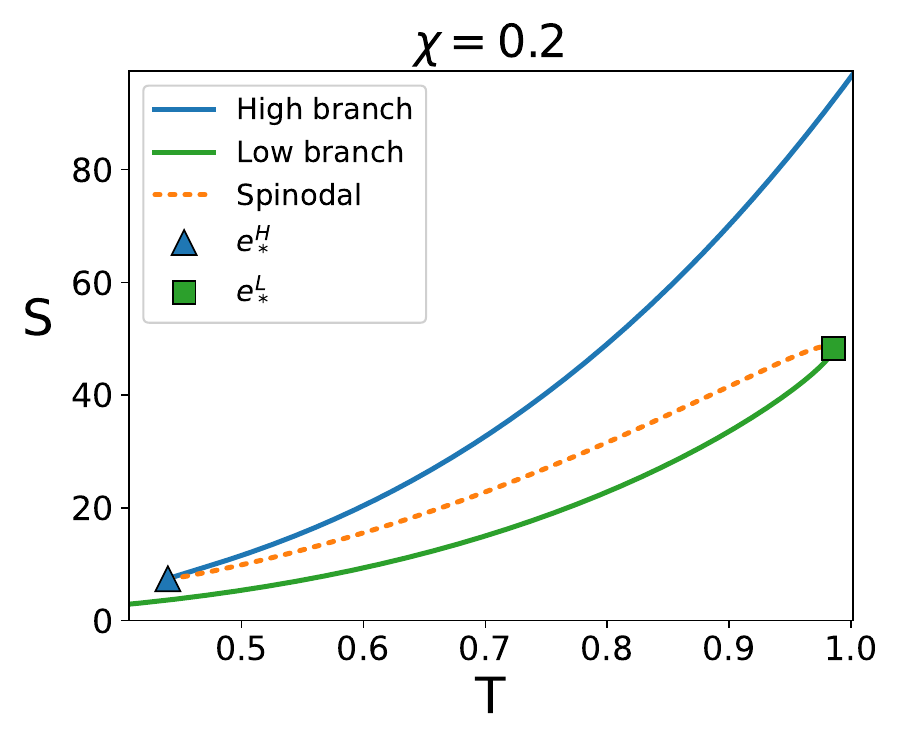}
    \includegraphics[width=0.49\textwidth]{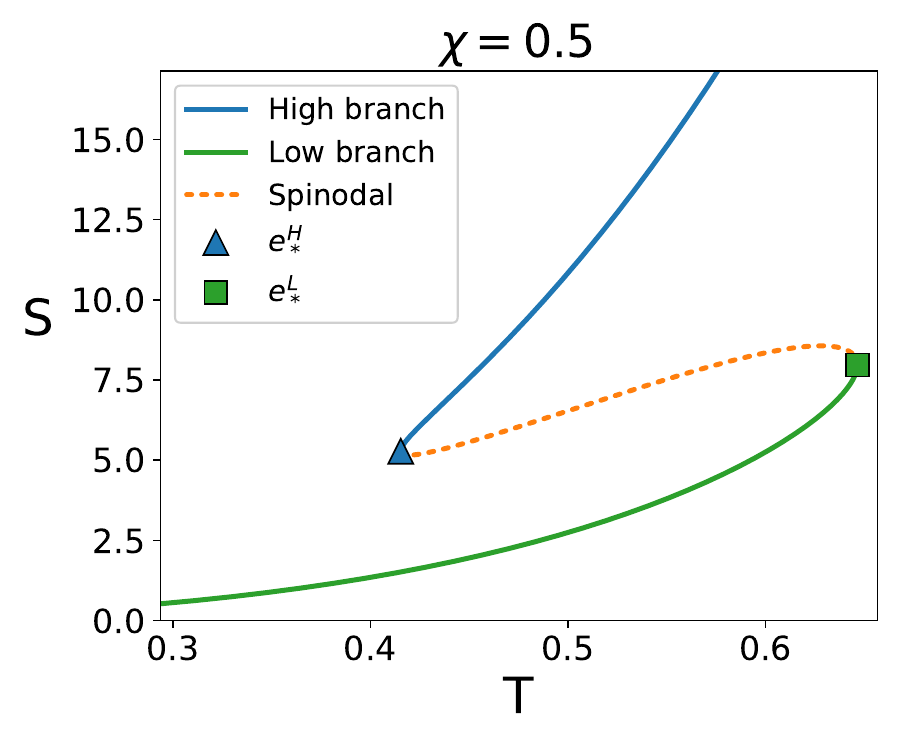}\\
    \includegraphics[width=0.49\textwidth]{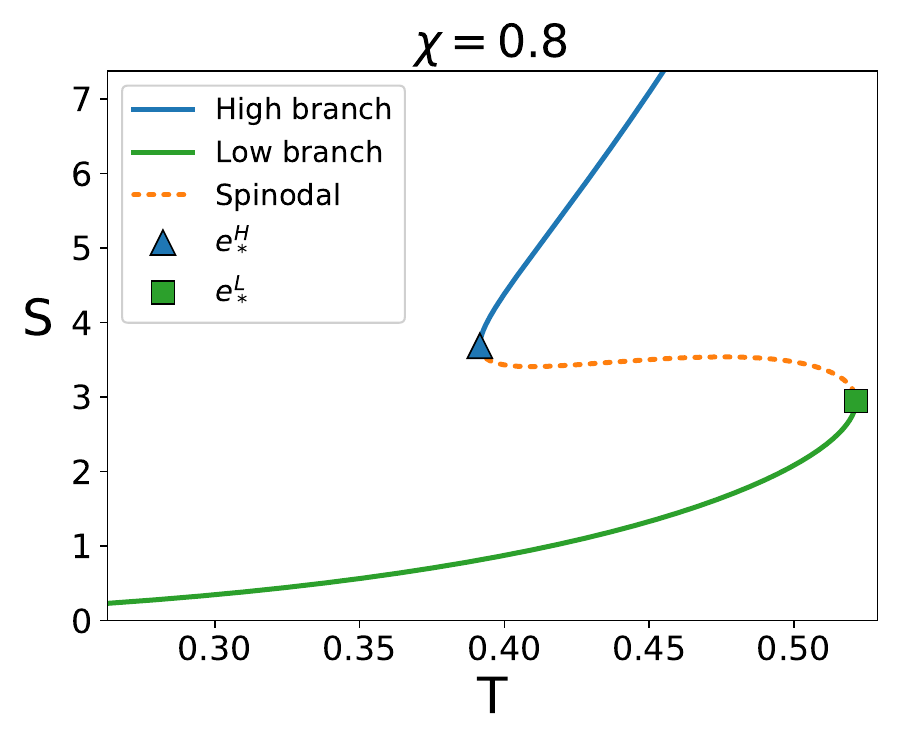}
    \includegraphics[width=0.49\textwidth]{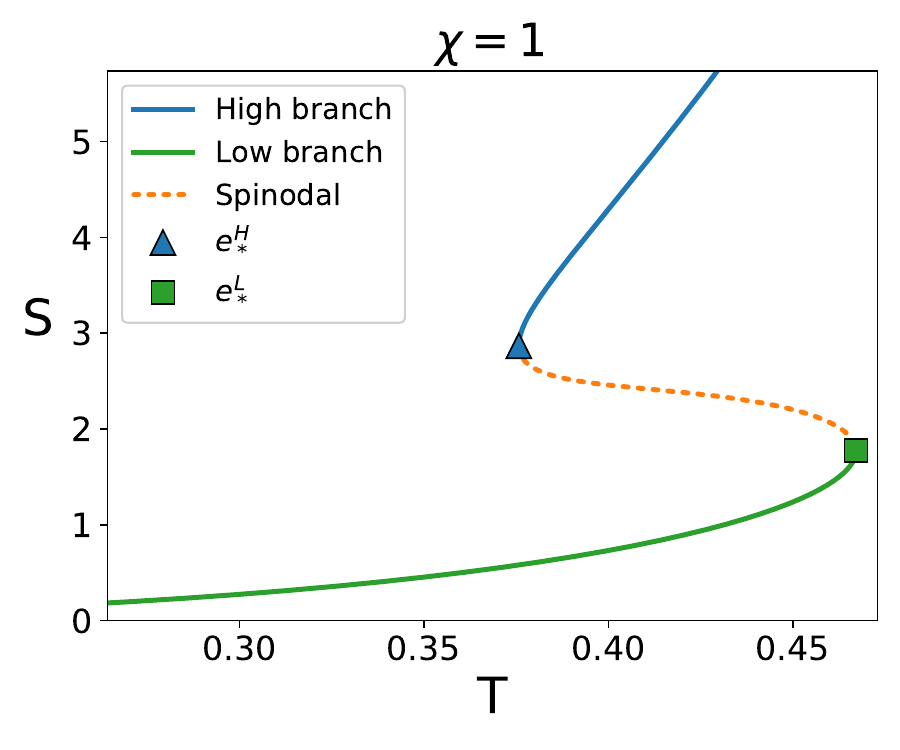}\\
    \includegraphics[width=0.49\textwidth]{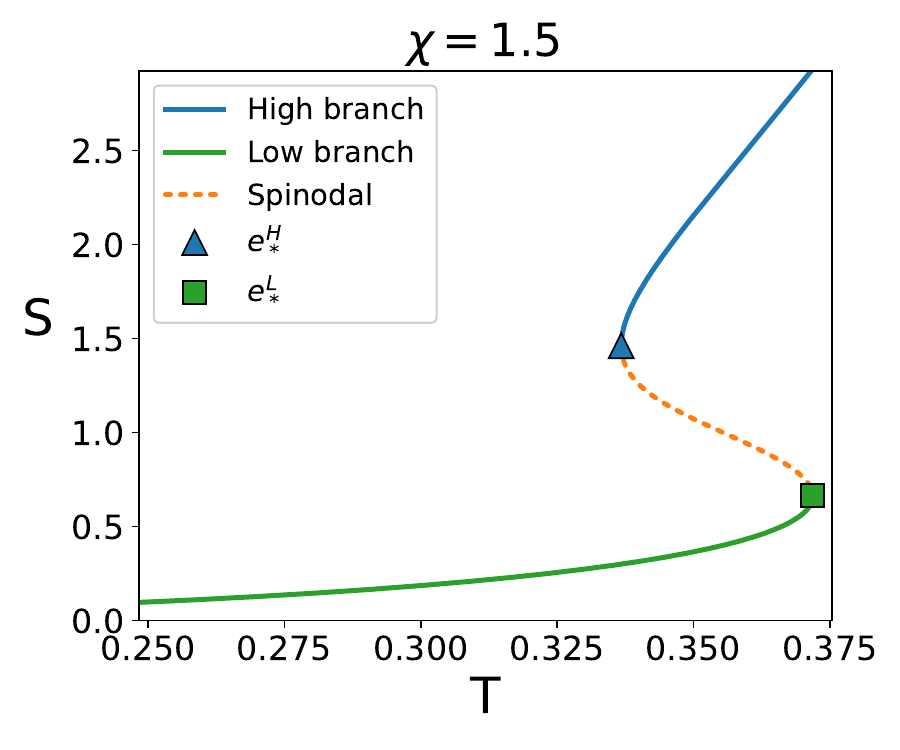}
    \includegraphics[width=0.49\textwidth]{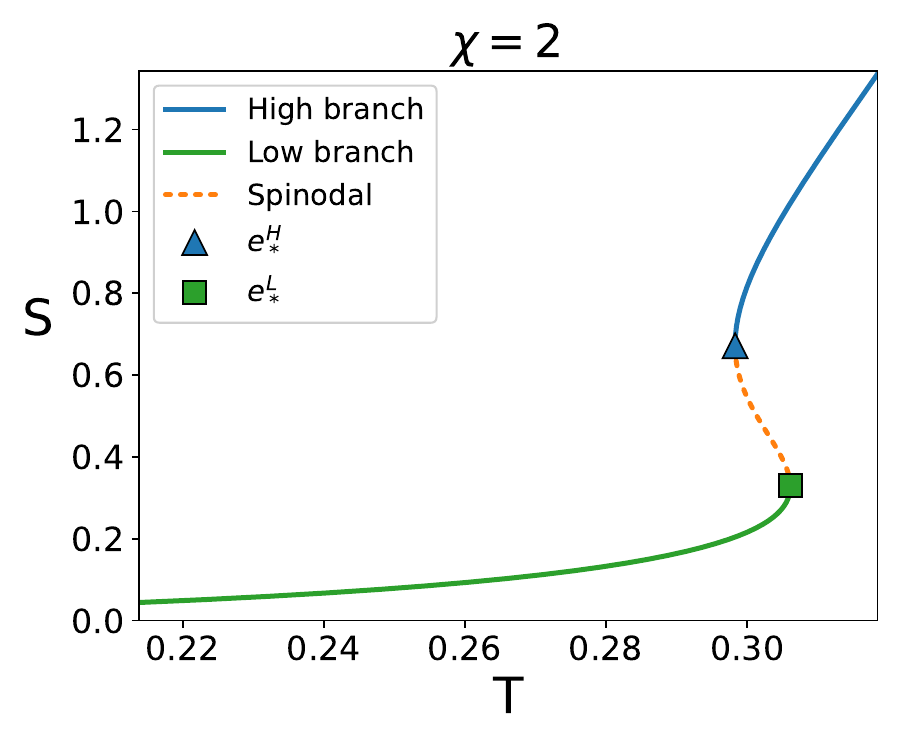}
    \caption{Entropy density as a function of temperature for several values of the backreaction parameter $\chi$ in the model of \cite{Bea2018}, with $\phi_M=0.88$ and $\phi_Q=10$.
    }
    \label{fig:diff_backreaction}
\end{figure}

We now compute the entropy density and temperature for different values of $\chi$. In \fig{fig:diff_backreaction} we show $s(T)$ for six representative values of the backreaction parameter. As $\chi$ is decreased, the matter sector backreacts less strongly on the geometry, and the thermodynamics approaches the probe limit. Correspondingly, the entropy curve becomes increasingly similar to that of a bag-model-like EoS, with branches that are closer to conformal behaviour.

\section{Code explanation: \texttt{SNOBEX}}\label{app: snobex}

In this appendix we present \texttt{SNOBEX} (\textbf{S}elf-similar \textbf{N}on-conformal \textbf{O}bstructions to \textbf{B}ubble \textbf{EX}pansion), the numerical tool used throughout this work to explore self-similar solutions in cosmological phase transitions with a general EoS.

\texttt{SNOBEX} is written in Python 3 and depends only on the standard scientific packages (\texttt{NumPy}, \texttt{SciPy}, \texttt{Matplotlib}). The source code is organized as an importable package and is publicly available at the following GitHub repository: \href{https://github.com/pedrota2000/SNOBEX}{\texttt{SNOBEX}}. The package can be installed by cloning the repository and installing the required Python dependencies.

The package is designed in a highly modular fashion, consisting of five modules, each encapsulating a distinct part of the computation:

\begin{itemize}
  \item \texttt{snobex.eos}: defines the analytic EoS for the high- and low-energy branches, $p_H(e)$ and $p_L(e)$ (implemented as \texttt{pplus} and \texttt{pminus}), as given in expressions \eqref{eq: EoS-p}. In addition, the module can bypass the analytic parametrization entirely and load a tabulated EoS from a plain-text file through \texttt{load\_custom\_eos}: given columns $(e, p)$ it builds spline interpolants for $p(e)$ and, if a third column is absent, derives the speed of sound $c_s^2(e)=\mathrm{d}p/\mathrm{d}e$ numerically via \texttt{speed\_of\_sound\_squared}.
  \item \texttt{snobex.thermodynamics}: constructs the entropy density $s(e)$, temperature $T$, and free energy $\mathcal{F}=-p$ for a given EoS and constants $s_*^H$ and $s_*^L$, by numerically solving the differential equation \eqref{eq: s(e)-ODE}. This module also computes the critical temperature $T_c$, defined as the temperature at which the free energies of the high- and low-energy branches coincide, $\mathcal{F}_L(T_c) = \mathcal{F}_H(T_c)$.
  \item \texttt{snobex.hydrodynamics}: implements the conservation of the fluid energy-momentum tensor in self-similar coordinates \eqref{eq:self-similar-eoms}, following the formulation of \cite{Espinosa2010}, which avoids divergences during numerical integration. The final form of the equations is
    \begin{equation}\label{eq:flow_tau}
        \begin{split}
          \frac{dv}{d\tau} \;&=\; 2\,c_s^2(e)\,v\,(1-v^2)\,(1-\xi v),\\[4pt]
          \frac{d\xi}{d\tau} \;&=\; \xi\,\bigl[(\xi-v)^2 - c_s^2(e)\,(1-\xi v)^2\bigr],\\[4pt]
          \frac{de}{d\tau} \;&=\; 2\,v\,(\xi-v)\,\bigl(e + p(e)\bigr).
        \end{split}
    \end{equation}
    This module also handles the junction conditions \eqref{eq:matching-usual} and includes functions for computing deflagration, detonation, hybrid, and shocked-detonation flows:
    \begin{itemize}
        \item \texttt{find\_def}: integrates deflagration flows for a given center energy $e_C$ and bubble wall velocity $\xi_w$. First, the energy is fixed at the center of the bubble. Then, the JCs are solved at the wall position given by $\xi_w$. The values right in front of the wall are then used as initial conditions for the flow integration, following the ODEs \eqref{eq:flow_tau}. This flow is integrated until a point where $\partial_\xi v \to \infty$ is reached. The code then searches for a point along the flow where the junction conditions (JCs) allow a transition to a rest state in the high-energy branch. If this position is found, then a deflagration solution exists for the given pair of values $e_C$ and $\xi_w$. The JCs are solved using \texttt{scipy} built-in function \texttt{least\_squares}. The flow is integrated using \texttt{scipy.solve\_ivp} functionality.
        \item \texttt{find\_deto}: this function integrates detonation flows for a given bubble wall velocity $\xi_w$ and nucleation energy $e_N$. To perform the integration, the function first solves the JC at the bubble wall position (eq. \eqref{eq:matching-usual}) using the built-in function \texttt{scipy.least\_squares}. This is done by initializing the system at the nucleation energy $e_N$. Then, the point right behind the wall is used as initial condition to integrate the flow equations \eqref{eq:flow_tau}. The flow is integrated using \texttt{scipy.solve\_ivp} functionality. The solution is accepted if the flow smoothly reaches $v \to 0$, and the resulting energy at the center of the bubble $e_C$ is returned.
        \item \texttt{find\_hyb}: this function computes the corresponding hybrid self-similar solution for a given bubble wall velocity $\xi_w$, and energy density at the Jouguet point $e_{JG}$. The code first constructs the rarefaction wave immediately behind the wall by integrating the flow equations \eqref{eq:flow_tau} using \texttt{scipy.solve\_ivp}. This rarefaction wave is discarded if it does not reach $v = 0$ smoothly. Then, this function solves the JCs \eqref{eq:matching-usual} and gets the state right after the wall. This is done by using the function \texttt{scipy.least\_squares}. The point right after the wall is then used as initial condition to integrate the flow. Finally, the code checks whether it is possible to introduce a shock along this flow to reach a rest state outside of the bubble. If it succeeds, the solution is allowed as a valid hybrid.
        \item \texttt{find\_shocked\_detonation\_jouguet}: this function computes the shocked-detonation solution for a given nucleation energy $e_N$ and bubble wall velocity $\xi_w$, i.e.\ a detonation whose trailing rarefaction is terminated by a second shock inside the low-energy fluid, producing a profile with two discontinuities. The wall must be super-sonic, $\xi_w \geq c_s^-(e_N)$; otherwise no solution is returned. The pipeline is as follows. First, the junction conditions \eqref{eq:matching-usual} are solved at the wall, initialized at the nucleation energy $e_N$ with $v_+ = -\xi_w$, to obtain the low-energy state $(v_-, e_-)$ right behind it, using \texttt{scipy.least\_squares}. The flow equations \eqref{eq:flow_tau} are then integrated from this point (using \texttt{scipy.solve\_ivp}) until the profile becomes multi-valued, i.e.\ $\partial_\xi v \to \infty$; only the injective portion is retained and interpolated to yield $v(\xi)$ and $e(\xi)$. Along this portion the code first attempts a shock that jumps directly to a state at rest ($v_- = -\xi_{sh}$), solving the corresponding junction conditions with \texttt{scipy.least\_squares}. If such a shock is not found, it instead solves the junction conditions supplemented with the Jouguet condition on the downstream state, $v_- = -c_s^-(e_-)$, so that the fluid leaves the shock exactly at the local sound speed. This closes the system and selects a single shock position $\xi_{sh}$, from which the remaining rarefaction is integrated down to $v = 0$. Since both discontinuities lie within the low-energy fluid, all junction conditions are evaluated with the low-energy branch on either side. Imposing the Jouguet condition yields the unique, physically-selected shocked detonation which maximizes entropy production at the shock position.
    \end{itemize}

  \item \texttt{snobex.separators}: computes the limiting contours that constrain solution construction. There are four types:
  \begin{itemize}
      \item Wall obstructions: arise from limiting conditions at the bubble wall. Functions \texttt{compute\_def\_separator} and \texttt{compute\_det\_separator} handle deflagrations and detonations, respectively.
      \item Flow obstructions: occur when flow profiles become multi-valued due to a non-constant speed of sound. These are computed with \texttt{compute\_hyb\_separator} and \texttt{limiting\_detonation\_contour\_finder}.
      \item Entropy obstructions: arise from imposing entropy production ($\Delta J_s \geq 0$) across the discontinuities. These are computed with \texttt{compute\_entropy\_separator\_det} and \texttt{compute\_entropy\_separator\_def}, while \texttt{entropy\_checker} verifies the entropy condition at every discontinuity of a given solution.
      \item Jouguet velocity: strictly speaking this curve is not an obstruction, but it separates hybrids and detonations. This is implemented by the following function \texttt{jouguet\_detonation\_contour\_finder}.
  \end{itemize}
  This module also provides the utilities that convert nucleation energies to the strength factor $\alpha_N$ and trace constant-$e_C$/$e_N$ contours in phase space. These computations are the most computationally expensive part of the code, as they require solving a large number of flow configurations.
  \item \texttt{snobex.plotting}: assembles interactive phase-space figures, manages progress bars, and handles click-to-inspect callbacks. Standard plots include nucleation energy $e_N$, bubble center energy $e_C$, and strength factor $\alpha_N$ versus bubble wall velocity $\xi_w$. Clicking on a point in phase space displays the corresponding velocity and energy profiles as functions of $\xi$.
\end{itemize}
Two self-contained entry points are provided: \texttt{main\_eos\_explorer.py}, which launches the interactive EoS GUI, and \texttt{main\_bubble\_solver.py}, which runs the full phase-space computation from a parameter file (and optionally from a tabulated custom EoS).

\bibliographystyle{JHEP}

\providecommand{\href}[2]{#2}\begingroup\raggedright\endgroup

\end{document}